\documentclass[twocolumn]{aastex6}
\usepackage{amsmath}
\usepackage{tensor}
\usepackage{graphicx}
\usepackage{bm}
\usepackage{tabularx}
\usepackage{booktabs}
\usepackage[hang,flushmargin]{footmisc}
\usepackage{array}

\newcommand{\adsurl}[1]{\href{#1}{ADS}}
\providecommand{\url}[1]{\href{#1}{#1}}
\newcommand{\ra}[1]{\renewcommand{\arraystretch}{#1}}
\newcommand{\ee}[1]{\mbox{${} \times 10^{#1}$}}

\newcommand{\degree}{\mbox{$^{\circ}$}}

\newcommand{\co}{\mbox{$\rm CO$}}
\newcommand{\thirteenco}{\mbox{$\rm ^{13}CO$}}
\newcommand{\water}{\mbox{$\rm H_{2}O$}}

\newcommand{\ammonia}{\mbox{{\rm NH}$_3$}}

\newcommand{\owater}{\mbox{o-$\rm H_{2}O$}}
\newcommand{\pwater}{\mbox{p-$\rm H_{2}O$}}
\newcommand{\coul}[2]{\mbox{${\rm CO}~J\rm =#1 \rightarrow#2$}}
\newcommand{\tcoul}[2]{\mbox{${\rm ^{13}CO}~J\rm =#1 \rightarrow#2$}}

\newcommand{\ohul}[3]{\mbox{$\rm ~^{2}\Pi_{#1,~#2}^{#3}$}}

\newcommand{\oi}{\mbox{$\rm [O~\textsc{i}]$}}

\newcommand{\oiul}[2]{\mbox{$\rm [O~\textsc{i}]~\tensor*[^{3}]{P}{_{#1}} \rightarrow\tensor*[^{3}]{P}{_{#2}}$}}
\newcommand{\ciul}[2]{\mbox{$\rm [C~\textsc{i}]~\tensor*[^{3}]{P}{_{#1}} \rightarrow\tensor*[^{3}]{P}{_{#2}}$}}
\newcommand{\hy}{\textsc{hyperion}}

\newcommand{\herschel}{{\it Herschel}}
\newcommand{\spitzer}{{\it Spitzer}}
\newcommand{\kms}{\mbox{$\rm km~s^{-1}$}}
\newcommand{\lsun}{\mbox{$\rm L_{\odot}$}}
\newcommand{\msun}{\mbox{$\rm M_{\odot}$}}
\newcommand{\tcol}{\mbox{$t_{\rm col}$}}
\newcommand{\lstar}{\mbox{$L_{\star}$}}
\newcommand{\rstar}{\mbox{$R_{\star}$}}
\newcommand{\tstar}{\mbox{$T_{\star}$}}
\newcommand{\mstar}{\mbox{$M_{\star}$}}
\newcommand{\rinf}{\mbox{$R_{\rm inf}$}}
\newcommand{\rcen}{\mbox{$R_{\rm c}$}}
\newcommand{\trot}{\mbox{$T_{\rm rot}$}}
\newcommand{\msunyr}{\mbox{M$_\odot$ yr$^{-1}$}}
\newcommand{\csef}{\mbox{$c_{\rm s,eff}$}}

\usepackage{natbib}
\usepackage{etoolbox}

\makeatletter
\patchcmd{\NAT@citex}
  {\@citea\NAT@hyper@{%
     \NAT@nmfmt{\NAT@nm}%
\hyper@natlinkbreak{\NAT@aysep\NAT@spacechar}{\@citeb\@extra@b@citeb}%
     \NAT@date}}
  {\@citea\NAT@nmfmt{\NAT@nm}%
   \NAT@aysep\NAT@spacechar\NAT@hyper@{\NAT@date}}{}{}

\patchcmd{\NAT@citex}
  {\@citea\NAT@hyper@{%
     \NAT@nmfmt{\NAT@nm}%
\hyper@natlinkbreak{\NAT@spacechar\NAT@@open\if*#1*\else#1\NAT@spacechar\fi}%
       {\@citeb\@extra@b@citeb}%
     \NAT@date}}
  {\@citea\NAT@nmfmt{\NAT@nm}%
\NAT@spacechar\NAT@@open\if*#1*\else#1\NAT@spacechar\fi\NAT@hyper@{\NAT@date}}
  {}{}
\makeatother

\newcolumntype{L}[1]{>{\raggedright\let\newline\\\arraybackslash\hspace{0pt}}m{#1}}
\newcolumntype{C}[1]{>{\centering\let\newline\\\arraybackslash\hspace{0pt}}m{#1}}
\newcolumntype{R}[1]{>{\raggedleft\let\newline\\\arraybackslash\hspace{0pt}}m{#1}}

\begin{document}
\title{The Class 0 Protostar BHR71: {\it Herschel} Observations and Dust Continuum Models}
\author{Yao-Lun Yang\altaffilmark{1}, Neal J. Evans II\altaffilmark{1,2}, Joel D. Green\altaffilmark{1,3}, Michael M. Dunham\altaffilmark{4,5}, and Jes K. J{\o}rgensen\altaffilmark{6}}

\affil{1.  Department of Astronomy, The University of Texas at Austin, Austin, TX 78712, USA \\
       2.  Korea Astronomy and Space Science Institute, 776 Daedeokdae-ro, Yuseong-gu, Daejeon, 34055 \\
       3.  Space Telescope Science Institute, Baltimore, MD 21218, USA \\
       4. Department of Physics, State University of New York at Fredonia, 280 Central Ave, Fredonia, NY 14063, USA \\
       5.  Harvard-Smithsonian Center for Astrophysics, Cambridge, MA 02138, USA \\
       6.  Centre for Star and Planet Formation, Niels Bohr Institute \&\ Natural History Museum of Denmark, University of Copenhagen, {\O}ster Voldgade 5–7, DK-1350 Copenhagen K., Denmark}
\setcounter{footnote}{0}

\begin{abstract}
We use \herschel\ spectrophotometry of BHR71, an embedded Class 0 protostar, to provide new constraints on its physical properties.  We detect 645 (non-unique) spectral lines amongst all spatial pixels.  At least 61 different spectral lines originate from the central region.  A \co\ rotational diagram analysis shows four excitation temperature components, 43~K, 197~K, 397~K, and 1057~K. Low-$J$ \co\ lines trace the outflow while the high-$J$ \co\ lines are centered on the infrared source.  The low-excitation emission lines of \water\ trace the large-scale outflow, while the high-excitation emission lines trace a small-scale distribution around the equatorial plane. We model the envelope structure using the dust radiative transfer code, \hy, incorporating rotational collapse, an outer static envelope, outflow cavity, and disk.  The evolution of a rotating collapsing envelope can be constrained by the far-infrared/millimeter SED along with the azimuthally-averaged radial intensity profile, and the structure of the outflow cavity plays a critical role at shorter wavelengths.  Emission at 20-40~\micron\ requires a cavity with a constant-density inner region and a power-law density outer region.  The best fit model has an envelope mass of 19~\msun\ inside a radius of 0.315 pc and a central luminosity of 18.8~\lsun.  The time since collapse began is 24630-44000 yr, most likely around 36000 yr.  The corresponding mass infall rate in the envelope (1.2\ee{-5} \msunyr) is comparable to the stellar mass accretion rate, while the mass loss rate estimated from the CO outflow is 20\%\ of the stellar mass accretion rate.  We find no evidence for episodic accretion.
\end{abstract}
\maketitle

\section{Introduction}

\subsection{Background}
Embedded protostars are the youngest protostars, emitting most of their light at far-infrared wavelengths.
After collapse begins, the protostar forms and accretes mass from the envelope,
and a wind and/or jet is formed, which sweeps up material into a molecular outflow.
The radiation from the protostar encounters a thick and dense envelope on its way out from the center.  Therefore, embedded protostars can only be observed at infrared and submillimeter wavelengths unless they are seen face-on.  Observationally, protostars are classified by their radiation at infrared wavelengths.  \citet{1987IAUS..115....1L} introduced the $\alpha$-index, classifying protostars into Class~I, II, and III based on the shape of the spectral energy distribution (SED) in the near-infrared to mid-infrared.  Observations at submillimeter wavelengths suggested that a subset of Class~I protostars were especially embedded, based on the ratio of the submillimeter luminosity ($\lambda>$350$\mu$m) to the bolometric luminosity, $L_{\rm submm}/L_{\rm bol}$.  \citet{1993ApJ...406..122A} defined Class~0 protostars as sources that have $L_{\rm submm}/L_{\rm bol}> 0.5$\%.
\citet{1995ApJ...445..377C} classified protostars by their bolometric temperatures: $T_{\rm bol}$ below 70~K for Class 0; 70-650~K for Class I; 651-2880~K for Class II; and above 2880~K for Class III.
The different methods of classification generally agree, but
some sources receive different classifications from different methods.

The distinction between SED classes and physical stages was emphasized by \citet{2006ApJS..167..256R}.  The detailed relation between the Classes defined by SEDs and the theoretical stages of protostellar evolution is still unclear \citep{2009A&A...498..167V}, but recent studies have shown a generally good, though imperfect, correlation between Class I sources and Stage I sources \citep{2015ApJ...806..231H, 2016A&A...586A..44C}.

The general path of star formation starts from the collapse of a dense core due to gravitational instability, forming a protostar at the center of the core.
\citet{1977ApJ...214..488S} showed that the collapse of a singular isothermal core follows a similarity solution.
This model begins with an isothermal core near hydrostatic equilibrium.  The collapse starts at the center; once the central region begins collapsing, the shell adjacent to the collapsing region loses pressure support and starts to fall in. A wave of infall propagates outward through the envelope at the sound speed. (We use the term “envelope” to describe the surrounding gas and dust that is not in the star, the disk, or the outflow.) The original model \citep{1977ApJ...214..488S} simplified the collapse process by neglecting rotation, turbulence, and magnetic fields.  Rotational motion has been widely observed in young stellar objects.  The conservation of angular momentum implies a small but non-zero rotational speed for the pre-collapse core.  \citet{1984ApJ...286..529T} (hereafter TSC) generalized the Shu model by including the effect of rotational motion, making the inner part of envelope flatten toward the equatorial plane. The angular momentum of a dense core is much larger than that of a main-sequence star, suggesting that angular momentum must be removed from the core by some mechanism, most likely involving magnetic fields.  A large magnetic field can suppress the fragmentation of the envelope \citep{2011ApJ...730...40P,2015MNRAS.451.3340K} or slow the
rotation \citep{2014prpl.conf..173L}.  While these effects may
be important, the TSC model used here does not consider the magnetic field.  Incorporating a magnetic field and turbulence is beyond the scope of this study.

When material falls inward, the centrifugal barrier becomes important at the centrifugal radius, where the centrifugal force is comparable to the gravitational force.  Bipolar outflows, which are widely observed around objects that have accretion activities, remove angular momentum from the system by ejecting material along the rotational axis at high velocity.  The outflow process involves a highly collimated atomic or molecular jet or wind ($v\sim 300$ \kms) and a less collimated molecular outflow ($v\sim$1-30 \kms) consisting of shells of gas swept up  by the high velocity jet \citep{1985Icar...61...36H}.
The molecular outflow in Class 0 protostars sweeps up the envelope and carves out an outflow cavity.  The low density in the outflow cavity makes the cavity wall more exposed to the ultraviolet radiation from the central protostar, producing numerous molecular and atomic fine-structure lines observed by \herschel\ Space Observatory \citep{2012A&A...537A..98V,2012A&A...538A..45S,2013A&A...549A..16N,2013ApJS..209....4L}.
Modeling the energetics of the \co\ and water lines  shows that different types of shocks contribute to most of the emission, but a general consensus has not been reached on the detailed structure of shocks among different Class 0 objects \citep{2012A&A...537A..55V,2013A&A...552A.141K,2014A&A...572A...9K}.

Synthetic SEDs of protostars calculated with 3-D radiative transfer simulations provide a direct comparison with observations to test the models of star formation \citep{2005ApJ...627..293Y,2010ApJ...710..470D,2006ApJS..167..256R,2009ApJ...703..131O,2012ApJ...753...98O}.
\citet{2006ApJS..167..256R}  modeled protostars with a fully collapsing envelope, a flared disk, an outflow cavity with a constant dust density, and a central luminosity source, and they simulated a grid of SEDs of the protostars with a wide variety of parameters.  This grid of SEDs serves as a tool to understand the structure of individual protostars before an in-depth modeling \citep{2015ApJS..217....6J,2015AJ....149..108C}.  \citet{2005ApJ...627..293Y} developed 1-D evolutionary models assuming
steady accretion and an inside-out collapse.
\citet{2010ApJ...710..470D} used a more complicated 2-D model setup that included episodic accretion and a rotating, collapsing envelope to obtain better agreement of the evolution of low-mass protostars throughout different stages with the observed population of protostars in different Classes.  Radiative transfer calculations also play an important role in hydrodynamic simulations of star formation to realize the synthetic images and SEDs to compare with measured properties in simulations or observations \citep{2012ApJ...753...98O}.

\subsection{BHR71}
BHR71 is a Bok globule, apparently associated with the Southern Coalsack. The
distance of the Coalsack is generally taken to be 200~pc \citep{1989A&A...225..192S,1994BaltA...3..199S},
but some recent studies suggest a closer distance of 150~pc \citep{1997A&A...326.1215C,1998A&A...338..897K,2008hsf2.book..222N}.  We adopt 200~pc as the distance of BHR71 for better comparison with other studies of this source.
\ammonia\ emission was found in BHR71 by \citet{1995MNRAS.276.1052B}. \citet{1997ApJ...476..781B} (hereafter B97) discovered a strong bipolar molecular outflow, estimated an outflow dynamical age of 10000 years, and concluded that the outflow was almost in the plane of sky, with an inclination angle of $84^{\circ}$.

Observations in the \herschel\ Key Program, \textit{Dust, Ice, and Gas In Time} (DIGIT; PI: N.~Evans), show substantial emission at far-infrared wavelengths, and suggest that BHR71 is a Class 0 protostar with an estimated bolometric temperature of 47~K \citep{2013ApJ...770..123G}.
Using a 1-D dust radiative transfer model, \citet{2012A&A...542A...8K} estimated an infall radius of 3500~AU with the mass accretion rate of $\rm 3\times10^{-5}~\msun~yr^{-1}$.
Shock activity has been observed in the SiO and H$_{\rm 2}$ knots along the outflow direction suggesting the presence of non-stationary shocks ($CJ$-type) \citep{2011A&A...532A..53G,2015A&A...575A..98G}.  Mass loss rates of 2-4\ee{-7} \msunyr\ are estimated from the [O~\textsc{i}]~63~\micron\ line \citep{2015ApJ...801..121N}.

We use spectroscopy data from DIGIT and archival \spitzer\ as our primary data in this paper.  We discuss the observations and reduction in Section~\ref{sec:obs}, the results of spectra analyses in Section~\ref{sec:results}, the continuum modeling in Section~\ref{sec:rad}, the discussion of the best fit model in Section~\ref{sec:discussion}, and finally we summarize our conclusions in Section~\ref{sec:con}.  A parameter study for the continuum modeling is described in the Appendix.

\section{Observations and Reduction}
\label{sec:obs}
\begin{figure}
    \centering
    \includegraphics[width=0.47\textwidth]{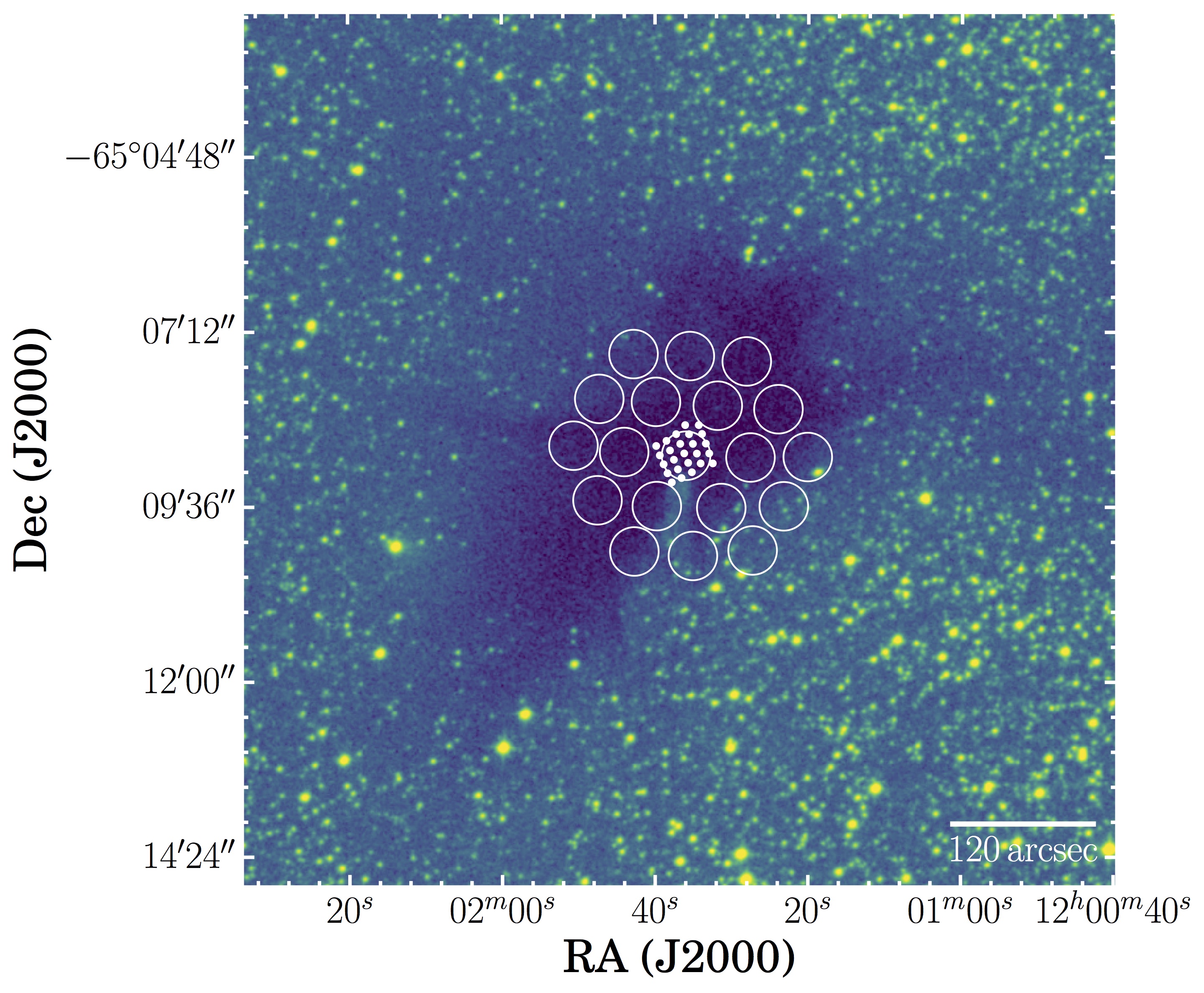}
    \caption{The footprint of PACS and SPIRE (SLW) spaxels overlaid with the optical image of BHR71 collected from the STScI Digitized Sky Survey.  The optical image is observed in AAO-SES/SERC-ER survey with IIIaF+OG590 filter centered around 6600\AA.  The circles show the SPIRE spaxels, while the dots show the PACS spaxels.}
    \label{fig:footprint}
\end{figure}

BHR71 was observed by the PACS and SPIRE instruments on \herschel. The PACS spectra come from the DIGIT \herschel\ Key Program, while the SPIRE spectra come from the {\it Herschel} Open Time Program, CO in Protostars (COPS, PI: J.~Green). The PACS instrument is a 5$\mathrm{\times}$5 array of 9.4\arcsec$\mathrm{\times}$9.4\arcsec\ spatial pixels (hereafter ``spaxels'') covering wavelengths from 50 to 210 \micron\ with $\mathrm{\lambda / \Delta \lambda \sim}$~1000-3000.
The spatial resolution of PACS ranges from $\mathrm{\sim}$~9\arcsec\ at 50 \micron\ to $\mathrm{\sim}$~18\arcsec\ at 210 {\micron}, corresponding to 1800-3600~AU at the distance of BHR71.  The regions observed with PACS and SPIRE are indicated in Figure~\ref{fig:footprint}.

The SPIRE instrument is a Fourier-Transform Spectrometer (FTS). There are two bands on SPIRE covering $194-313$ \micron\ (SSW; Spectrograph Short Wavelengths) and 303-671 \micron\ (SLW; Spectrograph Long Wavelengths) with $\mathrm{\lambda / \Delta \lambda \sim}$~$300-800$. The spatial resolution of SPIRE ranges from 17\arcsec-40\arcsec, corresponding to 3400-8000~AU at the distance of BHR71.
The on-source exposure times for PACS and SPIRE spectra are 7268 seconds and 3168 seconds respectively, achieving line flux RMS values of $\mathrm{17\times 10^{-18}-66\times 10^{-18}~W~m^{-2}}$ and $\mathrm{3.8\times 10^{-18}-36\times 10^{-18}~W~m^{-2}}$ respectively.

\subsection{Data Reduction}
\label{sec:reduction}

There are two types of data products used in this study: the datacube product is used for comparing the spatial distribution of emission lines; the extracted source spectrum, 1-D spectrum, is used for comparing to the radiative transfer simulation.  The datacube products are collected from the Herschel Science Archive, processed with SPG v14.0.1 (PACS) and SPG v14.1.0 (SPIRE).  The calibrations for PACS and SPIRE are version 72 and \texttt{spire\_cal\_14\_2}, respectively.

The 1-D spectrum is extracted from the PACS and SPIRE data cubes with two different methods that best represent the nature of the source and instruments.
The size of BHR71 is larger than a single spaxel of PACS and SPIRE, except for the SPIRE Long Wavelength Spectrometer Array (SLW).  Therefore, neither a standard point source calibration nor an extended source calibration is suitable for extracting the entire observation into a single spectrum.
For PACS, we aim to extract a 1-D spectrum that best matches the SPIRE spectrum.  The 1-D spectrum is calculated from all PACS spaxels (total 25 spaxels) weighted by their fraction of pixel area within a circular aperture centered at the central spaxel.  The size of the aperture is chosen to be 31.8\arcsec, yielding  a flux in a same area as the central 3x3 spaxels, to match the SPIRE spectrum.
For the photometry, we measure the fluxes within the same aperture in three PACS bands (70~\micron, 100~\micron, and 160~\micron) with the aperture correction described in the Data Analysis Guide (Section 4.21.2.4).  The images observed in the Open Time Program led by J. Tobin are collected from the Herschel Science Archive without any further reduction.  The SPIRE images are collected in the same fashion as the PACS images.

For SPIRE, we utilize the Semi-extended Corrector Tool (SECT) in Herschel Interactive Processing Environment (HIPE) v.14 in the spectral calibration.
The purpose of SECT is mitigate the difference between the two modules (SSW and SLW) due to the beam size dependence on wavelength and the fact that the source is semi-resolved.  The detailed procedures are illustrated in \citet{2016MNRAS.458.2150M}.  Here we highlight the critical steps.  SECT fits a source size for a given source model, which we choose to be a 2-D Gaussian, and scales the entire spectrum into a wavelength-independent Gaussian reference beam.  There are two SECT processes in our calibration pipeline.  The first one fits the source size with the default Gaussian reference beam (40\arcsec).  The second one runs with the fitted source size without further optimization, and scales the spectrum into a Gaussian reference beam with the same size as the fitted source size.  The photometry fluxes at three SPIRE bands (250~\micron, 350~\micron, and 500~\micron) are calculated by performing aperture photometry on the extended source corrected images with beam correction, color correction and aperture correction.  The aperture sizes are set to be the convolution of the fitted source size and the mean FWHM of the beam at each band (see Table~\ref{tbl:phot}).

\subsection{Line Fitting}
After the data reduction, we performed a comprehensive and robust line fitting process to the spectrum of each spaxel as well as the 1-D spectrum. To make the fitting process general, we aimed to fit every line that fell into the range of wavelength of the spectra for several molecular and atomic species. The line information was collected from the Leiden Atomic and Molecular Database \citep{2005A&A...432..369S} and the Cologne Database for Molecular Spectroscopy (CDMS; \citealt{2005JMoSt.742..215M}) to construct a comprehensive line list including \co, \thirteenco, \water, OH ladders and atomic fine-structure emission lines, except for the [N~\textsc{ii}] line centroids, which were adopted from the measurement in the CDF archive (Figure~6 in \citet{2016AJ....151...75G}).  For each region where a line could be, we performed multiple stages of calculations, in which baselines were fitted with second-order polynomials and lines were fitted with single or double Gaussian profiles from a continuum-subtracted spectrum. The core fitting method is {\tt mpfit}, Levenberg-Marquardt non-linear least squares minimization \citep{2009ASPC..411..251M}. The fitting process is described in the next paragraph.

At the wavelength of each possible line, the fitting routine first fits the baseline locally, then subtracts the fitted baseline from the spectrum, and finally fits the line with baseline-subtracted spectrum.  This procedure is denoted as ``the first fitting''.  Although the line profile is extracted by ``the first fitting'', its noise estimation might be contaminated by nearby lines.  A line-free spectrum derived from the subtraction of any significant line in ``the first fitting'' serves as the pure noise spectrum in the second fitting which is exactly the same as ``the first fitting'' except for the noise input.  Accurate noise estimation around the line centroid can be retrieved by combining the result of second fit and the noise spectrum from the first fit.  The third, final, fitting is performed with this estimated noise as the uncertainty of the spectrum to get the best fit line flux and uncertainty.

In the baseline fitting, we selected a region of $\pm$10 spectral resolution elements around the line centroid. Any wavelength that could possibly have a line is not considered as a valid data point for baseline fitting in order to get the proper baseline without being affected by nearby lines. Most of the emission lines are fitted with a single Gaussian profile to the baseline-subtracted spectrum. But sometimes multiple lines lie too close in wavelength so that a single Gaussian profile is insufficient to decouple them. Therefore, we perform a double Gaussian profile fitting for a few heavily blended regions.  To reduce the degrees of freedom, the line centroids and widths are fixed to theoretical values and the corresponding spectral resolution.  The line width was allowed to vary within $\pm$30\%\ of the spectral resolution for fitting of isolated lines in the SPIRE spectra but was fixed for PACS spectra.  One step in the reduction of the SPIRE spectra is apodization, which smooths the sinc function recorded by the FTS into a Gaussian-like profile, and it may result in a larger line width if some minor features are included in the apodization process.  Therefore, we give more flexibility on the line width of the fitting of the SPIRE spectra.
The signal-to-noise ratio (S/N) is calculated by simply dividing the integrated line strength by the product of the full width at half maximum (FWHM) and the noise level obtained from the residuals after removal of both lines and continuum times a constant, 1.064, appropriate for a Gaussian line. The line fitting results are summarized in Section \ref{sec:fitting}. In the following analysis, we consider a line as a detection if it has S/N greater than 3.  This criteria represents a false-positive rate of 0.3\%, yielding about 10 false detections over the whole PACS/SPIRE spectra.  Applying a higher S/N criteria will lose too many real detections.  If only the lines with S/N greater than 5 are retained, the total number of detections falls from 645 to 376.  Therefore, we further visually inspect the fitting results to avoid the false-positive detections.

\begin{figure}[htbp!]
  \centering
  \includegraphics[width=.47\textwidth]{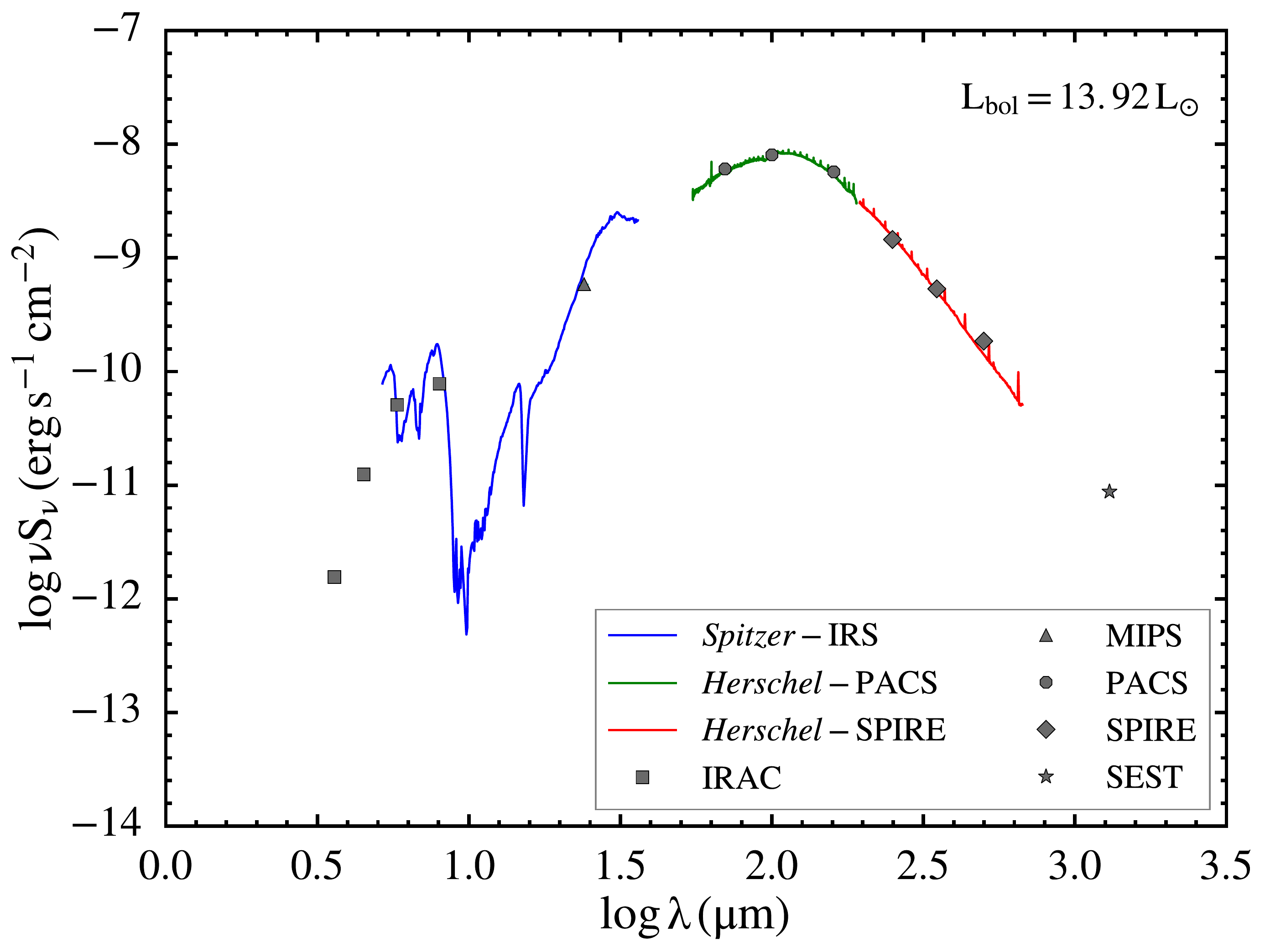}
  \caption{Spectral energy distribution (SED) of BHR71 including the {\it Spitzer}-IRS, {\it Herschel}-PACS/SPIRE and photometry measurements from archives. The observed bolometric luminosity is 13.92~\lsun.}
  \label{fig:bhr71_sed}
\end{figure}

\begin{figure*}[htbp!]
  \centering
  \includegraphics[width=.9\textwidth]{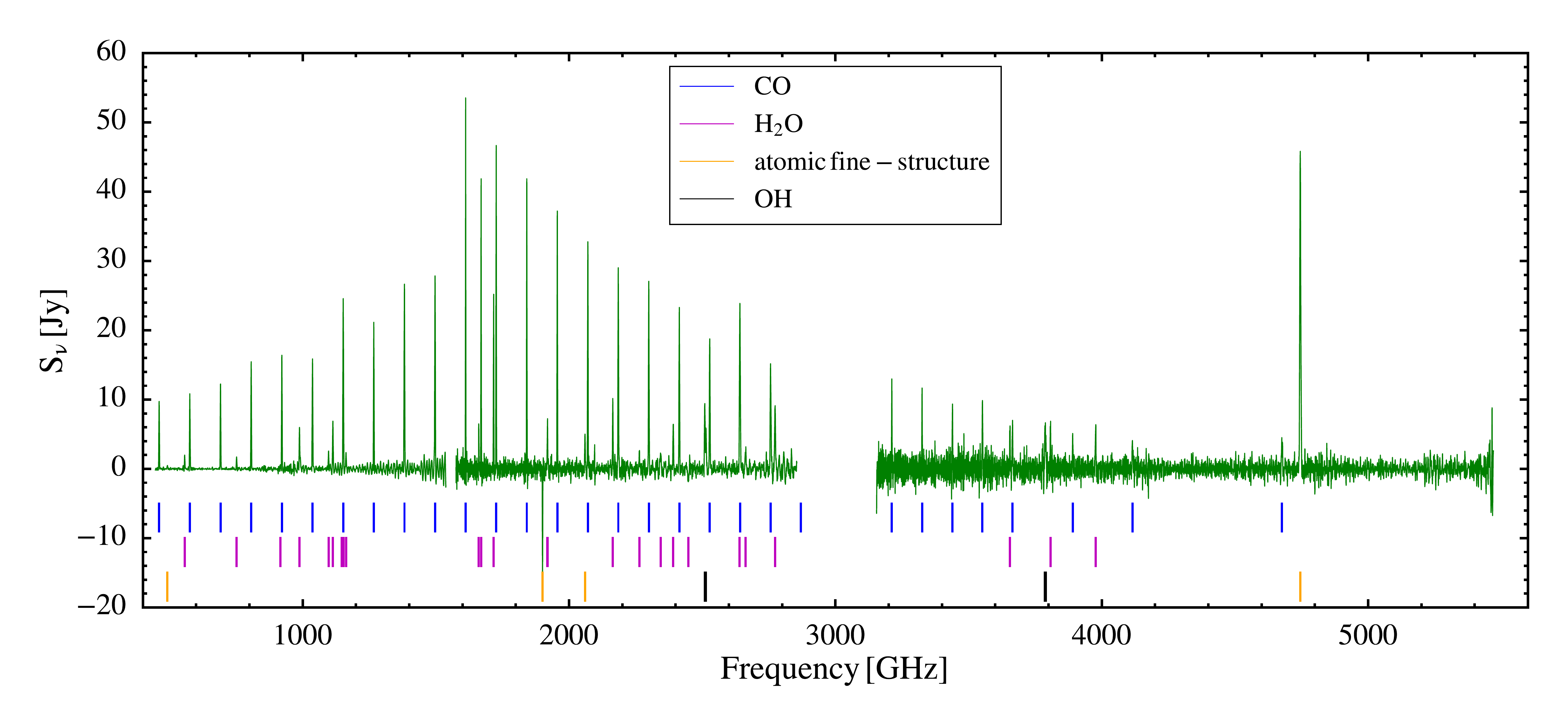}
  \caption{The flat spectra of BHR71 with PACS and SPIRE combined (green).  The continuum is removed by the line fitting process.  The \co\ rotational lines cluster at  low frequencies peaking around $J = 16$ to 18.  The \oi~63~\micron\ line is the strong line at the right side of the spectrum.  The location of lines of \co, \water, atomic fine-structure lines, and OH are indicated by the bars at the bottom of the figure (see legend).}
  \label{fig:bhr71_herschel}
\end{figure*}

In addition to the line fluxes, a line-free spectrum was produced by subtracting all the lines and smoothing the result.  This spectrum was compared to photometric data and the agreement was excellent ($\sim$10\%) for the photometric fluxes at 70~\micron, 100~\micron, 250~\micron, and 500~\micron\ and was close ($\sim 20 \%$) for the fluxes at 160~\micron\ and 350~\micron.  Together with data from other telescopes, we use the line-free spectrum to constrain the models of the SED (Figure~\ref{fig:bhr71_sed}).  A flat spectrum was then obtained by removing the continuum.  The 1-D continuum-removed spectrum of BHR71 shows abundant molecular lines and the variation of the noise across the PACS and SPIRE modules (Figure~\ref{fig:bhr71_herschel}).


\section{Results}
\label{sec:results}

In addition to the {\it Herschel} data, we obtained data from 2MASS, {\it Spitzer}-IRAC, {\it Spitzer}-IRS, and SEST. We describe these other data sets first, then the results of the {\it Herschel} observations.

\subsection{2MASS and {\it Spitzer}-IRAC Images}
\begin{figure*}[ht]
  \centering
  \includegraphics[width=0.41\textwidth]{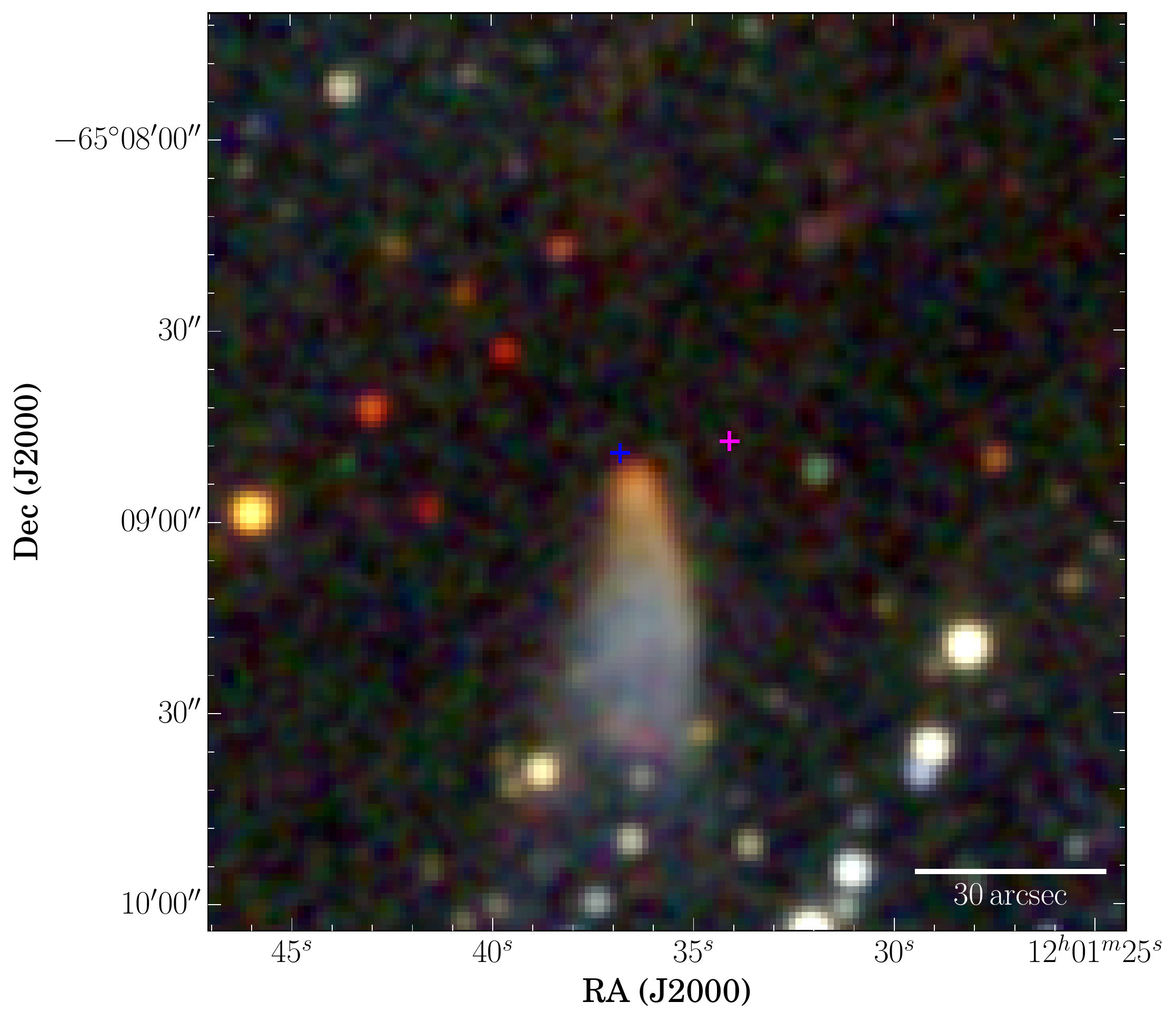}
  \includegraphics[width=0.4\textwidth]{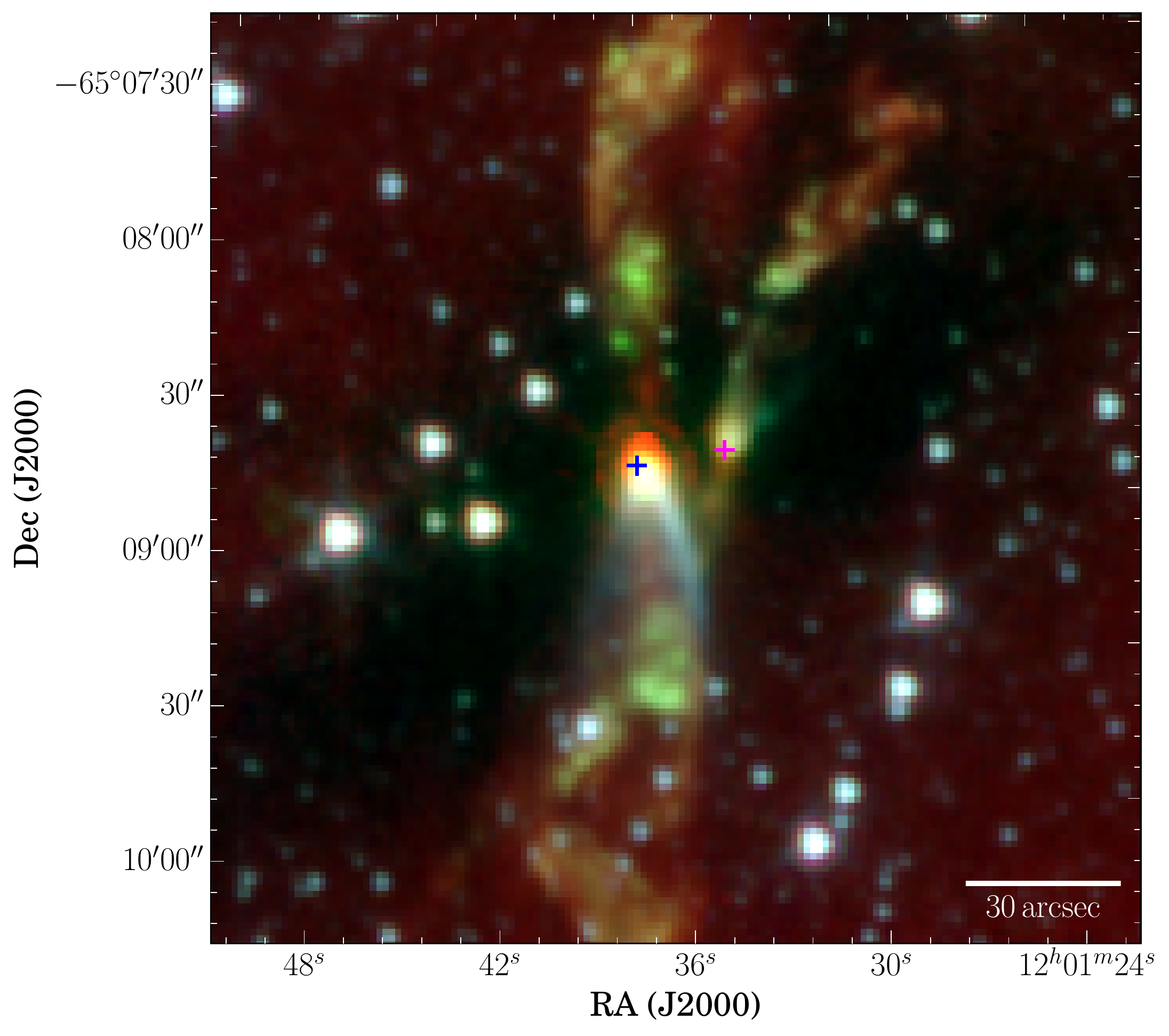}
  \caption{{\bf Left:} The false-color image with 2MASS {\sl J, H,} and {\sl K} bands in blue, green, and red.  The north lobe of the outflow is not seen in 2MASS image, while the south lobe shows a similar shape to the \spitzer-IRAC image on the right.  This image is about 145\arcsec$\times$145\arcsec, corresponding to 0.14 pc.  The blue and magenta crosses label the location of IRS1 and IRS2.  {\bf Right:} The false-color image with \spitzer-IRAC 3.6 $\mu$m, 4.5 $\mu$m, and 8.0 $\mu$m in blue, green, and red.  This image is 180\arcsec$\times$180\arcsec, corresponding to 0.17 pc.  The blue and magenta crosses are the continuum peaks measured in 3~mm continuum \citep{2008ApJ...683..862C}.  The primary source, IRS1, is at the center of image (blue cross), while there is a secondary source, IRS2, at the right from the center (magenta cross) that has $\sim 2 \%$ of the mass of the primary source.
  The coordinates of  IRS1 and IRS2 are 12$^{h}$01$^{m}$36.81$^{s}$ -65$^{d}$08$^{m}$49.22$^{s}$ and 12$^{h}$01$^{m}$34.09$^{s}$ -65$^{d}$08$^{m}$47.36$^{s}$, respectively.}
  \label{fig:irac}
\end{figure*}

We acquired the 2MASS $J$, $H$, and $K$ bands and \spitzer-IRAC images from the archives.  Figure~\ref{fig:irac} shows, on the left, the false-color image of BHR71 with $J$ band in blue, $H$ band in green, and $K$ band in red; on the right, the false-color image of BHR71 is shown with IRAC 3.6~\micron\ in blue, IRAC 4.5~\micron\ in green, and IRAC 8.0~\micron\ in red.  There are two infrared sources identified by \citet{2001ApJ...554L..91B}, IRS1 and IRS2, marked with blue and magenta crosses.
The  coordinates of IRS1 and IRS2 are 12$^{h}$01$^{m}$36.81$^{s}$ -65$^{d}$08$^{m}$49.22$^{s}$ and 12$^{h}$01$^{m}$34.09$^{s}$ -65$^{d}$08$^{m}$47.36$^{s}$, respectively, measured from observations at 3 mm \citep{2008ApJ...683..862C}; positions at other wavelength, such as near-infrared, may differ because of extinction and scattering.
The derived gas masses within about 7\arcsec\ of the two sources are different by a factor of 50, $M_{\rm IRS1}=2.12\pm0.41~\msun$ and $M_{\rm IRS2}=0.05\pm0.02~\msun$ \citep{2008ApJ...683..862C}.  Each infrared source has an associated outflow.   However there is not sufficient evidence to establish the relationship between the two infrared sources.  The bolometric luminosities of IRS1 and IRS2 are 13.5~\lsun\ and 0.5~\lsun, respectively \citep{2008ApJ...683..862C}.  In this study, the secondary source, IRS2, is not considered in detailed modeling, but it is discussed in Section~\ref{sec:binary}.

B97 fitted the \coul{1}{0} outflow observations with an edge-on biconical outflow cavity, and the {\it Spitzer}-IRAC image is roughly consistent with that picture.  The center of IRS1 locates outside of the tip of the scattered light observed in the 2MASS image, while it locates within the tip of scattered light seen in the \spitzer\ image.  Although \citet{1995MNRAS.276.1052B} deduced an inclination angle of 84$^{\circ}$ from the CO outflow, the fact that the center of IRS1 locates within the tip of scattered light in the \spitzer\ image hints that the inclination angle might not be so close to an edge-on view, which would make the center locate outside of the tip of the scattered light.  We later conclude from the simulated images that the inclination angle should be close to 50$^{\circ}$ (see Section~\ref{sec:incl}).  The arc shape in IRAC 3.6 $\mu$m (blue) at the south of IRS1 shows that the outflow cavity is irradiated by the light from the center.

\subsection{\spitzer-IRS Spectrum and Submillimeter Flux}
The \spitzer-IRS spectrum was observed in high spectral resolution mode (R$\sim$600) and extracted with the method described in \citet{2011ApJS..195....3F}.  The flux at 1.3~mm is measured by \citet{2010ApJS..188..139L} from the observation of B97 with the Swedish-ESO Submillimeter Telescope (SEST).  A total flux of 3.8~Jy is measured within a 9000~AU radius aperture \citep{2008ApJ...683..862C}.

\subsection{The PACS and SPIRE 1D Spectra}
\label{sec:1d_spec}

The SED of BHR71 including \herschel\ PACS and SPIRE 1-D line-free spectra (green and red, respectively) is shown in Figure~\ref{fig:bhr71_sed} along with the photometric data obtained from the archive.  The photometric fluxes observed by \herschel\ are calibrated as described in Section~\ref{sec:reduction}.

\begin{table}[htbp!]
  \ra{1.15}
  \centering
  \caption{Photometric Fluxes}
  \begin{tabular}{cccc}
      \toprule
      instrument/$\lambda$      &        flux & uncertainty &  aperture  \\
      ($\mu$m)              &        (Jy) &        (Jy) &  (arcsec)  \\
      \hline
      IRAC1 / 3.6             & 1.87\ee{-3} & 6.18\ee{-4} &      7.2 \\
      IRAC2 / 4.5             & 1.87\ee{-2} & 3.67\ee{-3} &      7.2 \\
      IRAC3 / 5.8             & 9.84\ee{-2} & 6.81\ee{-3} &      7.2 \\
      IRAC4 / 8.0             & 2.08\ee{-1} & 1.15\ee{-2} &      7.2 \\
      MIPS / 24               &   4.7       & 4.49\ee{-1} &     20.4 \\
      PACS / 70               &   141.6     &   7.7       &     31.8 \\
      PACS / 100              &   269.2     &   10.6      &     31.8 \\
      PACS / 160              &   305.1     &   7.5       &     31.8 \\
      SPIRE / 250             &   121.0     &   12.9      &     24.1 \\
      SPIRE / 350             &   62.4      &   9.1       &     29.6 \\
      SPIRE / 500             &   30.9      &   6.6       &     39.8 \\
      SEST / 1300             &   3.8       &   0.57      &     90.0 \\
      \bottomrule
  \end{tabular}
  \label{tbl:phot}
\end{table}

\subsection{Line Fitting Results}
\label{sec:fitting}
We detect 61 lines in the 1-D PACS and SPIRE combined spectrum, 430 lines in the PACS-cube, and 215 lines in the SPIRE-cube.  The line-fitting results for the 1-D spectrum are summarized in Table~\ref{fitting_bhr71}.  The complete fitting results including all spaxels in PACS and SPIRE are available from the Herschel Science Archive\footnote{\href{http://www.cosmos.esa.int/web/herschel/user-provided-data-products}{ http://www.cosmos.esa.int/web/herschel/user-provided-data-products}}.

\begin{table*}[htbp!]
\centering
\caption{Line fitting, PACS+SPIRE 1D spectra}
\begin{tabular}{llllllll}
\toprule
Transition & E$_{u}$ & Wave. & Flux\,\tablenotemark{a} & Transition & E$_{u}$ & Wave. & Flux\,\tablenotemark{a} \\
~ & (K) & (\micron) & ($\mathrm{10^{-18}W/m^2}$) & ~ & (K) & (\micron) & ($\mathrm{10^{-18}W/m^2}$) \vspace{3pt} \\
\toprule
\coul{41}{40} & 4737.13 & 64.11  & 134  [42.5] & \coul{18}{17} & 944.97  & 144.79 & 652 [26.0] \\
\coul{36}{35} & 3668.78 & 72.85  & 114  [33.0] & \coul{17}{16} & 845.59  & 153.27 & 631 [14.1] \\
\coul{34}{33} & 3279.15 & 77.05  & 93.2 [24.3] & \coul{16}{15} & 751.72  & 162.82 & 667 [23.9] \\
\coul{32}{31} & 2911.15 & 81.81  & 135  [26.6] & \coul{15}{14} & 663.35  & 173.63 & 637 [15.5] \\
\coul{31}{30} & 2735.28 & 84.41  & 209  [34.1] & \coul{14}{13} & 580.49  & 186.00 & 622 [20.6] \\
\coul{30}{29} & 2564.83 & 87.17  & 155  [29.3] & \coul{13}{12} & 503.13  & 200.28 & 654 [33.1] \\
\coul{29}{28} & 2399.82 & 90.16  & 165  [23.6] & \coul{12}{11} & 431.29  & 216.92 & 612 [34.1] \\
\coul{28}{27} & 2240.24 & 93.34  & 182  [23.0] & \coul{11}{10} & 364.97  & 236.61 & 459 [19.9] \\
\coul{25}{24} & 1794.23 & 104.47 & 423  [36.4] & \coul{10}{9}  & 304.16  & 260.25 & 467 [17.2] \\
\coul{24}{23} & 1656.47 & 108.75 & 508  [43.9] & \coul{9}{8}   & 248.88  & 289.10 & 356 [14.3] \\
\coul{23}{22} & 1524.19 & 113.46 & 489  [22.8] & \coul{8}{7}   & 199.11  & 325.24 & 365 [13.3] \\
\coul{22}{21} & 1397.38 & 118.59 & 530  [36.1] & \coul{7}{6}   & 154.87  & 371.64 & 349 [8.47] \\
\coul{21}{20} & 1276.05 & 124.19 & 595  [17.8] & \coul{6}{5}   & 116.16  & 433.54 & 279 [4.07] \\
\coul{20}{19} & 1160.20 & 130.37 & 634  [33.7] & \coul{5}{4}   & 82.97   & 520.14 & 247 [5.94] \\
\coul{19}{18} & 1049.84 & 137.22 & 662  [27.5] & \coul{4}{3}   & 55.32   & 650.23 & 220 [5.70] \\

\midrule
\tcoul{9}{8}  & 237.93  & 302.42 & 43.9 [13.2] & & & & \\
\midrule
OH\ohul{1/2}{1/2}{-}$\rightarrow$\ohul{3/2}{3/2}{+} & 181.9 & 79.12 & 128 [34.3] & OH\ohul{3/2}{5/2}{-}$\rightarrow$\ohul{3/2}{3/2}{+} & 120.7 & 119.34 & 165 [25.7] \\
OH\ohul{1/2}{1/2}{+}$\rightarrow$\ohul{3/2}{3/2}{-} & 181.7 & 79.18 & 128 [34.3] & OH\ohul{3/2}{5/2}{+}$\rightarrow$\ohul{3/2}{3/2}{-} & 120.5 & 119.44 & 280 [25.7] \\
\midrule
$\owater~3_{21}\rightarrow2_{12}$ & 305.3 & 75.38 & 155   [26.5] & $\owater~5_{23}\rightarrow4_{32}$ & 642.4 & 156.27& 57.4  [17.1] \\
$\owater~4_{23}\rightarrow3_{12}$ & 432.2 & 78.75 & 123   [36.7] & $\owater~3_{03}\rightarrow2_{12}$ & 196.8 & 174.63 & 347$^{b}$  [16.3]$^{b}$ \\
$\owater~6_{16}\rightarrow5_{05}$ & 643.5 & 82.02 & 110   [28.2] & $\owater~2_{12}\rightarrow1_{01}$ & 114.4 & 179.53 & 516  [14.6] \\
$\owater~2_{21}\rightarrow1_{10}$ & 194.1 & 108.10& 324   [44.7] & $\owater~2_{21}\rightarrow2_{12}$ & 194.1 & 180.48 & 89.0 [14.0] \\
$\owater~7_{43}\rightarrow7_{34}$ & 1339.9& 112.5 & 97.7  [23.1] & $\owater~3_{21}\rightarrow3_{12}$ & 305.3 & 257.82 & 54.0 [17.8] \\
$\owater~4_{14}\rightarrow3_{03}$ & 323.5 & 113.54& 427   [22.8] & $\owater~3_{12}\rightarrow2_{21}$ & 249.4 & 259.99 & 148  [17.2] \\
$\owater~7_{25}\rightarrow7_{16}$ & 1125.7& 127.91& 81.0  [24.0] & $\owater~3_{12}\rightarrow3_{03}$ & 249.4 & 273.20 & 58.0 [15.2] \\
$\owater~4_{23}\rightarrow4_{14}$ & 432.2 & 132.43& 58.0  [18.4] & $\owater~1_{10}\rightarrow1_{01}$ & 61    & 538.21 & 41.7 [5.90] \\
\midrule
$\pwater~4_{04}\rightarrow3_{13}$ & 319.5 & 125.38 &  171  [19.0] & $\pwater~1_{11}\rightarrow0_{00}$ & 53.4  & 269.30 & 168 [15.9]  \\
$\pwater~3_{13}\rightarrow2_{02}$ & 204.7 & 138.52 &  207  [23.1] & $\pwater~2_{02}\rightarrow1_{11}$ & 100.8 & 303.46 & 146 [13.2]  \\
$\pwater~3_{22}\rightarrow3_{13}$ & 296.8 & 156.20 &  79.6 [17.1] & $\pwater~2_{11}\rightarrow2_{02}$ & 136.9 & 398.63 & 35.6 [5.95] \\
\midrule
$\rm [O~\textsc{i}]~\tensor*[^{3}]{P}{_{1}}\rightarrow\tensor*[^{3}]{P}{_{2}}$ & 227.71 & 63.18 & 2009 [74.0] &
$\rm [C~\textsc{ii}]~\tensor*[^{2}]{P}{_{3/2}}\rightarrow\tensor*[^{2}]{P}{_{1/2}}$ & 91.21 & 157.75 & -227$^{c}$ [17.7] \\
$\rm [O~\textsc{i}]~\tensor*[^{3}]{P}{_{0}}\rightarrow\tensor*[^{3}]{P}{_{1}}$ & 326.58 & 145.54 & 202 [21.8]  &
$\rm [C~\textsc{i}]~\tensor*[^{3}]{P}{_{1}}\rightarrow\tensor*[^{3}]{P}{_{0}}$ & 23.62 & 609.15 & 13.2 [4.02] \\
\bottomrule
\multicolumn{8}{l}{$^{\rm a}$The uncertainty of the line flux is labeled in the bracket.} \\
\multicolumn{8}{p{\textwidth}}{$^{\rm b}$The $\owater~3_{03}\rightarrow2_{12}$ line may be contaminated by $\pwater~5_{33}\rightarrow6_{06}$ line, but the separation between two lines is too small to decouple their contribution.} \\
\multicolumn{8}{p{\textwidth}}{$^{c}$The absorption found at $\rm [C~\textsc{ii}]~\tensor*[^{2}]{P}{_{3/2}}\rightarrow\tensor*[^{2}]{P}{_{1/2}}$ is likely to be the artifact of the subtraction of the off-source position during the observation.}
\label{fitting_bhr71}
\end{tabular}
\end{table*}

\subsection{CO Rotational Diagram}
There are 30 rotational transition lines detected in the 1-D spectrum, ranging from $J$=41$\rightarrow$40 to $J$=4$\rightarrow$3.  Here we focus on the 1-D spectrum for the best constraint on the excitation temperatures for the whole object.  The relative strengths of transitions of the same molecule probe the excitation environment of the molecule. By constructing a rotational diagram for \co, we can constrain the excitation temperature and the total number  of \co\ molecules.  We follow the procedures described by \citet{2013ApJ...770..123G} to construct the rotational diagrams.

\begin{figure}[htbp!]
  \centering
  \includegraphics[width=.5\textwidth]{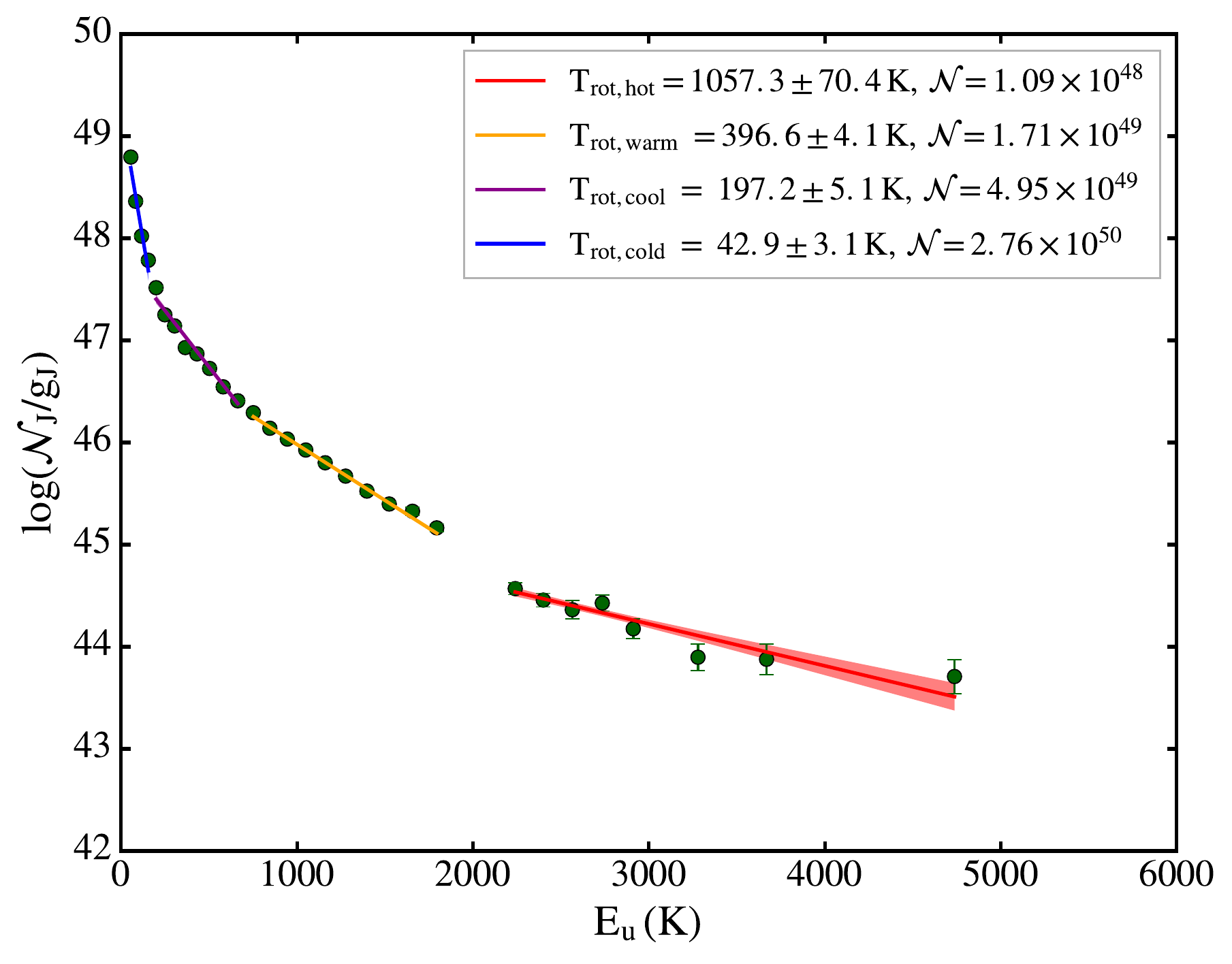}
  \caption{\co\ rotational diagram with a four-temperature fit.  The $\mathcal{N}_{J}$ is the total number of CO molecules at $J$ level, g$_{J}$ is the multiplicity at $J$ level, and E$_{\rm{u}}$ is the energy of the upper level of the corresponding transition.  The best-fit is plotted as a solid line, while the uncertainty of the fit is shown in the corresponding shaded area. }
  \label{fig:co_four}
\end{figure}

A single rotational temperature does not fit the observed population distribution (Figure~\ref{fig:co_four}).  Fitting the rotational diagram with multiple temperature components is necessary.  We recover four temperature components, 43 K, 197 K, 397 K, and 1057 K.  We let the break points between different temperature components be flexible during the $\chi^{2}$ minimization.
The ranges of the CO $J$-levels defined by the best-fitted break points are $J_{\rm up}=4$ to $J_{\rm up}=7$ (cold), $J_{\rm up}=8$ to $J_{\rm up}=15$
(cool), $J_{\rm up}=16$ to $J_{\rm up}=25$ (warm), and $J_{\rm up}=26$ to $J_{\rm up}=41$ (hot), where $J_{\rm up}$ is the rotational angular momentum quantum number of the upper level of transitions.
The highest and second highest temperature components are consistent with the ``warm'' and ``hot'' temperature components found in \citet{2013ApJ...770..123G} with the PACS spectrum only.  The need for the ``cool'' ($\trot \sim 200$ K) and ``cold'' ($\trot \sim 50$ K) components emerges from the SPIRE data.

\subsection{Distribution of CO Emission}

\begin{figure*}[htbp!]
\centering
  \includegraphics[width=.32\textwidth]{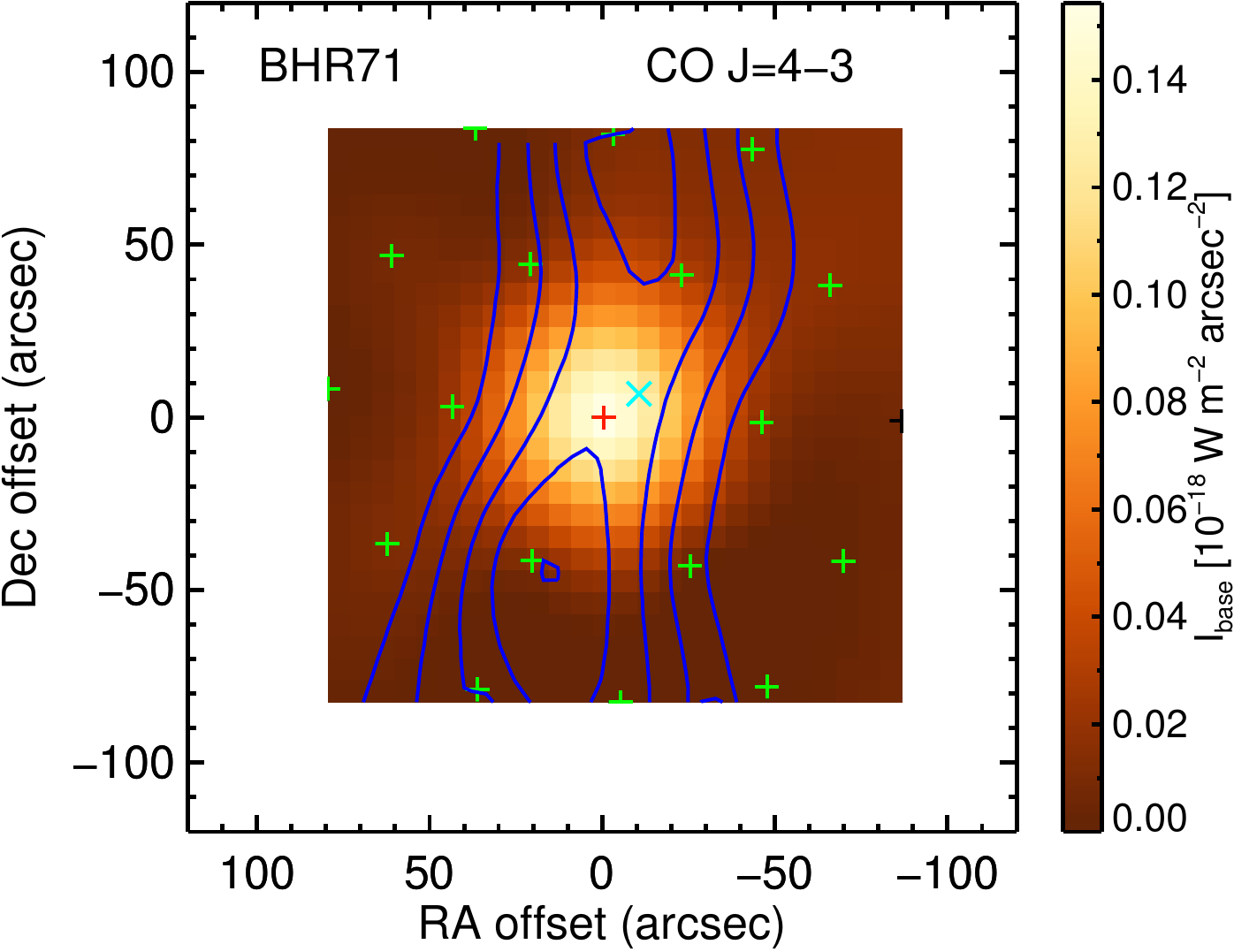}
  \includegraphics[width=.32\textwidth]{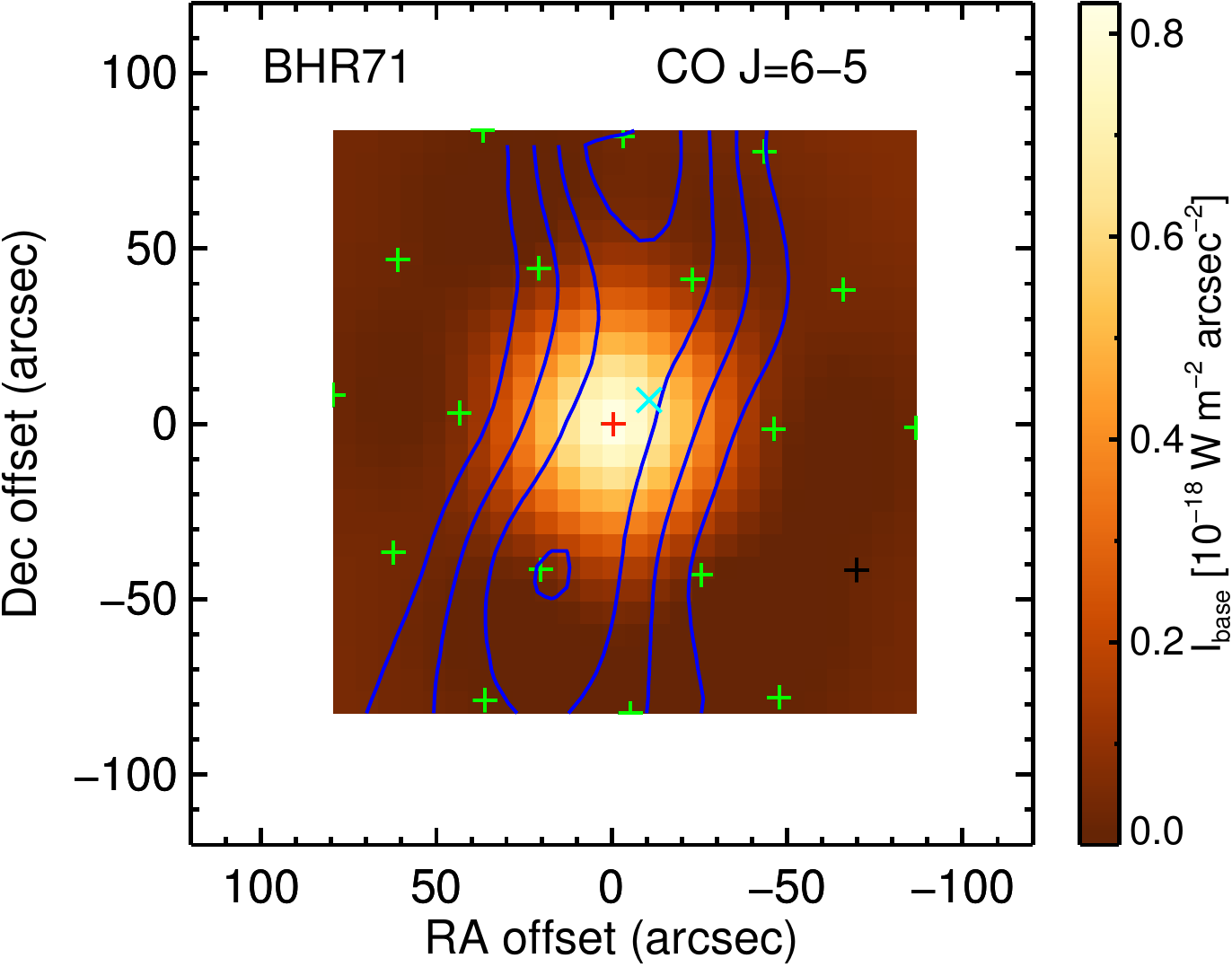}
  \includegraphics[width=.32\textwidth]{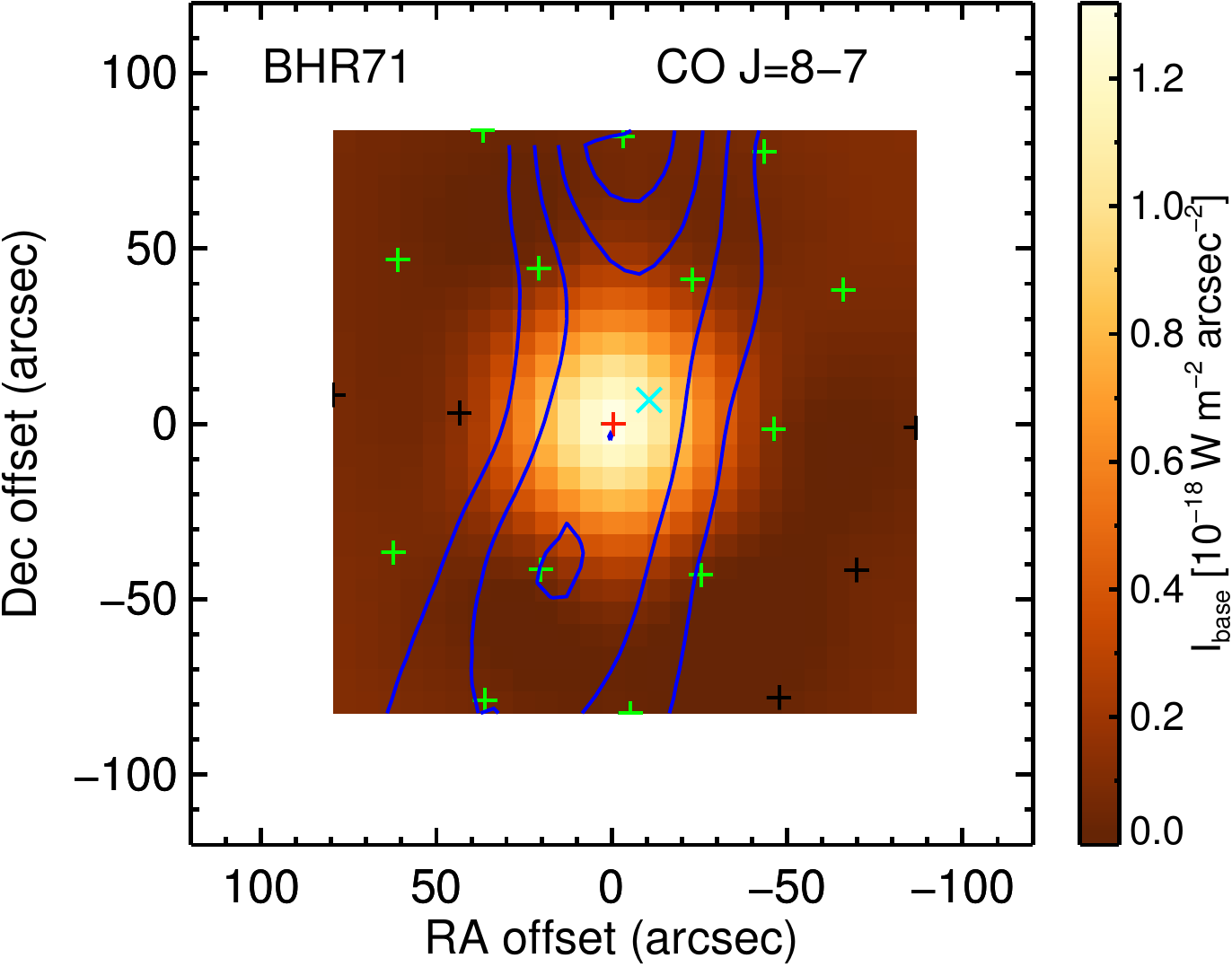}
  \\
  \includegraphics[width=.32\textwidth]{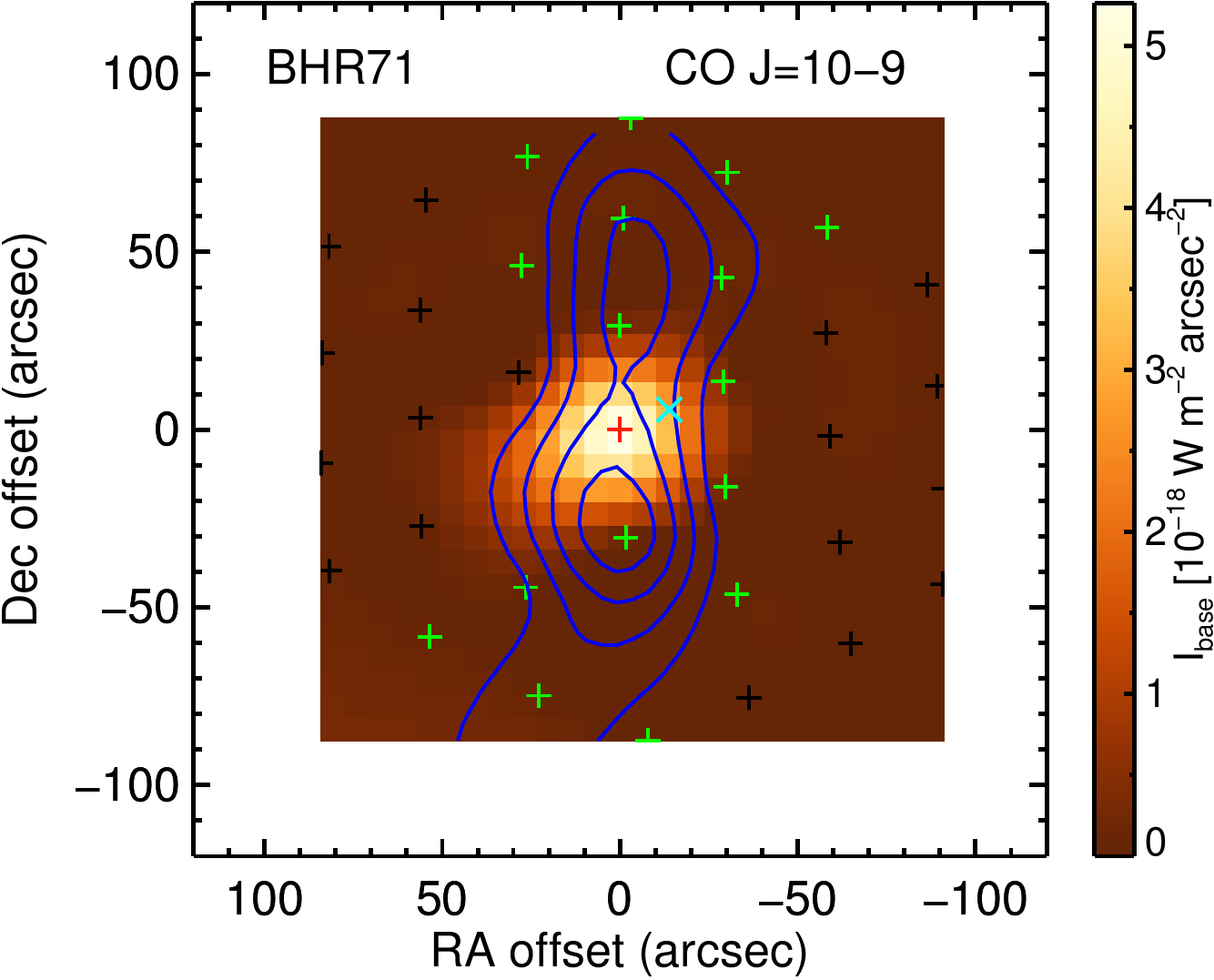}
  \includegraphics[width=.32\textwidth]{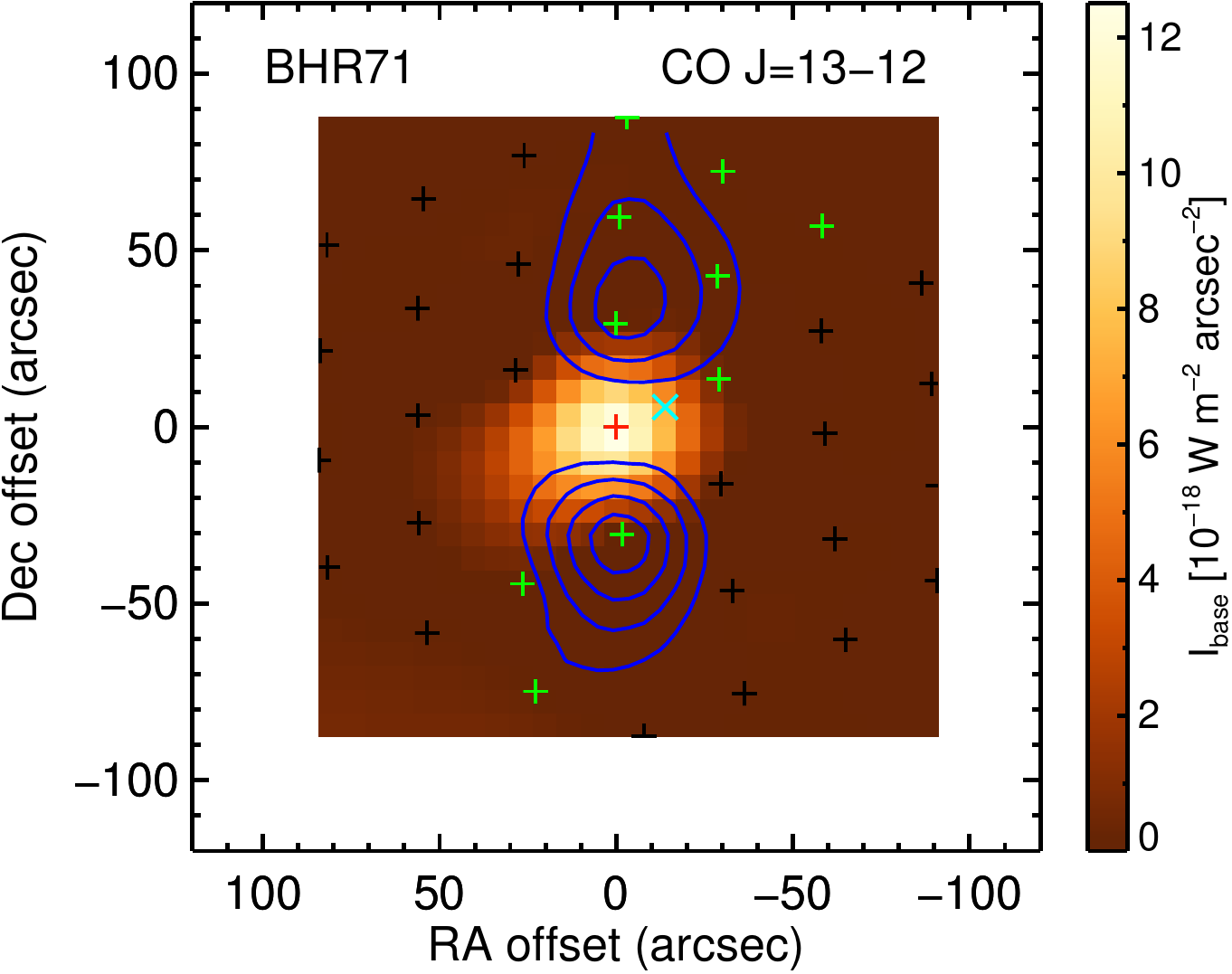}
  \includegraphics[width=.32\textwidth]{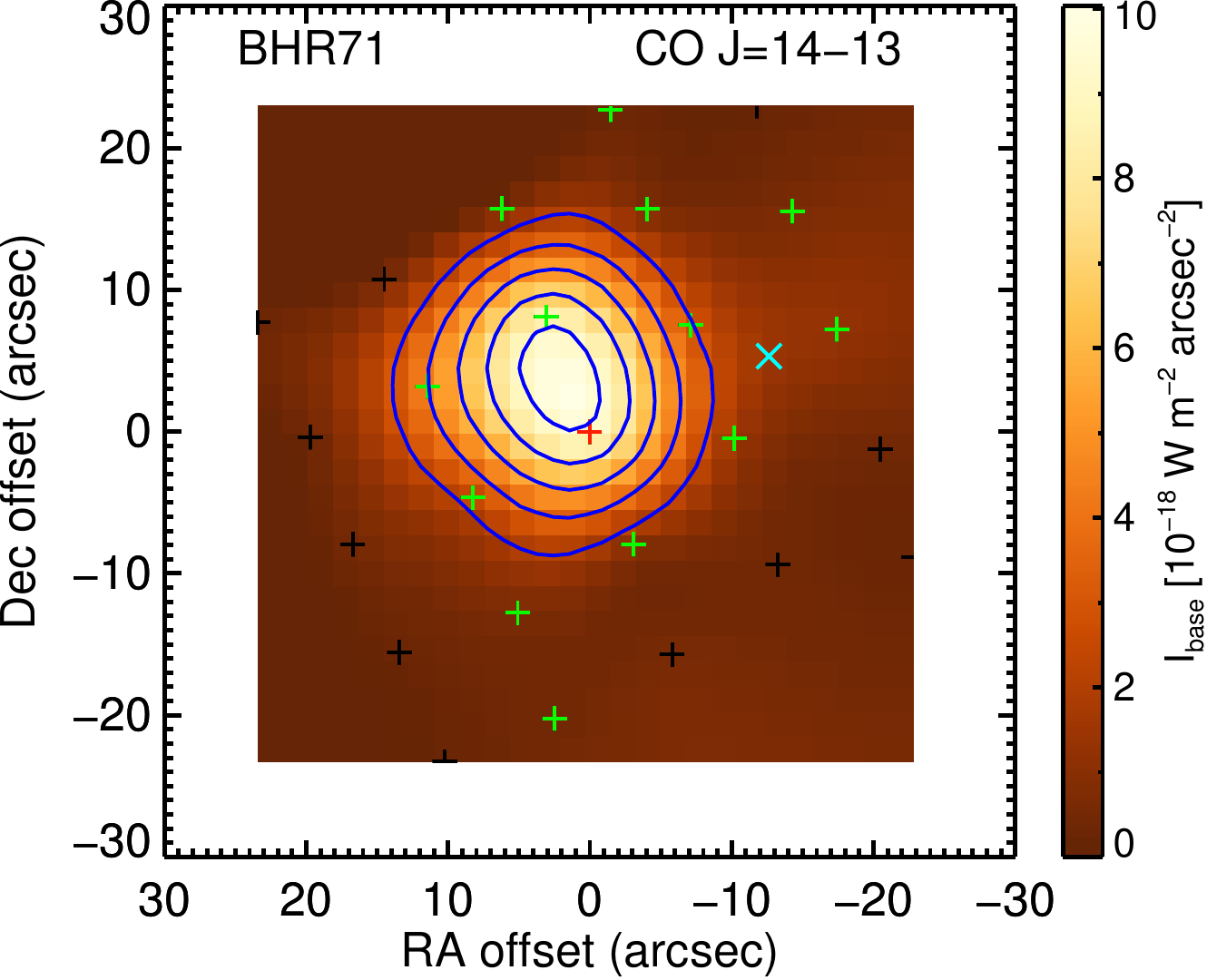}
  \\
  \includegraphics[width=.32\textwidth]{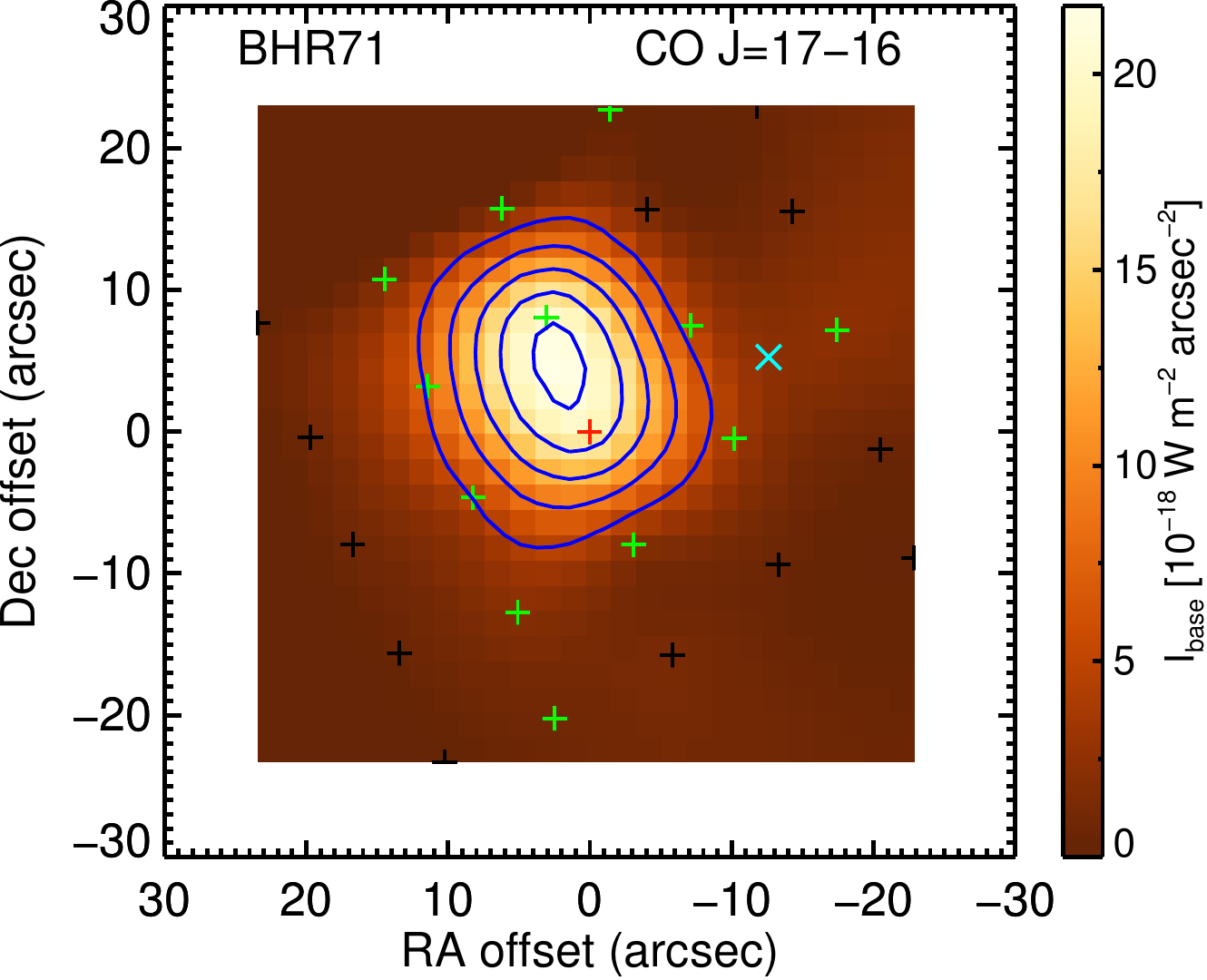}
  \includegraphics[width=.32\textwidth]{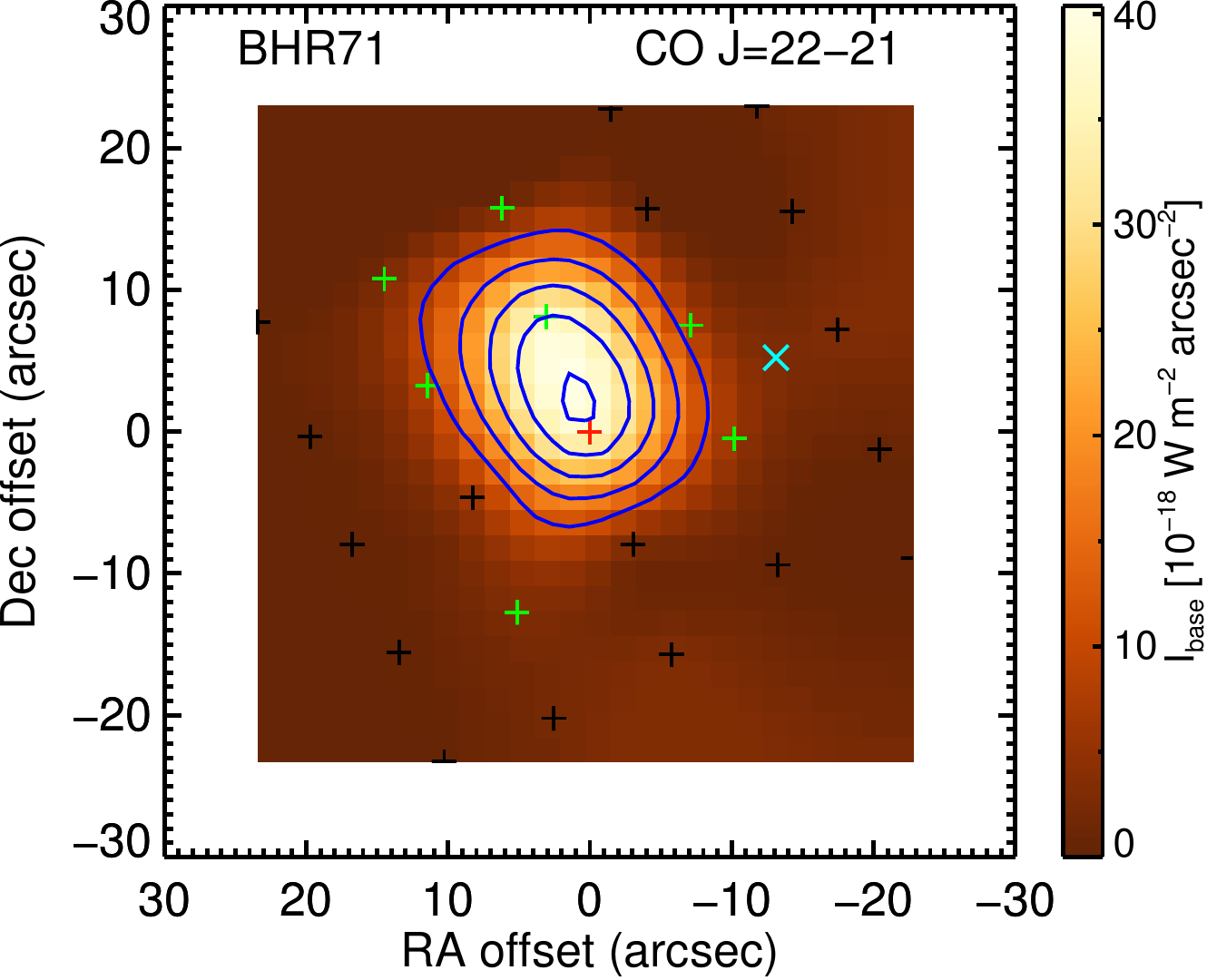}
  \includegraphics[width=.32\textwidth]{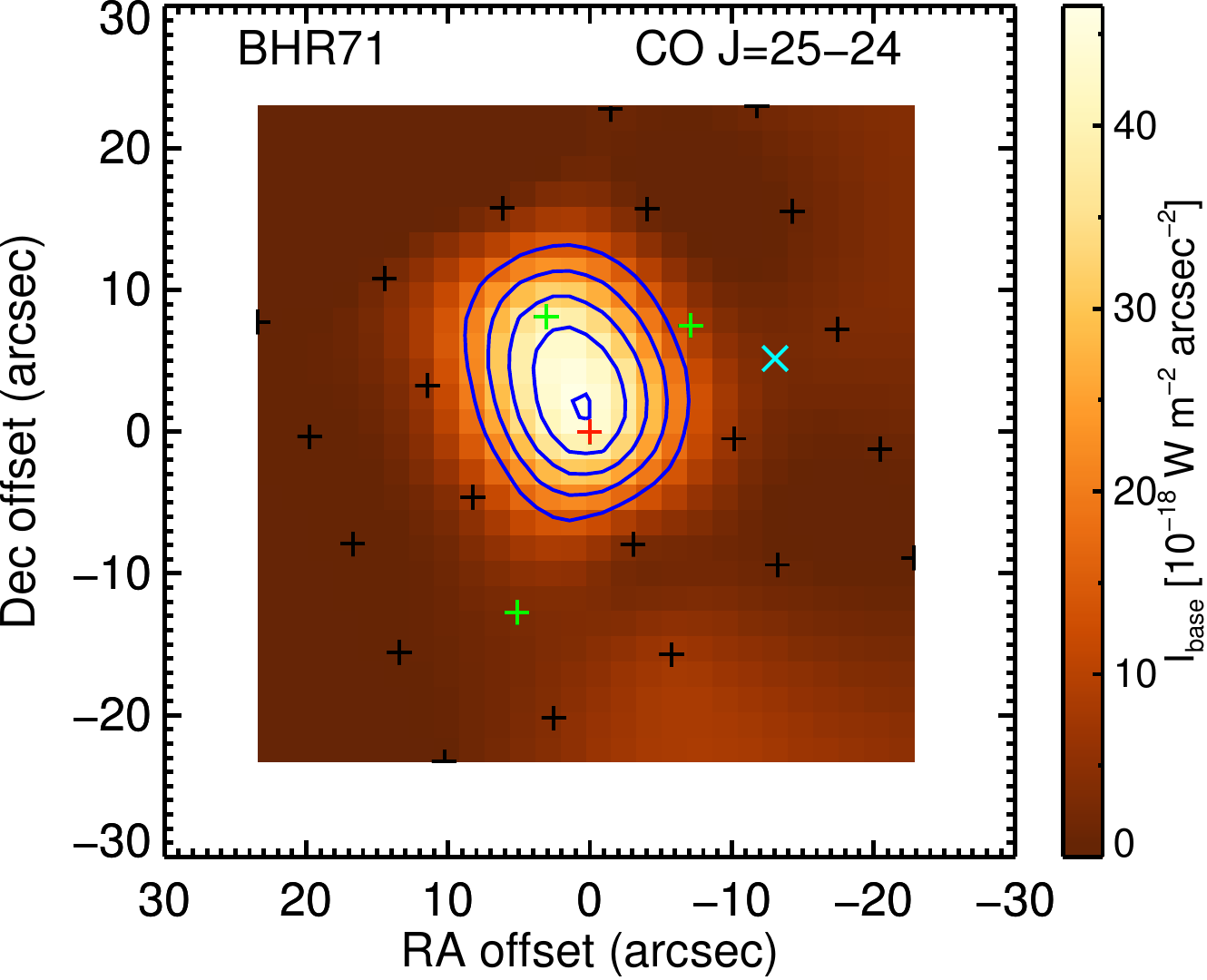}

\caption{\co\ contours of different transitions with the local continuum in color. The blue contours are plotted at 20\%\ intervals down to 20\%\ of the maximum line strength.  The black crosses are the configuration of spaxels, and green crosses indicate line detection.   The red cross indicates the center of the field of view and the position of IRS1, while the cyan ``X'' indicates the position of IRS2.  The contours of \coul{4}{3} to \coul{13}{12} are observed with a larger beam of SPIRE, while the contours of \coul{14}{13} and the lines from higher upper energy levels are observed with a smaller beam of PACS.  Therefore, these two groups have different spaxel configuration and field of view.}
\label{fig:co_contour}
\end{figure*}

With the spaxel configuration of {\it Herschel}, the spatial distribution of the emission lines can be extracted from the spectra.  Figure~\ref{fig:co_contour} shows a subset of  the \co\ maps (blue contours) on top of continuum images (false color).  The low-$J$ \co\ ($J \leq 13$) lines are extended along the north-south direction  of the outflow mapped by  \coul{1}{0} and $\mathrm{H_{2}}$ (B97, \citealt{2011ApJ...738...80G,
2011A&A...532A..53G}).
The common characteristic of low-$J$ \co\ lines is that the peak intensity is located at off-center positions.  The \coul{13}{12} line is actually undetected in the central spaxel, but detected in off-center positions, similar to the \coul{1}{0} line wing distribution (B97).

In contrast, the high-$J$ \co\ lines ($J_{\rm u} > 13$) are well-centered on the continuum, tracing the hot gas component near the central protostar.  An extension toward north-east direction is found for high-$J$ levels (e.g. last row in Figre~\ref{fig:co_contour}).  The reason for such effect is inconclusive from the data presented here, but it may be due to the occurrence of spot shocks in the outflow cavity.  Note that the difference between low-$J$ and high-$J$ distributions is partly caused by the switch from SPIRE to PACS at $J_{\rm up} = 14$, because the PACS observations do not cover the extended outflow.

\subsection{Distribution of OH Emission}
\begin{figure*}[ht]
\centering
    \includegraphics[width=.33\textwidth]{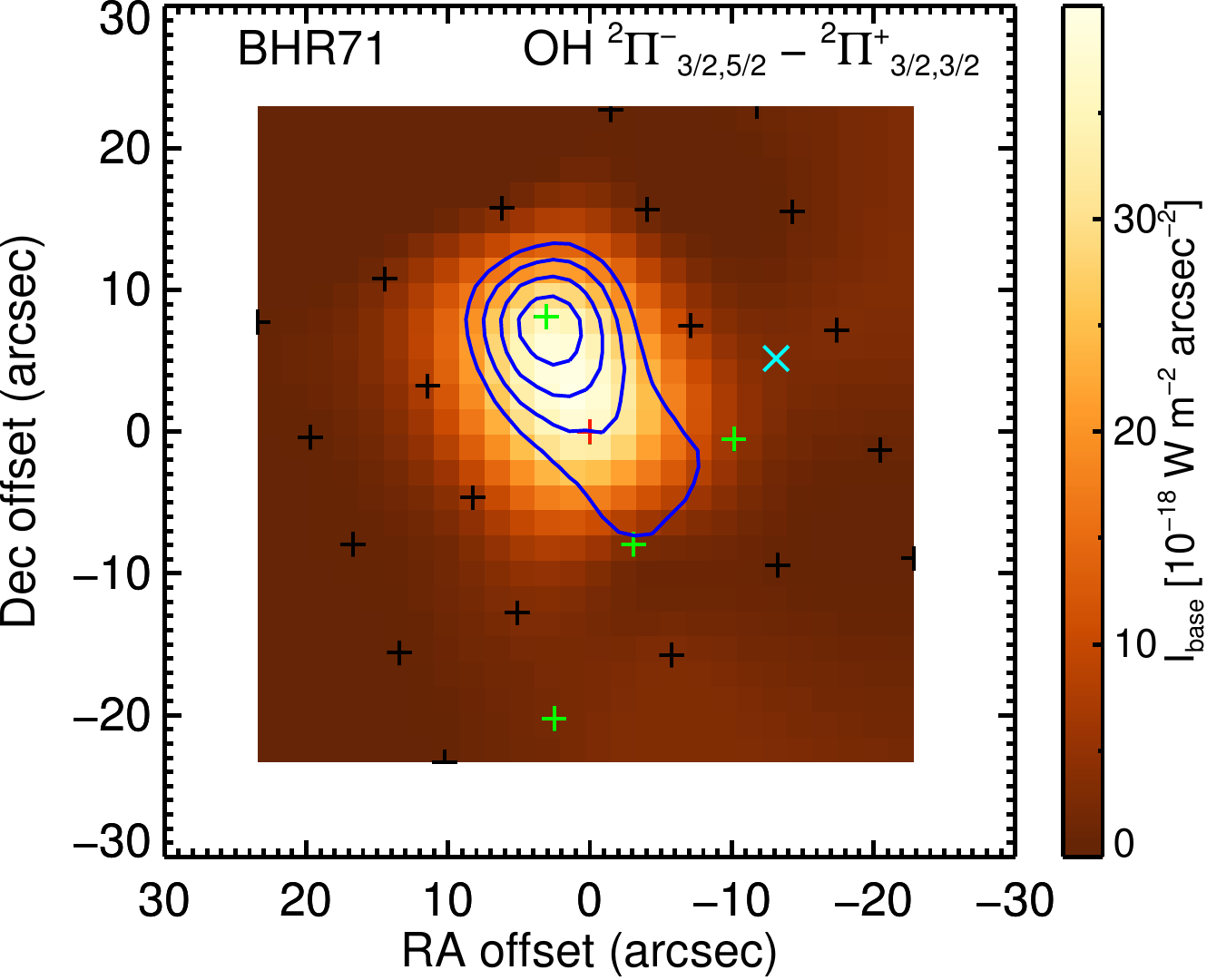}
    \includegraphics[width=.33\textwidth]{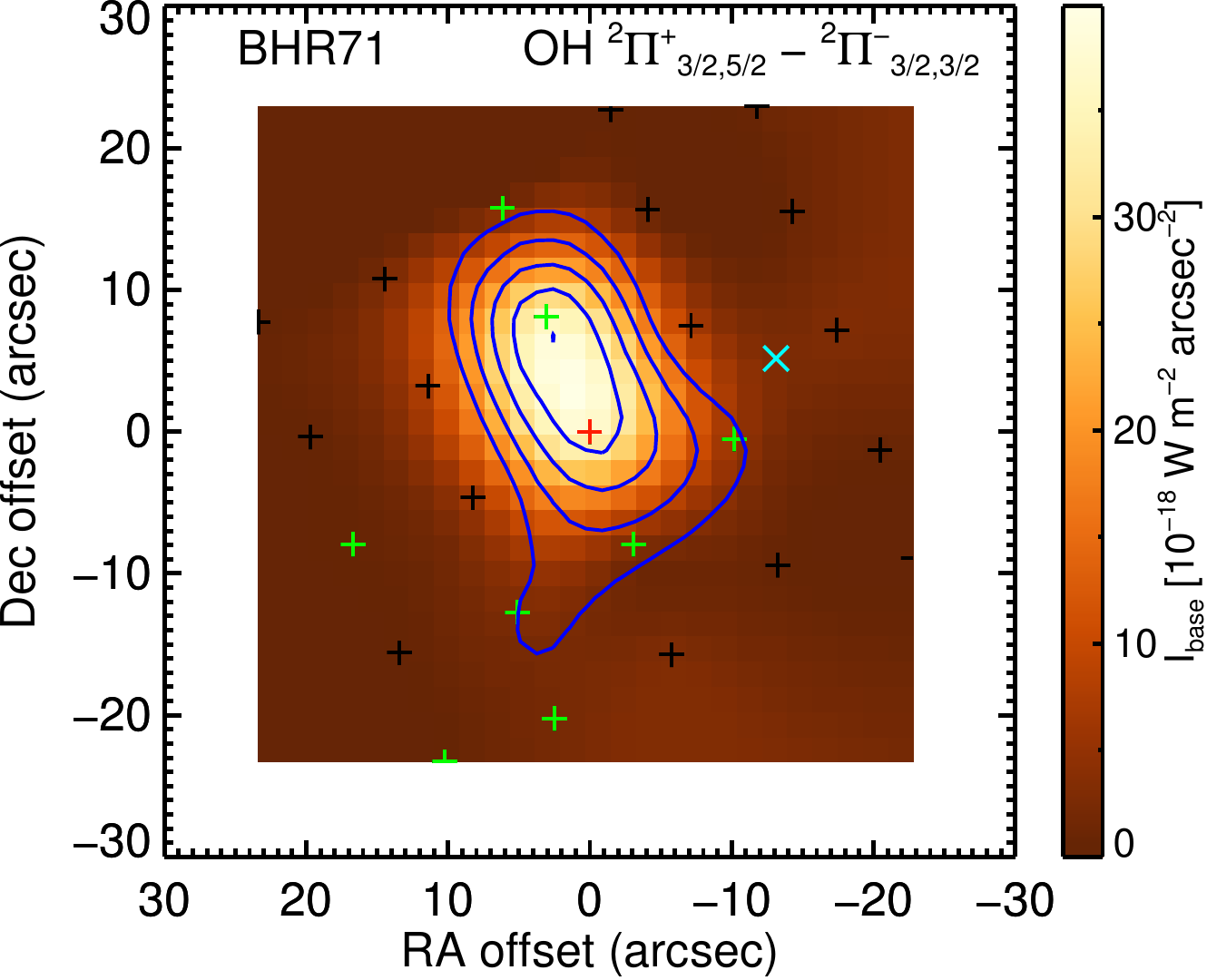}
    \caption{The contours of OH doublet lines at 119.23~\micron\ (left) and 119.46~\micron\ (right).  The color code, contour levels, and symbols are the same as Figure~\ref{fig:co_contour}}
\label{fig:oh}
\end{figure*}

The OH contours do not exhibit the bipolar morphology seen in the low-$J$ \co\ contours (Figure~\ref{fig:oh}).  However, two lines that have multiple detections across the field of view show extension toward the north-east.  Most of the spaxels have no detections of OH lines except for the doublet lines with the lowest upper state energy. These two lines also show extension toward north-east similar to the morphology found in high-$J$ CO distributions, which may be due to the occurrence of spot shocks in the outflow cavity (Figure~\ref{fig:oh}).  The spot shocks are the J-type shocks that occur when the wind first impacts  the envelope or outflow cavity \citep{2015A&A...574C...3M}.

\subsection{Distribution of Water Emission}
\subsubsection{Large Scale: SPIRE}

Lines of both ortho-water (\owater) and para-water (\pwater) were detected.
The difference in the field of view between PACS and SPIRE modules makes the emission line distributions hard to  compare directly.  This section focuses on the large scale distribution, observed by SPIRE, while the small scale distribution, observed by PACS, is presented in the next section.

\begin{figure*}[htbp!]
\centering
  \includegraphics[width=.32\textwidth]{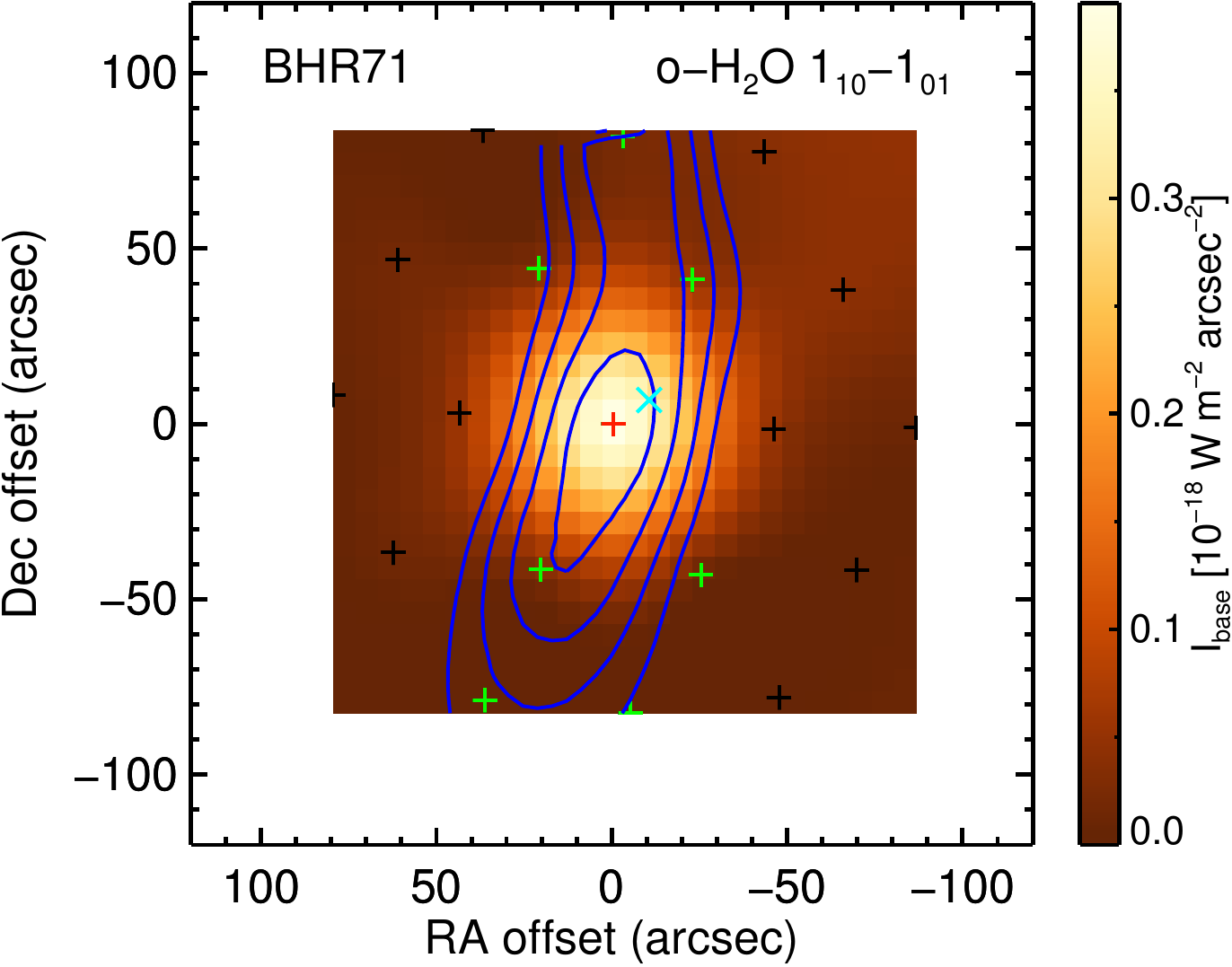}
  \includegraphics[width=.32\textwidth]{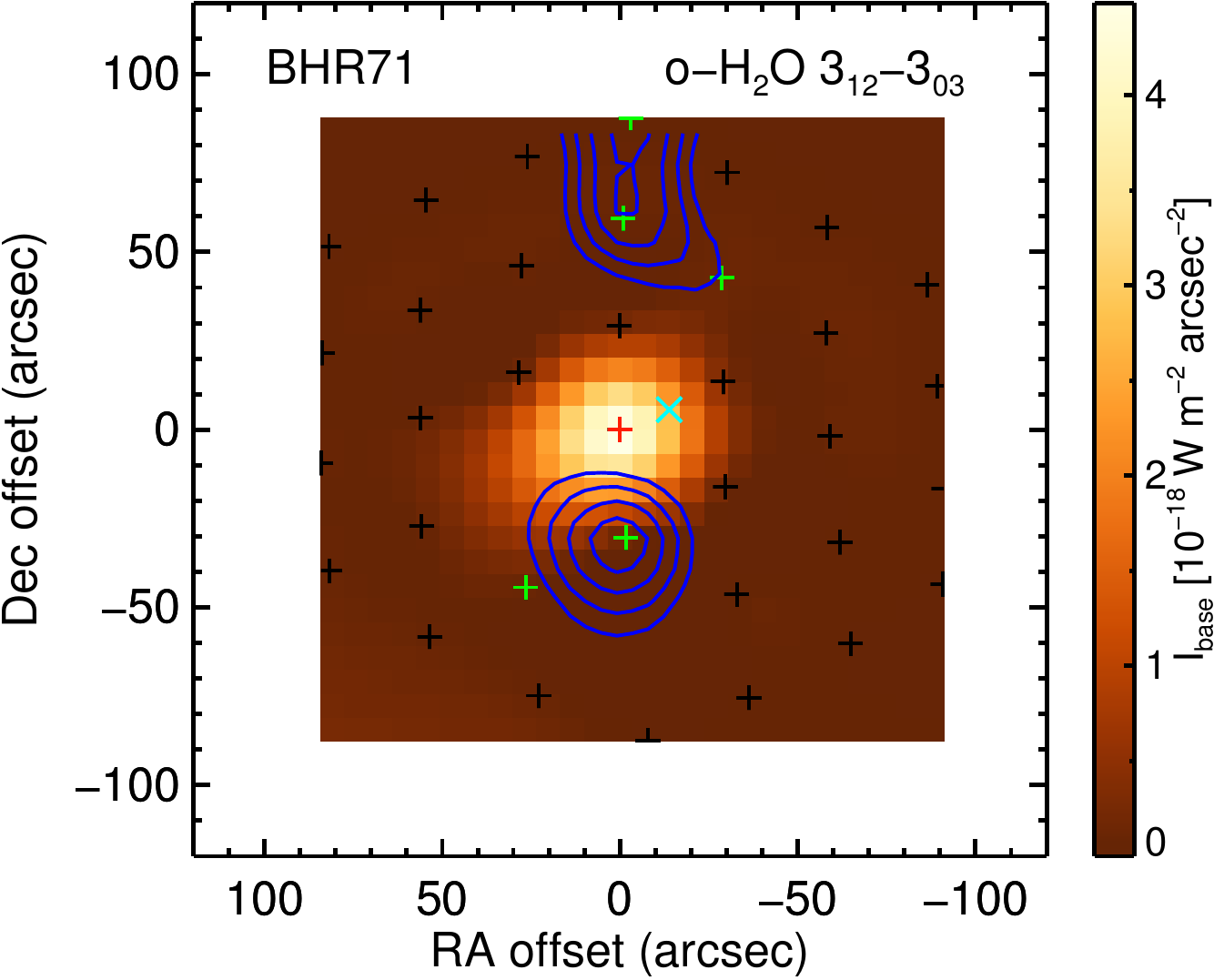}
  \\
  \includegraphics[width=.32\textwidth]{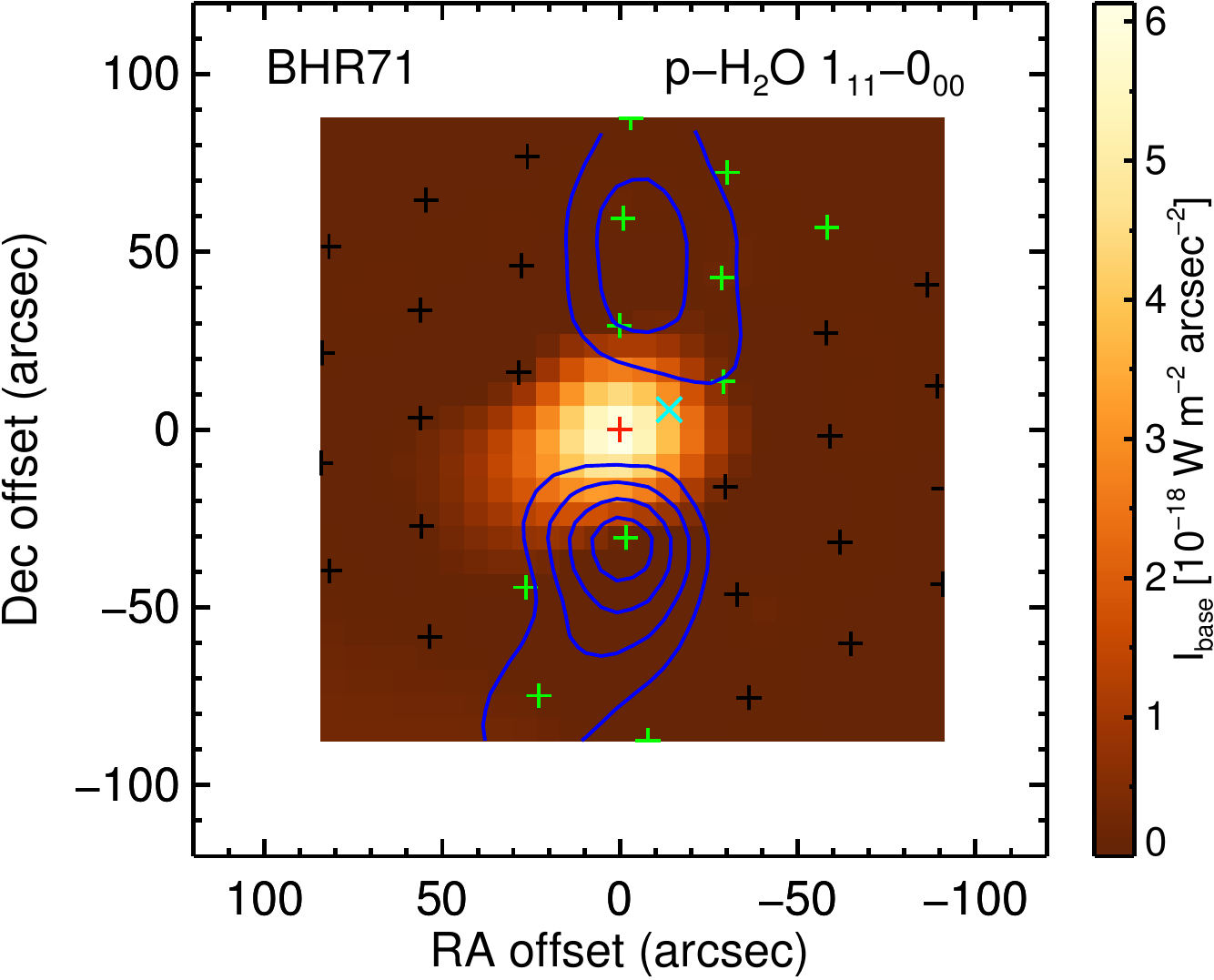}
  \includegraphics[width=.32\textwidth]{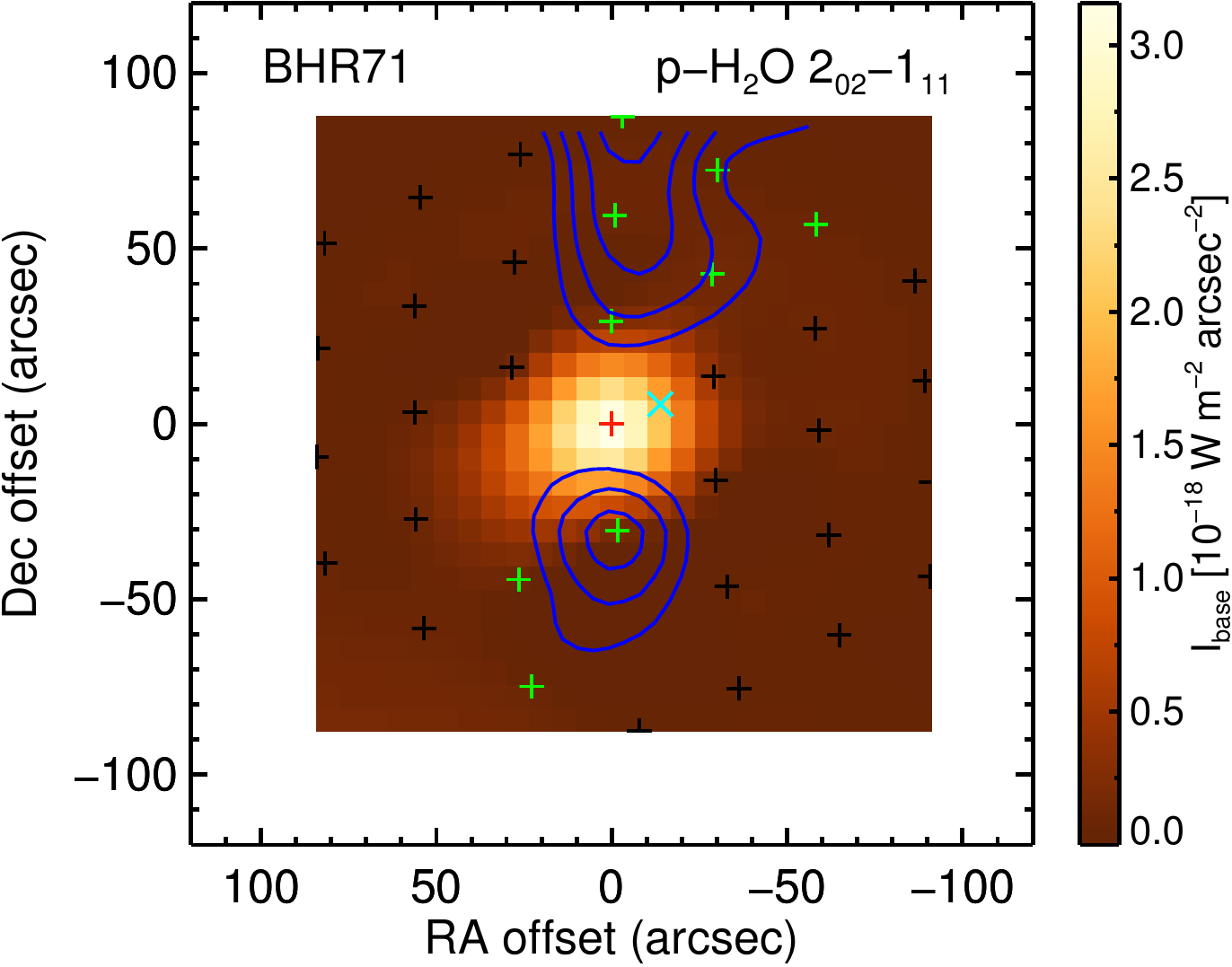}
  \includegraphics[width=.32\textwidth]{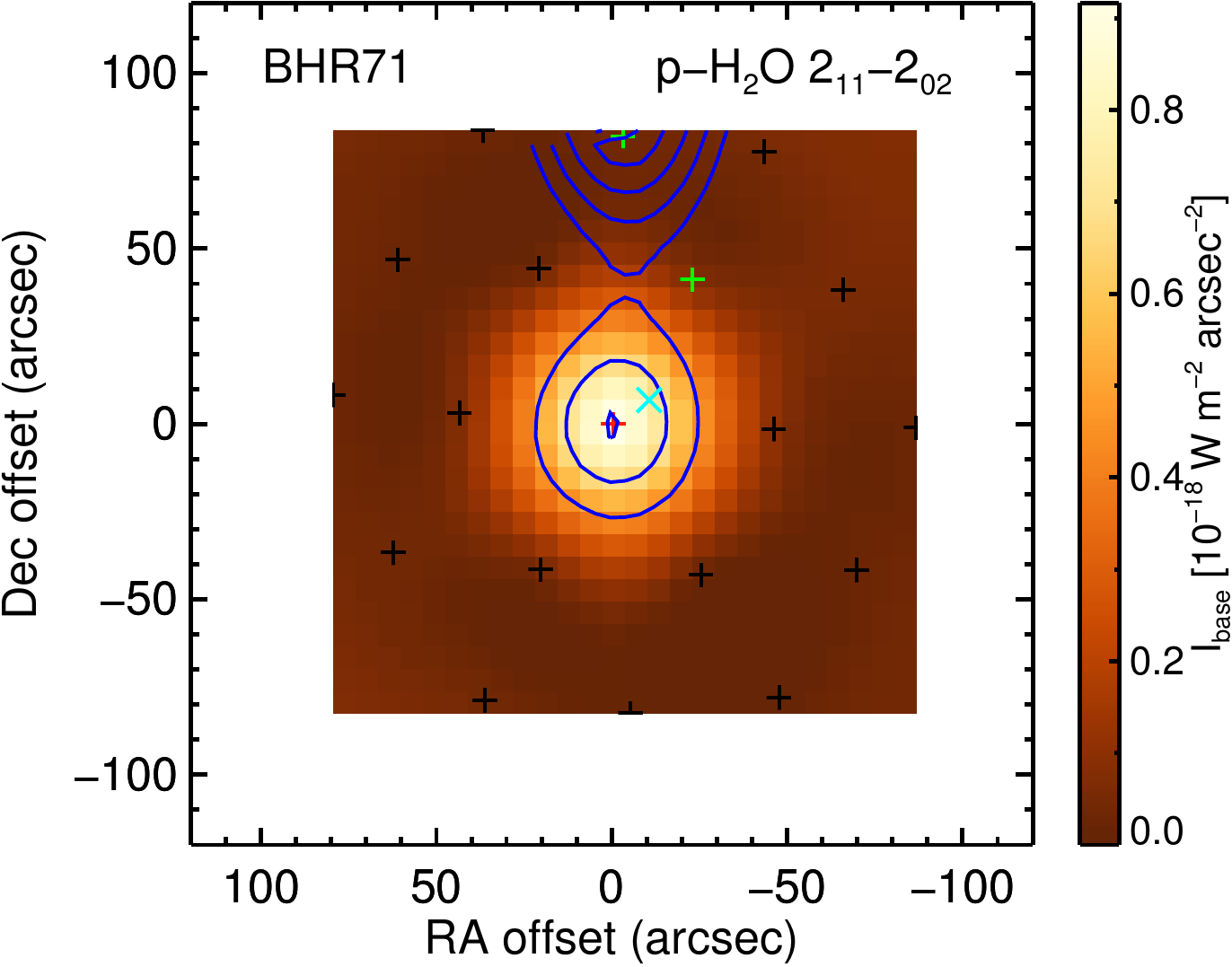}

\caption{The water emission lines distribution on a large scale, observed by SPIRE.  The color code, contour levels, and symbols are the same as Figure~\ref{fig:co_contour}.}
\label{fig:water_spire}
\vspace{10pt}
\end{figure*}

Figure~\ref{fig:water_spire} shows the detected water line distributions with an approximate 200\arcsec$\times$200\arcsec\ field of view and the spatial resolutions about 19\arcsec\ and 35\arcsec\ for SSW and SLW modules respectively.  The roughly north-south distribution of water lines, which is along the outflow direction, is the common feature of lines detected in the SPIRE module.  Two water lines, \owater$~1_{10}\rightarrow1_{01}$ and \pwater$~2_{11}\rightarrow2_{02}$ have detections at the center of dust continuum with an emission distribution extended north-south.  The other three lines are not detected at the center, but the emission distributes along the outflow direction with a similar size and morphology to the other two lines.  This morphology is similar to the distribution seen in \coul{13}{12} lines.

\subsubsection{Small Scale: PACS}
\begin{figure*}[htbp!]
    \centering
    \includegraphics[width=.32\textwidth]{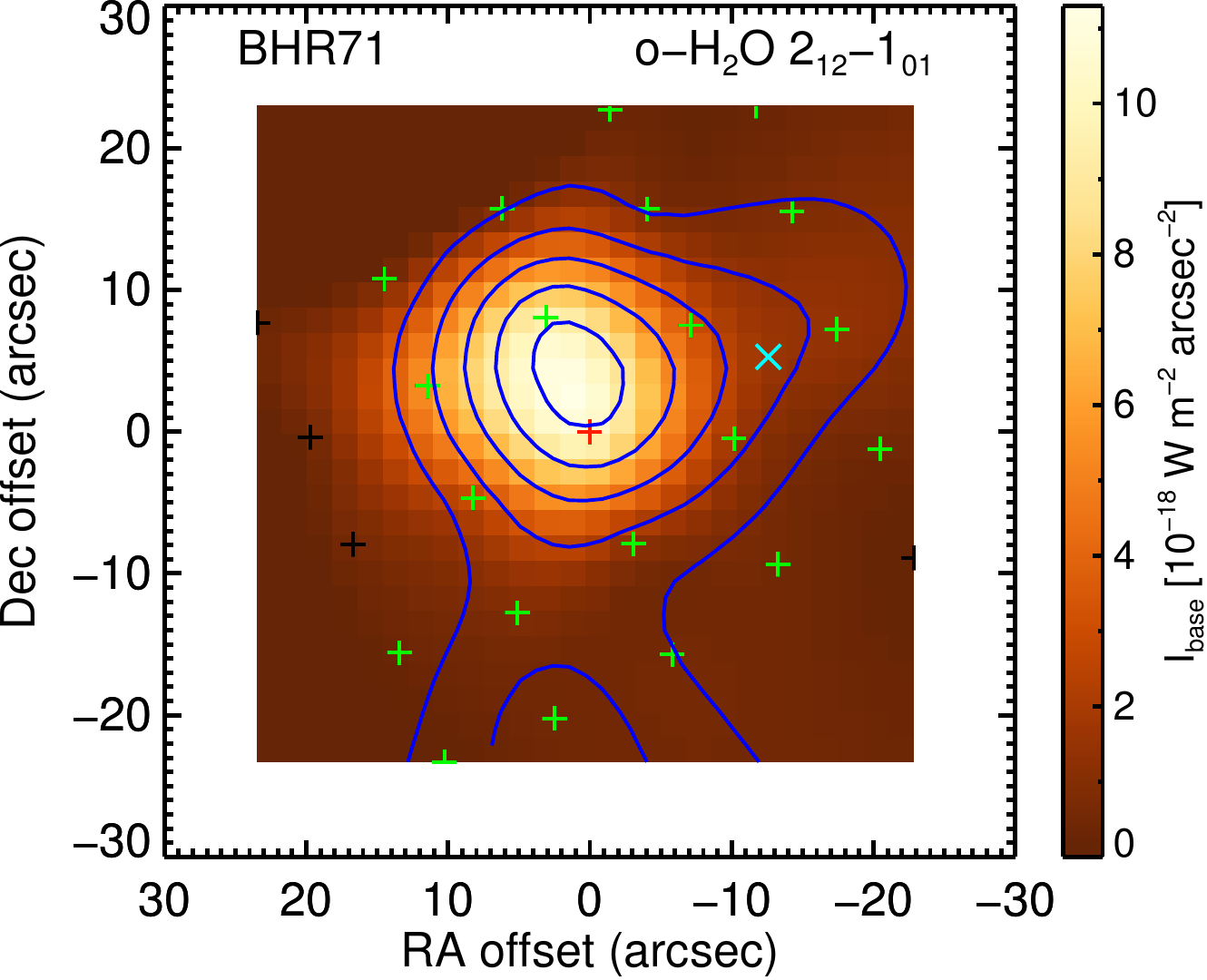}
    \includegraphics[width=.32\textwidth]{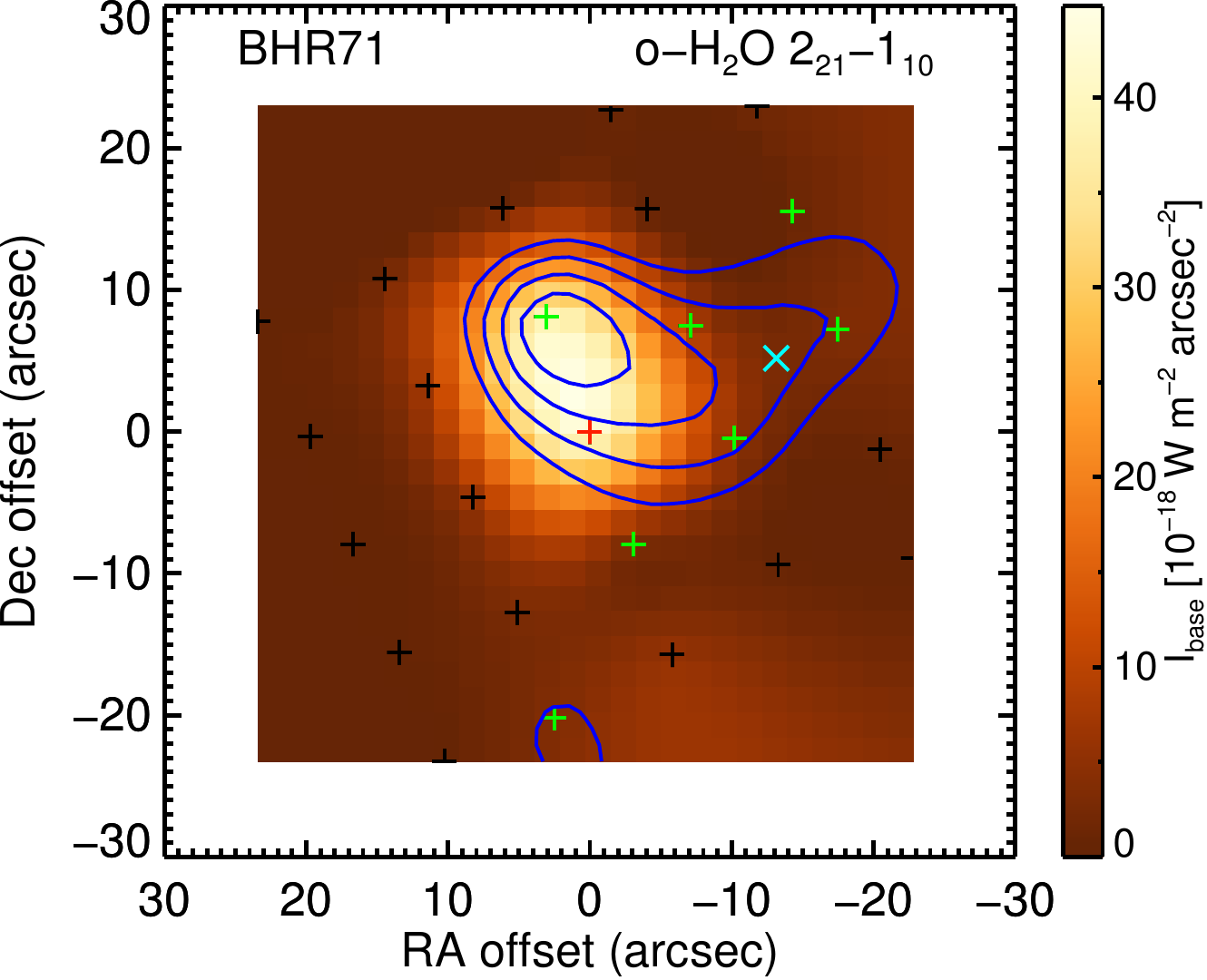}
    \includegraphics[width=.32\textwidth]{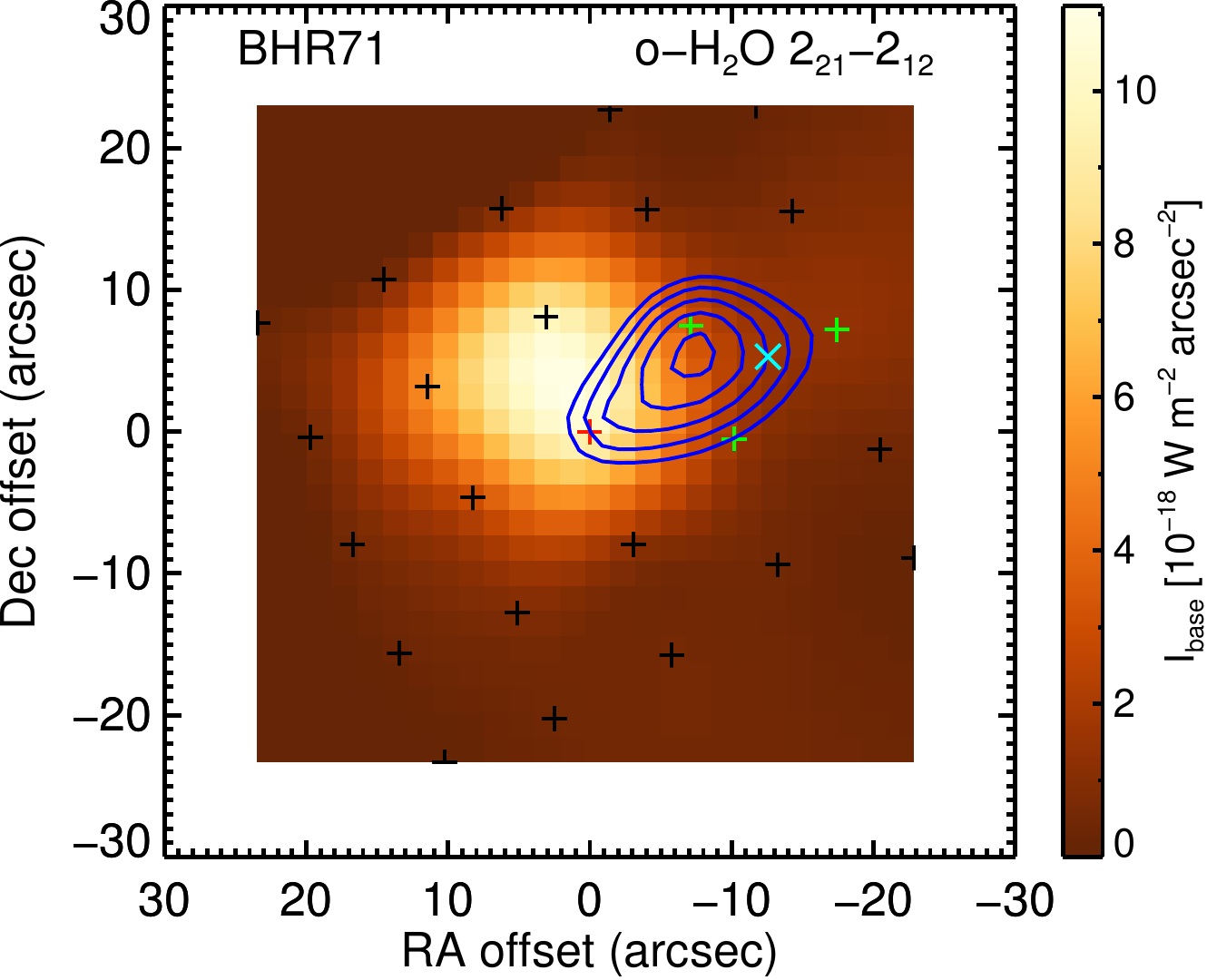}
    \\
    \includegraphics[width=.32\textwidth]{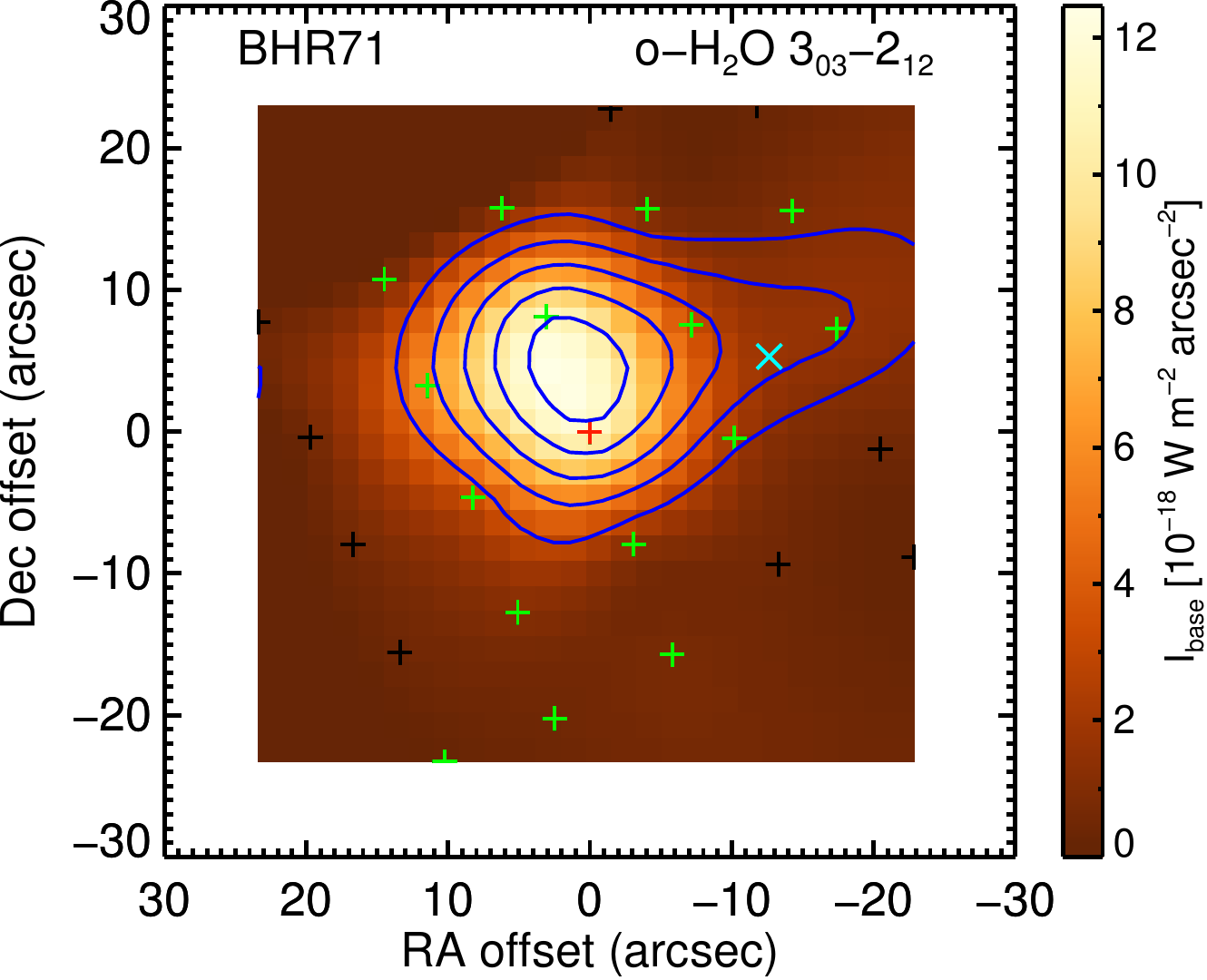}
    \includegraphics[width=.32\textwidth]{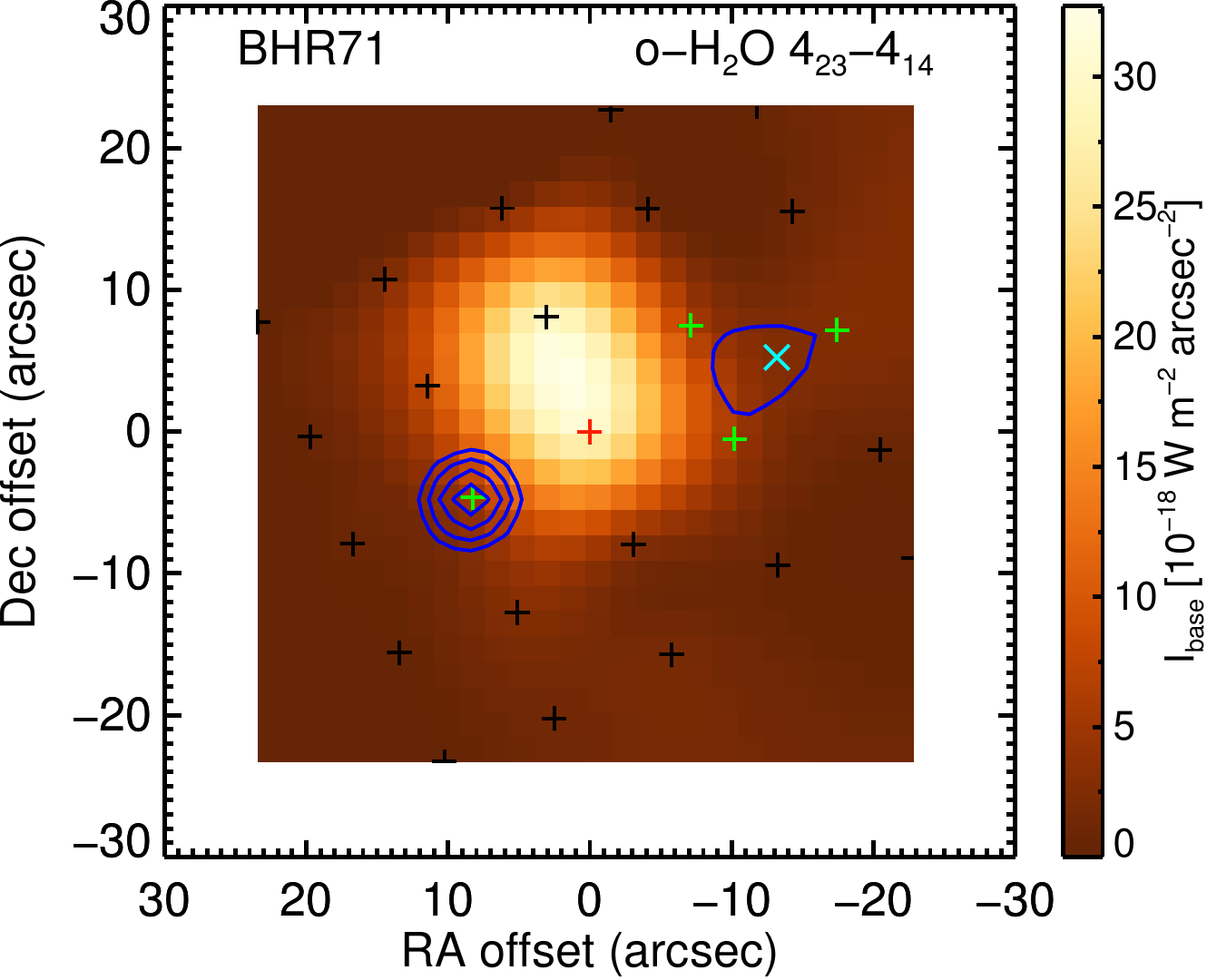}
    \includegraphics[width=.32\textwidth]{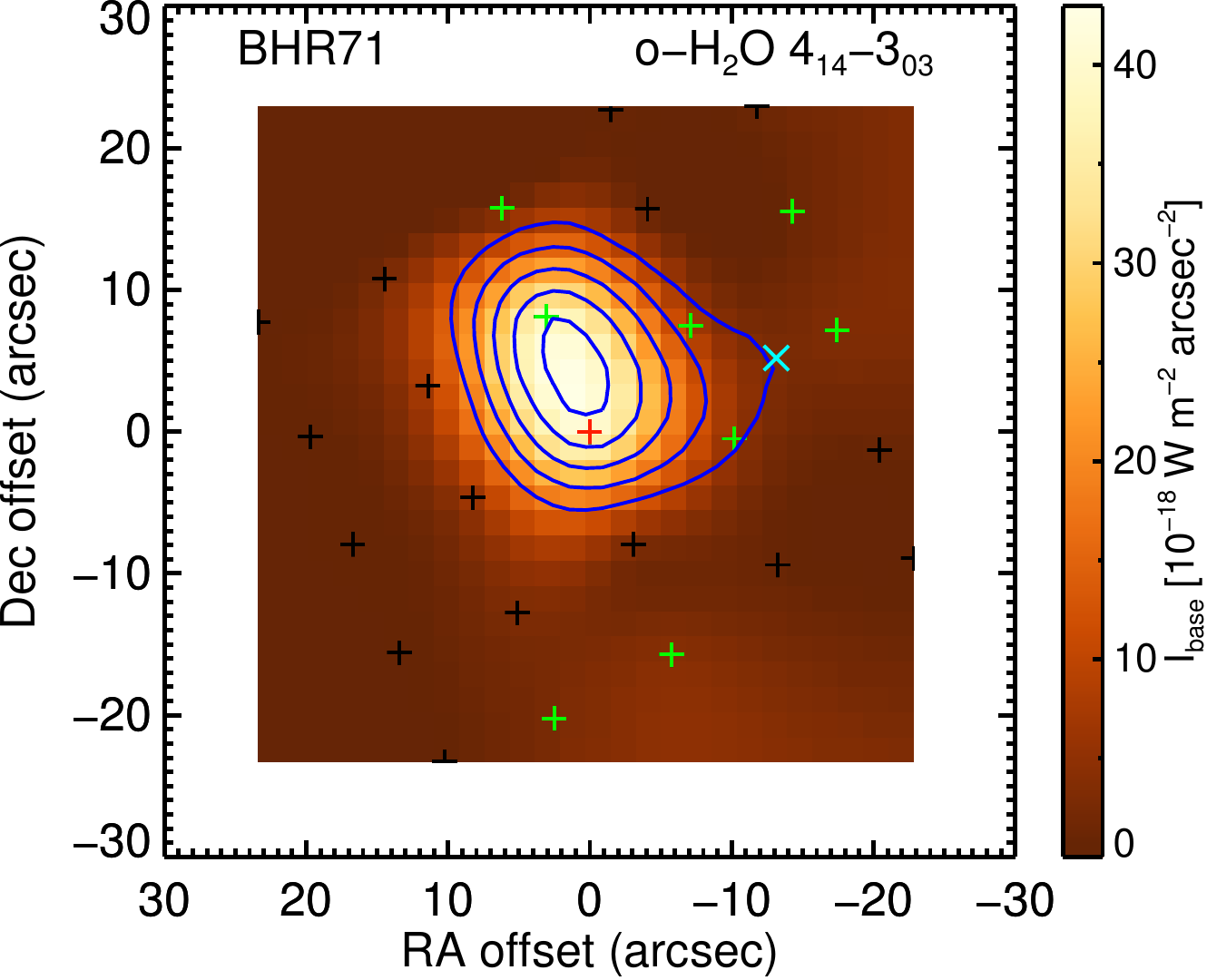}
    \\
    \includegraphics[width=.32\textwidth]{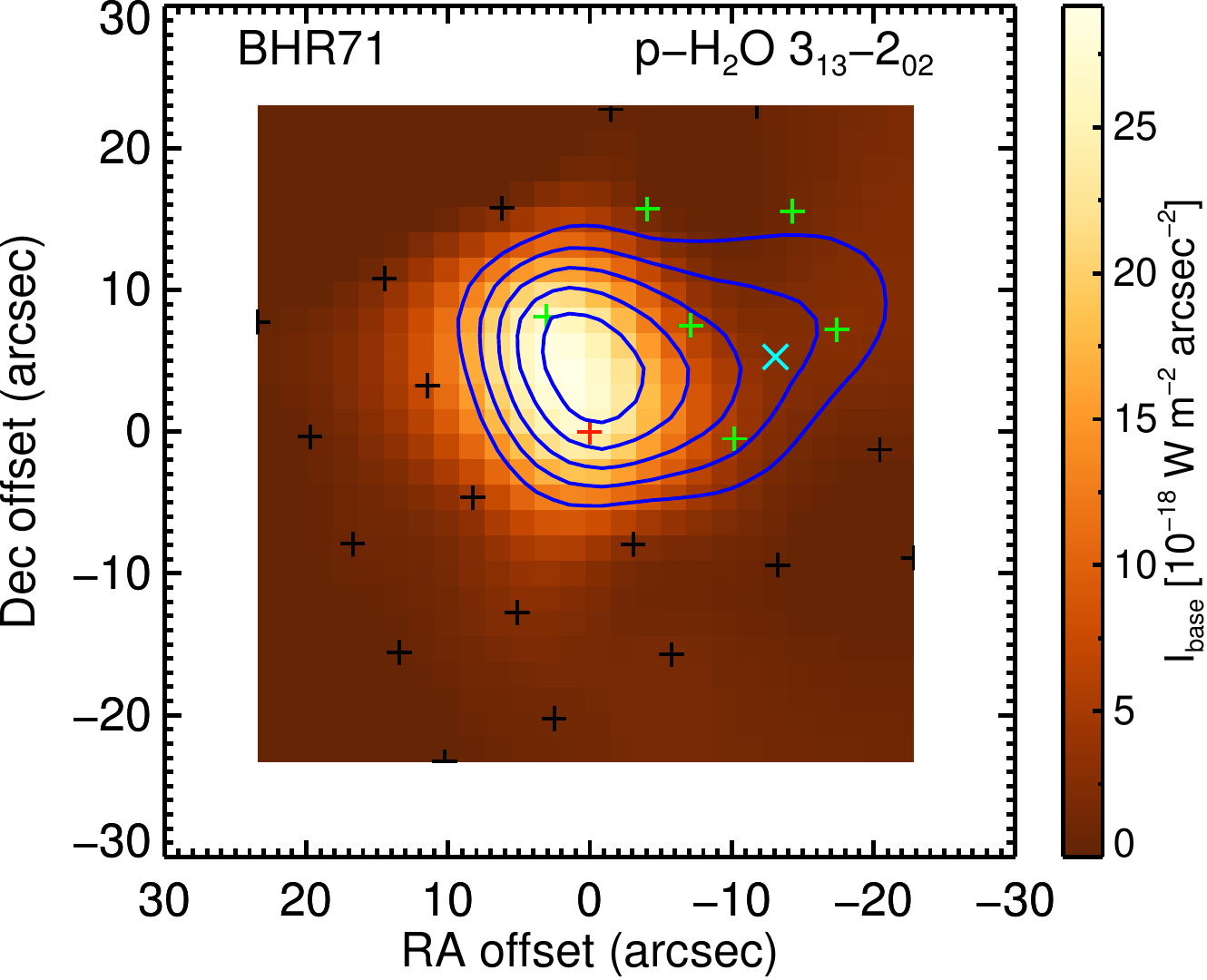}
    \includegraphics[width=.32\textwidth]{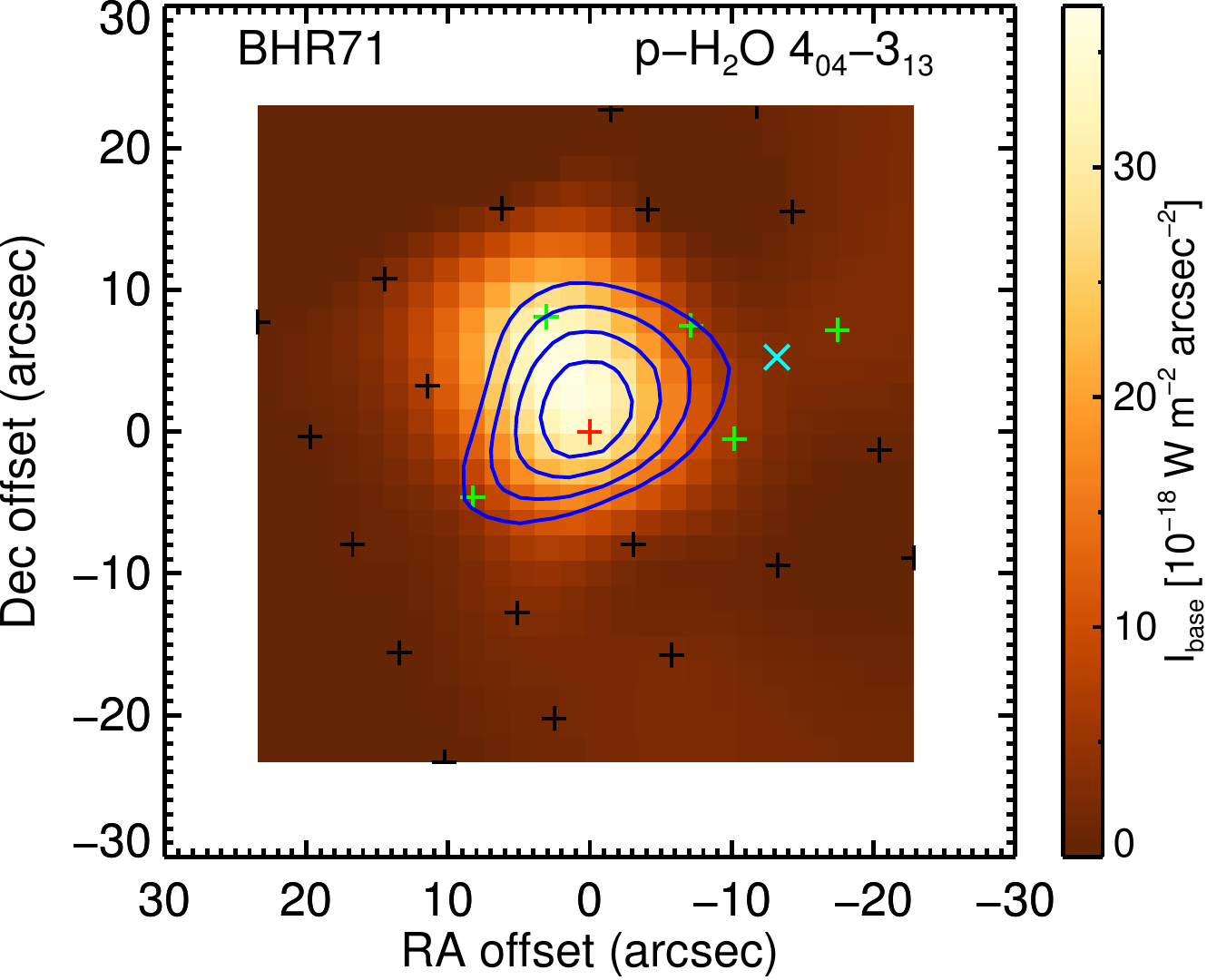}
    \caption{The water emission lines distribution, observed by PACS.  The color code, contour levels, and symbols are the same as Figure~\ref{fig:co_contour}.}
    \label{fig:water_pacs}
    \vspace{10pt}
\end{figure*}

The spaxel distribution of PACS provides a smaller, 40\arcsec$\times$40\arcsec, field of view with better spatial resolution of 9\arcsec\ to 18\arcsec.  Figure~\ref{fig:water_pacs} shows the detected water emission line distributions observed by PACS.  Unlike the characteristic north-south distribution seen in large scale SPIRE contours, the distributions are either centered or extended toward the east-west direction at small scales.  Some lines are not detected at the central position.

The distributions of several water lines extend toward IRS2, marked in cyan ``X''.  The \owater~$2_{21}-2_{12}$ and \owater~$4_{23}-4_{14}$ lines peak around IRS2 with only little or no emission detected at IRS1.
Other emission lines, including \owater~$2_{12}-1_{01}$, \owater~$2_{21}-1_{10}$, \owater~$3_{03}-2_{12}$, \pwater~$3_{13}-2_{02}$, and \pwater~$4_{04}-3_{13}$ show clear extension from the center toward IRS2, but peaking at IRS1.  The correlation of the water emission with IRS2 indicates that IRS2 may contribute an outsized portion of the water emission given that its luminosity is only $\sim 5\%$ of the luminosity of IRS1.

\subsection{Distribution of Atomic Fine-structure Transitions}

\begin{figure*}[htbp!]
    \centering
    \includegraphics[width=.315\textwidth]{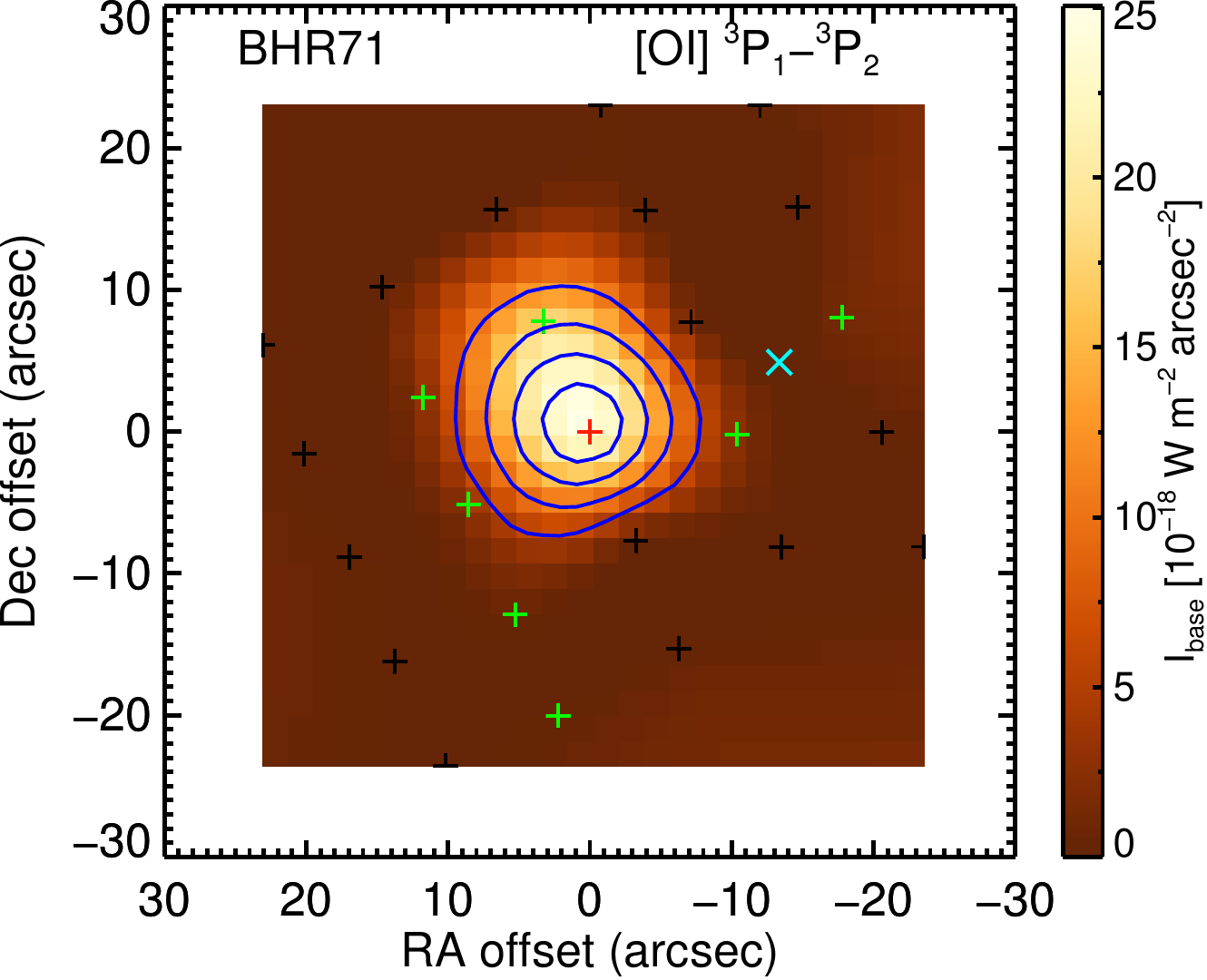}
    \includegraphics[width=.33\textwidth]{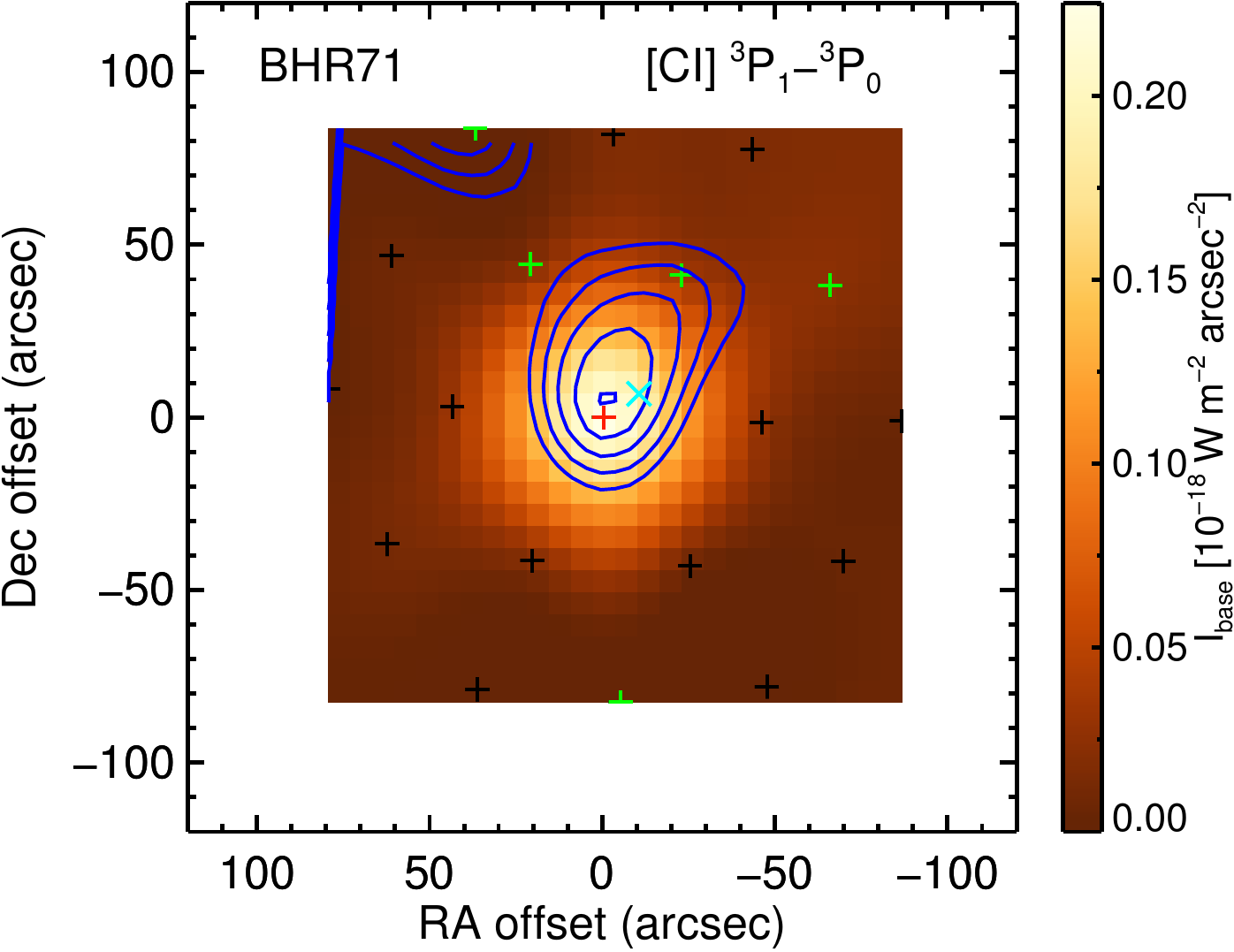}

    \caption{The \oiul{1}{2} emission line distribution is shown at the left, and the distribution of \ciul{1}{0} is shown at the right. The color code, contour levels, and symbols are the same as Figure~\ref{fig:co_contour}.}
    \label{fig:atomic}
\end{figure*}

The atomic fine-structure lines of \oi\ trace shocked regions in the protostellar environment \citep{1985Icar...61...36H,1989ApJ...342..306H},
especially the low density environment ($n\lessapprox \rm{10^{5}~cm^{-3}}$), where O~\textsc{i} is the dominant coolant. In the \oi\ contours (Figure~\ref{fig:atomic}), the distribution of \oiul{1}{2}\ at 63 \micron\ is fairly symmetric, while the \oiul{0}{1}\ at 145 \micron\ is only detected at the central spaxel.  Figure~\ref{fig:atomic} shows that \ciul{1}{0} emission lines are extended toward the north, while the continuum is well-centered.
We observed $\rm [C~\textsc{ii}]~\tensor*[^{2}]{P}{_{3/2}}\rightarrow\tensor*[^{2}]{P}{_{1/2}}$ in absorption, likely the result of off-source emission.  Due to regional widespread $\rm [C~\textsc{ii}]$ emission, the telescope nodding during the observations results in absorption when the off-source position has more $\rm [C~\textsc{ii}]$ emission than BHR71.

\subsection{Relation between IRS1 and IRS2}
\label{sec:binary}

IRS2, first observed with ISOCAM by \citet{2001ApJ...554L..91B}, is separated by $\sim$17\arcsec\ ($\rm \sim 3400~AU$) in projection from IRS1 toward the west.  Two sources can be easily seen in the \spitzer-IRAC image (Figure~\ref{fig:irac}) with a bipolar outflow associated with each source.
Observations of the \coul{3}{2} line \citep{2006A&A...454L..79P} reveal the kinematics of the CO outflows of each source; the outflow driven by IRS1 has a PA of $\rm \sim165^{\circ}-170^{\circ}$, while the outflow driven by IRS2 has a PA between $-35^{\circ}$ and $-30^{\circ}$.
\citet{2000prpl.conf..355M} summarized that two protostellar sources with their own independent envelope are likely to be separated by $\geq$6000~AU; and two sources are likely to share a common envelope if the separation is between 150-3000~AU.  However, the projected separation of 3400~AU  between IRS1 and IRS2 does not provide a clear indication of the relation between two sources.  Note that the separation of 3400~AU is the minimum distance between two sources.
\citet{2008ApJ...683..862C} found that the kinematics are dominated by IRS1 from their observation with ATCA.  Therefore, it is still uncertain whether the two sources are formed by initial core fragmentation or rotational fragmentation during the core collapse.

\begin{figure}[htbp!]
  \centering
  \includegraphics[width=0.47\textwidth]{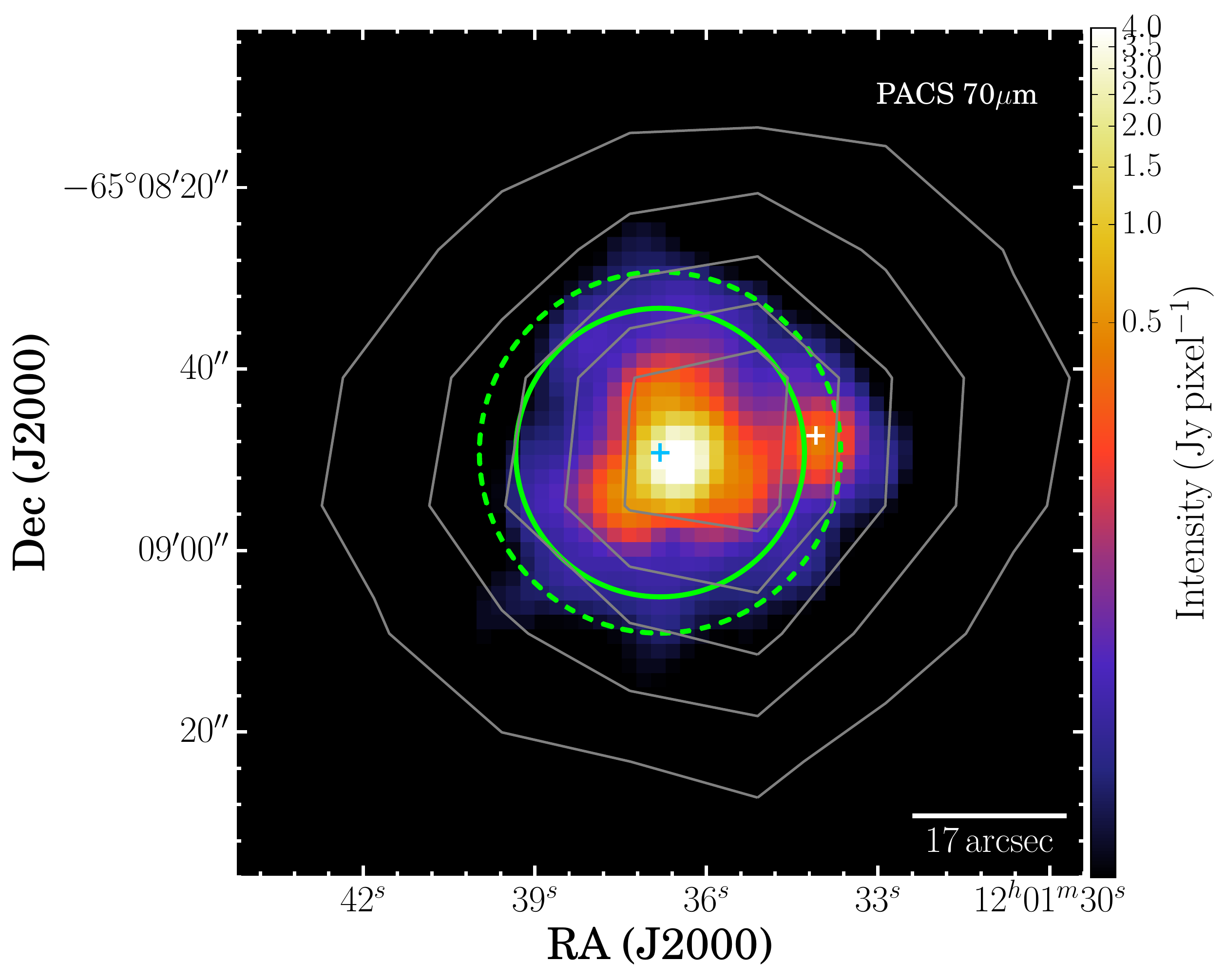}
  \includegraphics[width=0.47\textwidth]{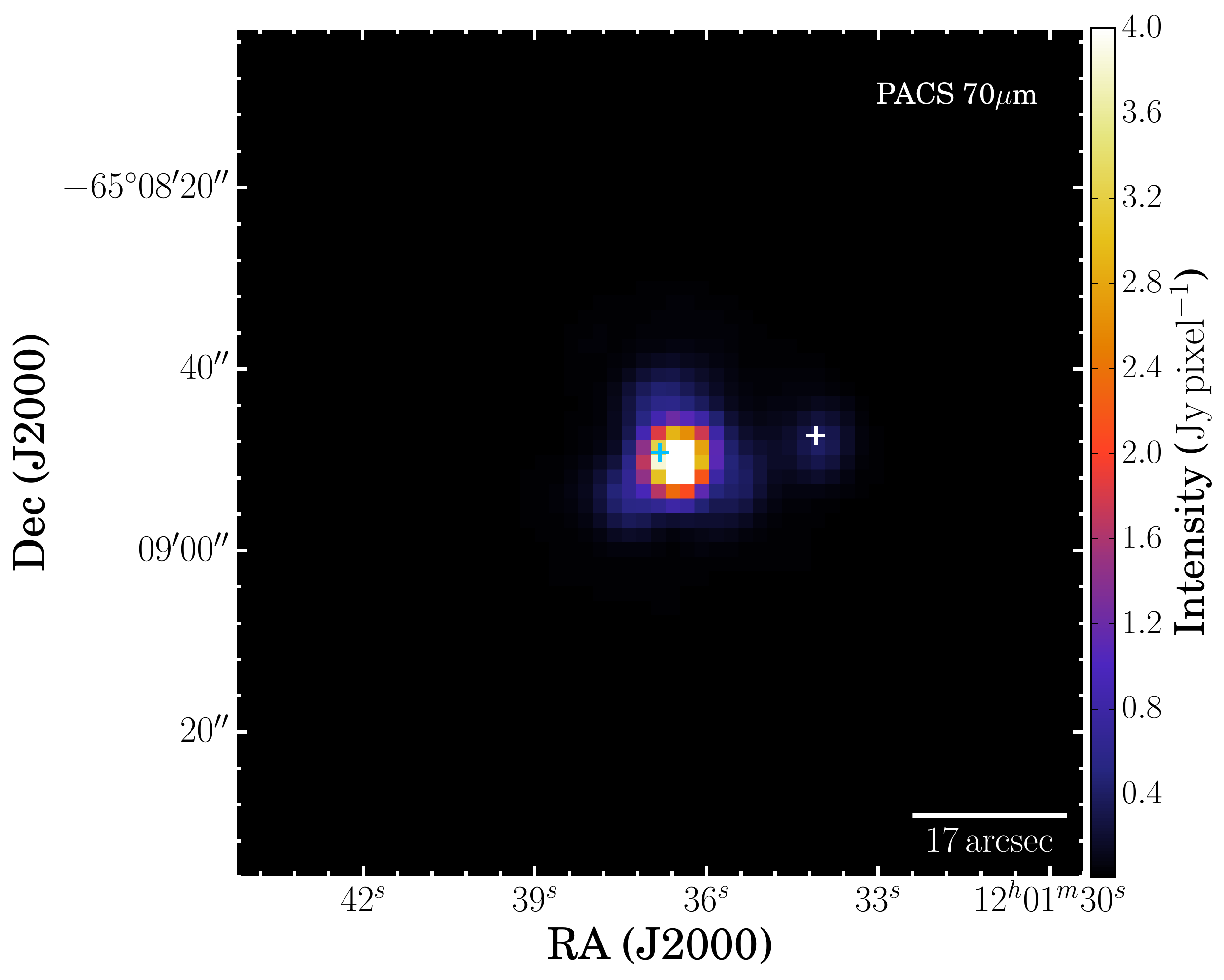}
  \caption{\herschel\ 70~\micron\ image of BHR71 acquired from the Herschel Science Archive.  The center of the image is at the center of BHR71 IRS1.  The flux is plotted with the same maximum and minimum values in logarithm scale on the top and linear scale on the bottom.  The blue and white crosses indicate the positions of IRS1 and IRS2 measured by \citet{2008ApJ...683..862C}.  The solid and dashed circles shown in the top figure indicate the region used for extracting 1-D PACS photometry and SPIRE 500~\micron\ photometry, respectively.  The grey contour shows the SPIRE 500~\micron\ emission.}
  \label{fig:pacs70}
\end{figure}

\herschel-PACS has a spatial resolution of $\sim9\arcsec-18\arcsec$, with which IRS2 may be resolved from IRS1;  \herschel-SPIRE has a spatial resolution of $\sim17\arcsec-40\arcsec$, barely separating the two sources at the shortest wavelength.  Our \herschel\ observations show that emission from IRS1 dominates the continuum emission at all wavelengths.  The continuum emission from IRS2 can be seen only in the PACS 70~\micron\ image with a logarithmic stretch (Figure~\ref{fig:pacs70}, top).  The flux ratio of IRS1 to IRS2 is about 17 at 70~\micron.  Two circles (solid and dashed) show the regions used for extracting the 1-D spectrum at PACS wavelengths and SPIRE 500~\micron.  The peak of IRS2 is not included for the PACS spectrum and photometry, and the flux contributed by IRS2 is at most 6\%.  The spatial resolution at SPIRE wavelengths cannot separate IRS1 and IRS2.  Assuming the same flux ratio measured from PACS 70~\micron\ image, the flux contributed from IRS is $\sim$6\%, similar to the assumption made in \citet{2008ApJ...683..862C}.  IRS2 alone
would produce a dust temperature equal to that produced by IRS1 within a radius of 137~AU, or 0\farcs7, a very small fraction of the region we
model.  Without further constraints on the nature of IRS2, we focus on IRS1 in the following continuum simulations (see Section~\ref{sec:rad}).

\section{Modeling the Continuum Emission}
\label{sec:rad}

A radiative transfer simulation can reveal the contribution and the structure of central protostar, disk, outflow, and envelope in detail. BHR71 is formed within an isolated Bok globule \citep{1995MNRAS.276.1052B}, making it a good candidate for a detailed dust radiative transfer simulation.
With a very complete SED, ranging from 3.6 to 1300 \micron\ and spatial
information at some wavelengths, we can constrain a fairly complex and
realistic model for the source structure. We use information from
previous observations for the distance (200 pc).
 The integral over all wavelengths in Figure~\ref{fig:bhr71_sed} yields
an observed luminosity of 13.92 \lsun. The actual luminosity is
likely to be higher because not all the beams cover the entire source
and because of inclination effects, so we constrain models to match
the observed luminosity by simulating observations into the actual
beam sizes. In optimizing the model, we use azimuthally averaged radial
profiles of intensity at 160~\micron, the ratio of emission at IRAC bands
from the north and south sides, and the SED. How each of these constrains
different model parameters is explained in the relevant sections.

The simulated flux densities into the observed apertures are compared to the observations in Table~\ref{tbl:phot}. For the \spitzer\ spectrum, we adopt the size of four pixels for each module of \spitzer\ (Equation \ref{eq:aper_spitzer}), which is the averaged aperture size adopted in the reduction.
\begin{align}
    d_{\rm aper} & = 7.2\arcsec  & \text{\rm for $\rm 5~\mu m<\lambda \leq 14~\mu m$} \nonumber \\
                 & = 20.4\arcsec & \text{\rm for $\rm 14~\mu m<\lambda \leq 40~\mu m$}
    \label{eq:aper_spitzer}
\end{align}
For the \herschel\ spectrum, we use a constant aperture size, 31.8\arcsec\ for PACS wavelength, and use aperture sizes that increase with wavelength for the three SPIRE bands (see Table~\ref{tbl:phot}).

\subsection{Dust Radiative Transfer Simulation}
\label{sec:dust}

We use \hy, a publicly available three-dimensional radiative transfer calculation package \citep{2011A&A...536A..79R}, as our tool to compare the synthetic observations of different models to the observed spectra.  The simulation uses a two-step procedure, calculating the dust temperature with the Monte Carlo method and ray-tracing photons from each cell.  We set the convergence criteria of the Monte Carlo  calculation with three parameters described in \citet{2011A&A...536A..79R}, $p$, $Q$ and $\Delta$, as 95, 2, and 1.02 in \hy.  In other words, the convergence is achieved when 95\%\ of the differences in specific energy absorption rate between two iterations are less than a factor of 2 and the differences of the 95th percentile of the value difference are less than a factor of 1.02.  The simulation is able to produce reliable results with shorter computational time with this convergence setup.  A model takes $\sim20$ minutes to run, using twenty 2.6~GHz processors.  Each simulation is run with a million photons for determining dust temperature, imaging, and raytracing.  We increase the number of photons for imaging to 50 million for the best fit model and the same model with the geometry suggested by B97 for better image quality (see Section~\ref{sec:incl}).

An input density distribution (Section~\ref{sec:dust_setup}), assuming a gas-to-dust ratio of 100, serves as the input of the  Monte~Carlo dust radiative transfer calculation, together with the dust properties, which are taken from Table~1 Col.~5 in \citet{1994A&A...291..943O} and extended with anisotropic scattering (Section~2.1 in \citet{2005ApJ...627..293Y}), hereafter
referred to as OH5 dust.

\subsection{Physical Model Setup}
\label{sec:dust_setup}
Envelope, disk, and outflow cavity are included in the dust density distribution. The three components are constructed separately but share some parameters.  We include the outflow cavity in our structure to make the simulation consistent with the outflow activities observed around BHR71.  As discussed later in Section~\ref{sec:othermods},  the outflow cavity is essential to achieving the best fit model for BHR71.

\subsubsection{Envelope Model}
\label{sec:envmodel}

We construct the envelope model based on the ``inside-out'' collapse model \citep{1977ApJ...214..488S} and its slowly rotating generalization.
Before the cloud collapse, the cloud can be approximated as an isothermal sphere with $\rho(r)\propto r^{-2}$ (Equation~\ref{eq:isothermal}).
\begin{equation}
  \rho(r) = \frac{c_{\rm s, eff}^{\rm 2}}{2 \pi G r^{\rm 2}}
  \label{eq:isothermal}
\end{equation}
where $c_{\rm s, eff}$ is the effective sound speed of the cloud, including the contribution of non-thermal velocity.  After the cloud starts to collapse, the inner regions of the cloud are infalling, while the outer regions are still in a static phase because the collapse propagates outward at the sound speed of the cloud. Protostars embedded in envelopes are mostly considered to be in this stage. The conservation of angular momentum requires that the outermost material can  fall to only the centrifugal radius instead of the center of the cloud. To take  the effects of rotation into account, we use the slowly rotating collapse model (\citet{1984ApJ...286..529T}; hereafter TSC model) to calculate the envelope density structure.  The  TSC model is specified
by three input parameters, the effective sound speed, the age (\tcol), and the initial rotation speed
($\Omega_{\circ}$). The age is the time since the collapse began and does not include
the pre-collapse evolution.
The relation between these three parameters and other properties are illustrated in Equation \ref{eq:tsc}.
\begin{align}
  \dot{M}_{\rm env} & = 0.975\frac{\csef^{3}}{G} \nonumber \\
  M_{\star} & = \dot{M}_{\rm env}\tcol \nonumber \\
  R_{\rm c} & = \frac{\Omega_{\circ}^{2}G^{3}M_{\star}^{3}}{16\csef^{8}} \nonumber \\
  R_{\rm inf} & = \csef \tcol ,
  \label{eq:tsc}
\end{align}
where $\dot{M}_{\rm env}$ is the mass infall rate of the envelope,
\mstar\ is the mass of the central star,
$R_{\rm c}$ is the centrifugal radius of the envelope,
and \rinf\ is the radius reached by the wave of infall.

It is necessary to include the static envelope beyond the infall radius in the envelope model for early stage protostars, because only the inner part of their envelopes are collapsing. The popular model grids of \citet{2006ApJS..167..256R} do not include the static envelope and assume the asymptotic solution, which applies only to the inner part of the infalling, rotating envelope.  Our envelope model setup has the full TSC model, which includes the transitional region between the infalling inner part and the static outer part \citep{2010ApJ...710..470D}.  In contrast, the collapse-only model is calculated from equation~\ref{eq:env}  \citep{1976ApJ...210..377U}.
\begin{align}
    \rho(r,\theta) = \frac{\dot{M}_{\rm env}}{4\pi(GM_{\star}R_{\rm c}^{3})^{1/2}}\left(\frac{r}{R_{\rm c}}\right)^{-3/2}\left(1+\frac{\mu}{\mu_{\circ}}\right)^{-1/2} \nonumber \\
    \quad \times\left(\frac{\mu}{\mu_{\circ}}+\frac{2\mu_{\circ}R_{\rm c}}{r}\right)^{-1}
    \label{eq:env}
\end{align}
where ($r,\theta$) is defined in spherical coordinate, $\mu$ is cos$\theta$ ($\theta$ is the polar angle with respect to the rotational axis), $\mu_{\circ}$ is the cosine of the angle of a streamline of infalling particles as $r \to~\infty$. We use the streamline equation to solve for the $\mu_{\circ}$ at each cell (Equation \ref{eq:mu_o}).
\begin{equation}
    \mu_{\circ}^{3}+\mu_{\circ}\left(\frac{r}{R_{\rm c}}-1\right)-\mu\left(\frac{r}{R_{\rm c}}\right) = 0
    \label{eq:mu_o}
\end{equation}

\begin{figure*}[htbp!]
  \centering
  \includegraphics[width=0.47\textwidth]{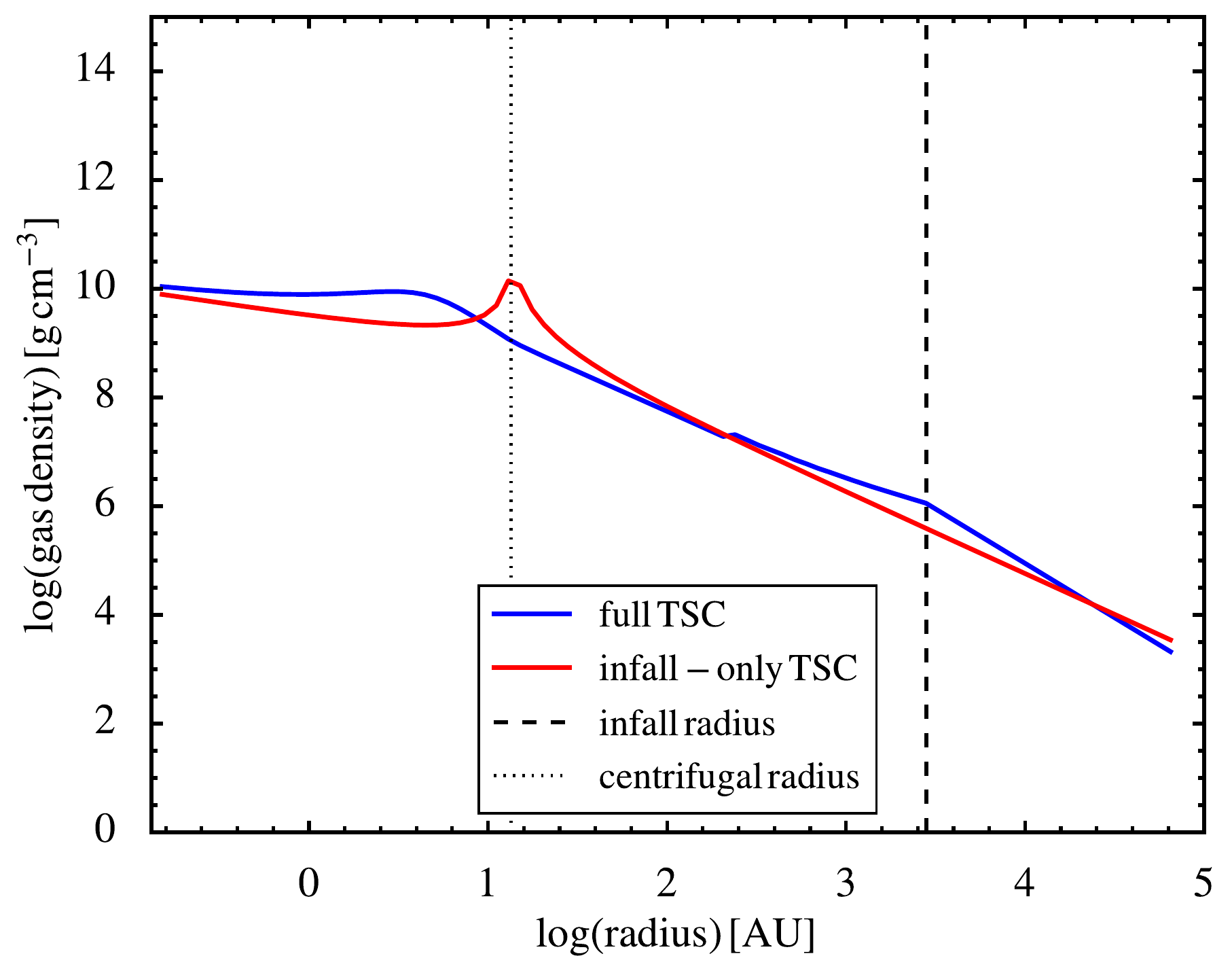}
  \includegraphics[width=0.47\textwidth]{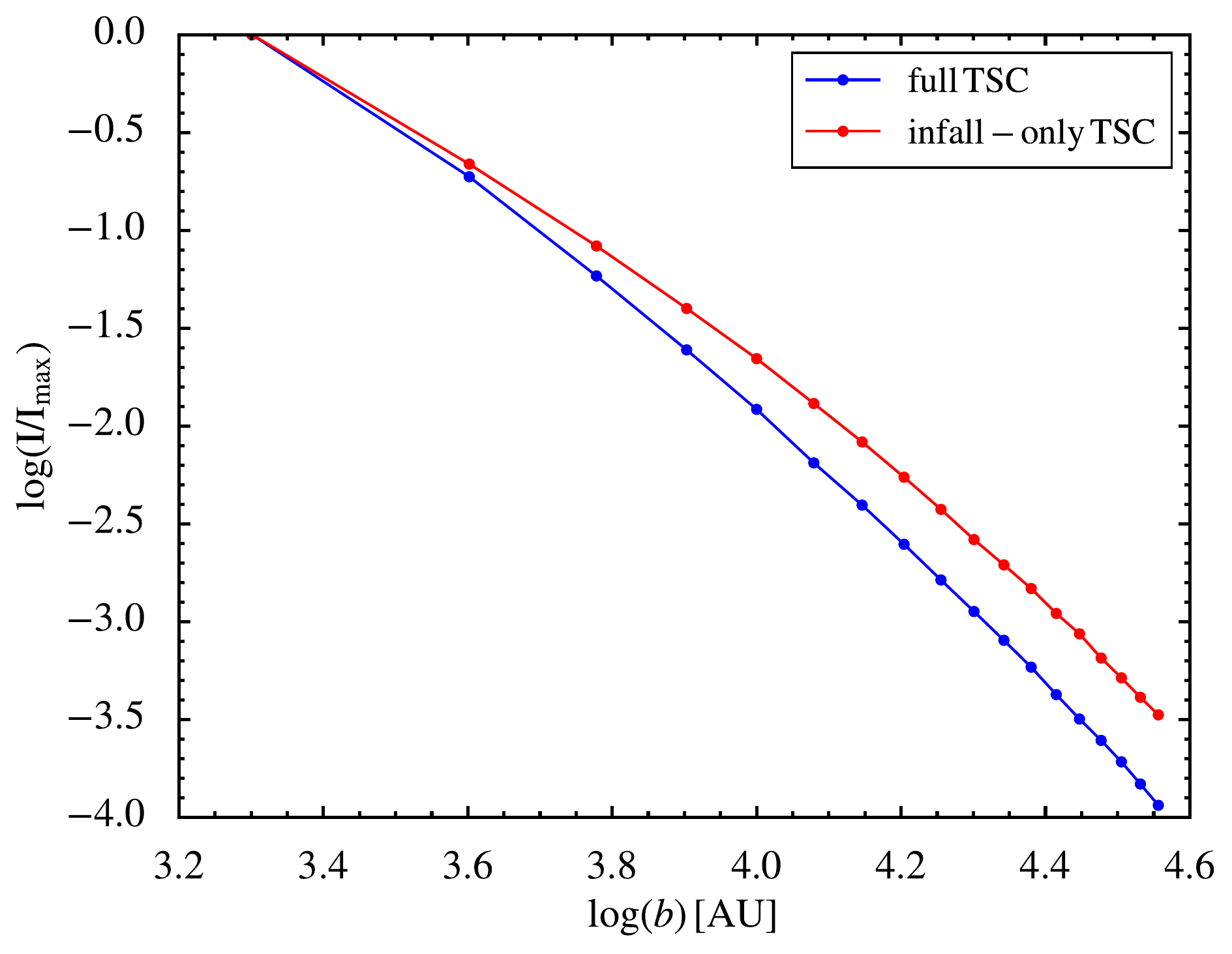}
  \caption{\textbf{Left:} The radial density profiles of the full TSC (blue) and infall-only TSC model (red).  Both models are calculated with the parameters listed in Table~\ref{best_fit}, except for the angular speed.  The parameters are adopted from Table~\ref{best_fit}.  Both profiles are density profiles along the equatorial plane without disks, illustrating the effect of the full TSC model within the centrifugal radius.  The centrifugal radius (black dotted lines) is 13~AU, and the infall radius (black dashed line) is 2800~AU.
  \textbf{Right:} The radial intensity profiles of the full TSC (blue) and infall-only TSC model (red).}
  \label{fig:tsc_com}
\end{figure*}

Figure~\ref{fig:tsc_com} (left) compares the density profiles of the TSC model including the static outer envelope to an infall-only model.  The infall-only model shows a sharp density condensation toward the centrifugal radius and  follows $\rho \propto r^{-1.5}$ relation everywhere beyond the centrifugal radius.  This model has been widely used in the simulations of protostars \citep{2006ApJS..167..256R}, etc.  On the other hand, the full TSC model has a smoother density profile around the centrifugal radius and has more density within the centrifugal radius.  In addition, the full TSC model does not follow a $r^{\rm -1.5}$ relation at all radii between the centrifugal radius and infall radius, and it includes the static profile beyond the infall radius, $\rho \propto r^{-2}$.
The effects of these differences on the SED and the radial intensity profile are substantial (Figure~\ref{fig:tsc_com}, right).

The inner radius is set to the dust sublimation radius. There is a dust-free region at the very inner radius, where the dust is destroyed by the radiation from the central protostellar source. To get an idea about the effect of dust sublimation on the inner radius of the envelope, we calculate the sublimation radius with a simple dust model, blackbody dust, so that $\langle Q_{\rm abs} \rangle_{\star} = \langle Q_{\rm abs} \rangle_{T_{\rm d}} = 1$, where the star in the subscript means the absorption of protostellar radiation and the $T_{\rm d}$ in the subscript means the absorption of dust at the temperature of $T_{\rm d}$.  Combining the equations, we can calculate the dust sublimation radius, $d_{\rm sub}$, with Equation \ref{eq:d_sub}, where $T_{\rm sub}$ is the dust sublimation temperature.
\begin{equation}
  \langle Q_{\rm abs} \rangle_{\star}\pi a^{2}u_{\star}c = 4\pi a^{2}\langle Q_{\rm abs} \rangle_{T_{\rm d}}\sigma T_{\rm d}^{4}
  \label{eq:dust_sub}
\end{equation}
where $u_{\star}$ is the energy density written in terms of the stellar
flux ($F_{\star}$) or luminosity ($L_{\star}$) and distance:
\begin{equation}
  u_{\star} = \frac{1}{c} F_{\star} = \frac{L_{\star}}{4 \pi d^{2}c}
\end{equation}
\begin{equation}
  \frac{d_{\rm sub}}{\rm AU} = 7.76\times10^{4} \left(\frac{L_{\star}}{\rm L_{\rm \odot}}\right)^{1/2}\left(\frac{T_{\rm sub}}{\rm K}\right)^{-2}
  \label{eq:d_sub}
\end{equation}
In the model setup, we set the dust sublimation temperature to 1600~K, similar to the value in \citet{2005ApJ...623..952E}, yielding $d_{\rm sub} = 0.11$ AU for an initial guess with $L_{\star}$=13.92~\lsun.  \hy\ calculates the radiative balance with each cell assuming local thermal equilibrium with the constraint of the dust sublimation temperature.  We use the ``slow'' mode of dust sublimation in \hy, which means that \hy\ decreases the dust density if the dust temperature of a certain cell is found greater than the dust sublimation temperature until the resulting dust temperature is equal to the dust sublimation temperature \citep{2011A&A...536A..79R}.

\subsubsection{The Disk Model}
\label{sec:diskmod}

We add a flared accretion disk model \citep{1973A&A....24..337S,2006ApJS..167..256R}  (Equation \ref{eq:disk}) to the envelope model.
\begin{eqnarray}
\rho(\varpi,z) = \rho_{0}\left(1-\sqrt{\frac{R_{\star}}{\varpi}}\right)\left(\frac{R_{\star}}{\varpi}\right)^{\alpha}  & & \nonumber \\
\times\ \mathrm{exp}\left\{-\frac{1}{2}\left(\frac{z/h_{\rm 100}}
{(\varpi/100~\text{\rm AU})^{\beta}}\right)^{2}\right\} & &
\label{eq:disk}
\end{eqnarray}
where $\varpi$ and $z$ in the equation are the variables in a cylindrical coordinate system, $\rho_{0}$ is the normalization constant for the given disk mass, $h$ is the disk scale height, $\alpha=\beta$+1, where $\beta$ is the flaring power. The disk scale height follows a simple power law with the flaring power $\beta$, where $h \propto \varpi^{\beta}$.  The offset of the disk scale height profile is determined by $h_{100}$ as the scale height at 100~AU.  The disk inner radius is set to the dust sublimation radius, while the outer radius is set to the centrifugal radius calculated by the collapse model.

\subsubsection{The Outflow Cavity Model}
\label{sec:cavmod}

We include an outflow cavity described in cylindrical coordinates by $z = c_{\circ}\varpi^{1.5}$, based on the model in \citet{2006ApJS..167..256R}, where the constant $c_{\circ}$ is determined by the cavity opening angle ($\rm \theta_{cav}$) in Equation~\ref{eq:cavity}.  We define the cavity opening angle as the angle of the edge of the cavity at 10000~AU with respect to the axis along the pole so that the full width opening angle is $2 \rm \theta_{cav}$.  Under this definition, the constant $c_{\circ}$ can be written as
\begin{equation}
 c_{\circ} = \frac{1}{(10000~{\rm AU})^{0.5}} \frac{{\rm cos}\theta_{\rm cav}}{{\rm sin}^{1.5}\theta_{\rm cav}}
 \label{eq:cavity}
\end{equation}

The ``cavity'' is not empty; we fill it with dust from the protostellar wind.  We modeled a variety of density distributions, including pure power laws and power laws modified to have a constant density region near the source.  To conserve the total mass in each slice of the outflow, the density decreases as $r^{-2}$, because the area of each slice increases approximately as $r^{\rm 2}$.  At the inner region of the outflow cavity, we expect a higher density region due to the shock into the ambient envelope, like the cavity shocks defined in \citet{2014A&A...572A..21M}. To quantify the property of this inner region, we simplify its structure as a region with a higher constant density.

First, we estimate the approximate density of the constant density region at the inner region of the cavity as an initial guess.  The idea of this estimation is to calculate the density at the innermost region of the cavity where the mass is distributed by outflows with a velocity of 300~\kms\ for a year.  An outflow travels about 60~AU in one year with a velocity of 300~\kms, resulting in a volume of 1.8$\rm \times 10^{45}~cm^{3}$ with the cavity opening angle of $20^{\circ}$.  For the mass loss rate in the outflow, we have 3.7$\times 10^{-6}~{\rm M_{\odot}~yr^{-1}}$ derived from the momentum and the dynamical age of CO outflow measured by B97, assuming a wind speed of 300~\kms.  This value is used as an initial guess as the inclination angle may change after the model optimization.
Combining the total mass in the outflow within a one year travel time and the corresponding volume, we estimate the gas density of the constant density region as 4$\rm \times 10^{-18}~g~cm^{-3}$.  We then further adopt the gas-to-dust ratio of 100, resulting in a dust density of 4$\rm \times 10^{-20}~g~cm^{-3}$, which is the initial guess of the density of the constant density region at the innermost part of outflow cavity.

We use a logarithmic grid on the $r$-axis, and a linear grid on the $\rm \theta$-axis and $\rm \phi$-axis. The logarithmic grid can give us enough resolution at the central region, where the inner part of the outflow cavity and the disk have relative smaller structures, while the lower resolution in the outer envelope reduces the computing time. The cell size is limited to 100~AU in $r$-axis to ensure a sufficient resolution in the outer envelope. Since the density profile is azimuthally symmetric, the regular grid is good enough to sample the density structure well.

Besides the density profile, the central luminosity (\lstar) is a free parameter in the model. Because of the asymmetric structure the observed luminosity is not the same as the central luminosity.  \hy\ takes two parameters, a stellar radius and a stellar temperature to specify the luminosity.   We assume the radius of the protostar (\rstar) is 3~$R_{\odot}$ \citep{1992ApJ...392..667P}.  We use the Stefan-Boltzmann law to calculate the effective temperature (\tstar) from the luminosity of the protostar.  Note that the actual luminosity is mostly due to accretion.  The effective temperature is simply a way to parameterize the central luminosity.
\begin{equation}
  L_{\rm cen}=4\pi R_{\star}^{2}\sigma T_{\star}^{4}
  \label{eq:central_lun}
\end{equation}

\subsection{Fixing Some Parameters}
\label{sec:bestfit}

There are 13 parameters in the model (Table~\ref{model_params}).  We focus on constraining the best fit model here, but the effect of each parameter is explored in the Appendix.

\begin{table}[htbp!]
  \ra{1.15}
  \caption{Model parameters}
  \centering
  \begin{tabular}{rp{2.5in}}
    \toprule
    \multicolumn{2}{c}{\bf Envelope parameters} \\
    \hline
    \tcol\ & Age of the protostellar system after the start of collapse. \\
    $\csef$$^{\rm a}$ & Sound speed of the envelope including the turbulent velocity. \\
    $\Omega_{\circ}$$^{\rm a}$ & The initial angular speed of the cloud. \\
    \multicolumn{2}{c}{\bf Disk parameters} \\
    \hline
    $M_{\rm disk}$$^{\rm b}$ & Total mass of the disk. \\
    $\beta$$^{\rm a}$ & The flaring power of the disk. \\
    $h_{\rm 100}$$^{\rm a}$ & The disk scale height at 100~AU. \\
    \multicolumn{2}{c}{\bf Outflow cavity parameters} \\
    \hline
    $\theta_{\rm cav}$$^{\rm a}$ & The cavity opening angle defined in Section \ref{sec:cavmod}. \\
    $\rho_{\rm cav,\circ}$ & The dust density of the inner cavity. \\
    $R_{\rm cav,\circ}$ & The radius where the cavity density starts to decrease. \\
    $\theta_{\rm incl.}$ & The inclination angle of the protostar, 0$^{\circ}$ for face-on and 90$^{\circ}$ for edge-on view. \\
    \multicolumn{2}{c}{\bf Other parameters} \\
    \hline
    $R_{\rm max}$$^{\rm a}$ & Outer radius of the envelope as well as the outer radius of the model. \\
    $T_{\star}$ & The temperature of the central protostellar source assuming blackbody radiation. \\
    $R_{\star}$$^{\rm a}$ & The radius of the central protostellar source. \\
    \bottomrule
    \multicolumn{2}{p{3in}}{$^{\rm a}$These parameters are fixed in the search of the best fit model.} \\
    \multicolumn{2}{p{3in}}{$^{\rm b}$The disk mass is set to be 25\%\ of the total accreted mass.}
  \end{tabular}
  \label{model_params}
\end{table}

With a computational time of $\sim$20 minutes per model, the time for a full grid over all 13 parameters would be prohibitive  ($\sim5^{13}$ models if 5 models for each parameter). We use the parameter studies described in the Appendix to guide our evaluation of the effect of parameter variations.

The search for the best fit model starts from fixing the parameters that have been derived from other observations to reduce the degrees of freedom and avoid degeneracies in simulations.  The sound speed in the envelope ($\csef$) and the initial rotational speed ($\Omega_{\circ}$) can be derived from the kinematics of molecular emission lines.  We fix the sound speed to avoid the degeneracy in the simulated SED between models with a smaller sound speed and a later age and models with a larger sound speed and an earlier age, discussed in Appendix~\ref{sec:env_para}.  We derive the effective sound speed from the linewidth measurement of $\rm NH_{3}$, which has a full width at half maximum of 0.74~\kms, contributed by the thermal velocity dispersion with T=13~K derived by \citet{1995MNRAS.276.1067B}, and 1-D turbulence velocity dispersion.  Using the formulae in Equation~\ref{eq:cseff}, we derive a 1-D turbulent velocity dispersion of 0.34~\kms\ and an effective sound speed of $0.37$ \kms.
\begin{align}
  \sigma_{\rm NH_{3}} & = \frac{\rm FWHM_{\rm NH_{3}}}{\sqrt{8ln2}} = \left(\frac{kT}{m_{\rm NH_{3}}}+\sigma_{\rm NT}^{2}\right)^{1/2} \nonumber \\
  c_{\rm s, eff}^{2} & = \left(\frac{kT}{\mu m_{\rm H}}+\sigma_{\rm NT}^{2}\right)
  \label{eq:cseff}.
\end{align}

The observations of $^{\rm 13}$CO (B97) show a deconvolved map size (FWHM) of 0.65$\times$0.4~pc, with a geometric mean of 0.525~pc. We adopt an envelope outer radius of 0.315~pc, which gives the same area as a two-dimensional Gaussian with a FWHM of 0.525~pc.  The resulting total gas mass of our best fit model is 19~\msun, while B97 derived a total molecular mass of $\sim$40~\msun\ from $\rm ^{13}CO$ emission.  An overestimated dust opacity by us or/and an underestimated abundance of $\rm ^{13}CO$ by B97 could explain the factor of two difference.  \citet{2011ApJ...728..143S} compared the dust opacity measured from infrared extinction and submillimeter continuum emission and found that the OH5 opacity at 850~\micron\ can be overestimated by a factor of two.  Therefore, the total envelope mass can be underestimated by a factor of two.  B97 assumed a [$\rm ^{12}CO/^{13}CO$] ratio of 89, close to the value found in our Solar system.  However, \citet{1993ApJ...408..539L} found an isotope ratio of 62$\pm$4 from four interstellar clouds.  The value often used for local ISM is around 77 \citep{1994ARA&A..32..191W}. The lower
isotope ratios would result in lower estimates for optical depth and lower
masses inferred from  $\rm ^{13}CO$, effects that approach a factor of two.
Some of the discrepancy in mass may also come from the extended structure, which is not included in our model but lies along the line of sight.  \citet{2008ApJ...683..862C} also estimated the gas mass from 3~mm continuum observation, from which they derive 2.12~\msun\ within 0.075$\times$0.069~pc.  If we scale down our total gas mass to the radius of a tophat beam that gives the same area, it would be 2.6~\msun, suggesting that our model setup sucessfully produces the observed envelope properties on small scales.
The observations of $\rm N_{2}H^{+}$ indicate a velocity gradient of 7.8~\kms~pc$^{\rm -1}$ \citep{2008ApJ...683..862C}, corresponding to a rotational speed, $\Omega_{\circ} = 2.5 \times 10^{-13}$~s$^{-1}$, which we adopt.

Additionally, the disk mass is set to be a certain fraction of the total infallen mass, calculated from the products of mass infall rate and age (Equation \ref{eq:tsc}).  We assume that 25\%\ of the total infallen mass forms the mass of the disk, about the maximum usually assumed for stable disks \citep{1990ApJ...358..495S,2009ApJ...692.1609V}.  The other disk parameters are fixed, as we find that the disk is not very significant for the young ages we find.   Also, we fix the radius of the protostar as 3~$\rm R_{\odot}$, suggested by \citet{1992ApJ...392..667P} and adjust the stellar temperature to change the luminosity to match the observations.

Lastly, we fix the outflow cavity opening angle as 20$^{\circ}$ based on the observation of B97.  The geometry of the outflow cavity was modeled by B97, assuming a cavity opening angle of $15^{\circ}\pm5^{\circ}$ to fit the \coul{1}{0} distribution with a biconical flow model, resulting in an inclination angle of 84$^{\circ}$.  The cavity opening angle measured from the \spitzer\ image shows a similar value (Figure~\ref{fig:irac}).  Because of the degeneracy between the cavity opening angle and the cavity inclination angle (see
Appendix~\ref{sec:cav_para}), we choose to fix the opening angle at its largest estimated value in simulations, leading to an upper limit to the inclination angle, defined so that 90$^{\circ}$ represents an edge-on view.

After fixing 8 out of 13 parameters in our model (see Table~\ref{best_fit}), we are left with the following
parameters: \tcol, $\rho_{\rm cav,\circ}$, $R_{\rm cav,\circ}$, $\theta_{\rm incl.}$, and $T_{\star}$.

\subsection{Best-Fit model}
We start with initial guesses for the remaining free parameters based on the best fit model extracted from the grid of SEDs from \citet{2006ApJS..167..256R}.  We then explore the parameter space around those values until a best fit model is found.  Among all free parameters, we first constrain the remaining flexible envelope parameter, the age.

\begin{figure}[htbp!]
  \centering
  \includegraphics[width=0.47\textwidth]{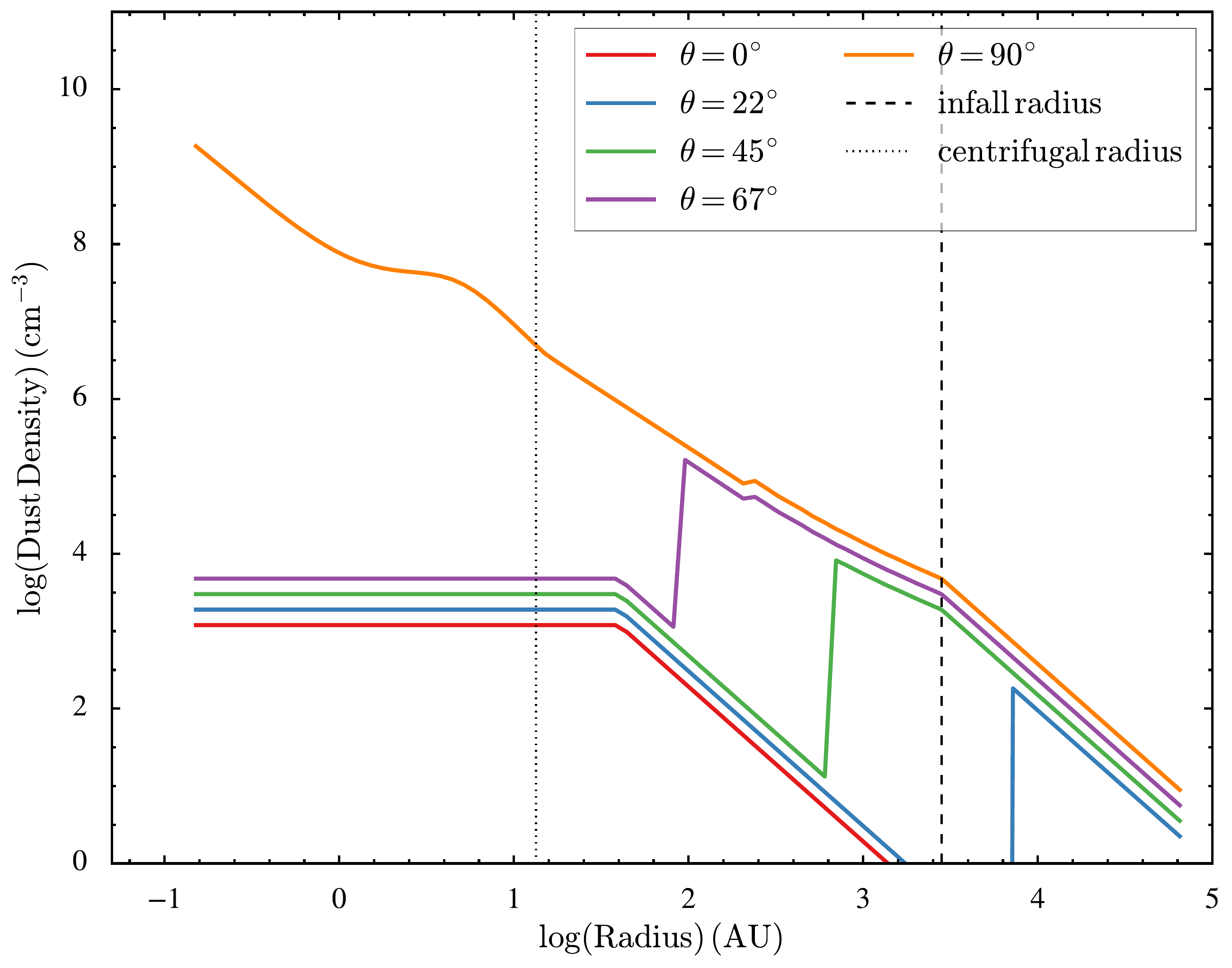}
  \caption{The radial dust density profile of the entire model setup at different polar angles ($\theta$).  The density profiles are offset by 0.2~dex between adjacent lines for a better visualization, while the dust profile at 90$^{\circ}$ (perpendicular to the outflow axis) is unchanged.}
  \label{fig:dust_radial}
\end{figure}

Figure~\ref{fig:dust_radial} shows the radial density profile of our hybrid cavity profile at different polar angles, 0$^{\circ}$, 22$^{\circ}$, 45$^{\circ}$, 67$^{\circ}$, and 90$^{\circ}$.  The flat plateau at the smaller radii represents the constant density region at the innermost cavity followed by a power law decrease, while the flared disk contributes to the density excess at polar angle of 90$^{\circ}$ at radii shorter than centrifugal radius.  The density jump seen at every line of sight except for the one with a polar angle of 90$^{\circ}$ occurs when the line of sight enters the envelope from the outflow cavity.  The glitch around ${\rm log}(r) = 2.2$ is an artifact from the calculations of TSC model.  We find no abrupt change in the simulated SEDs due to this glitch, while exploring the parameter space.

\begin{figure}[htbp!]
    \centering
    \includegraphics[width=0.47\textwidth]{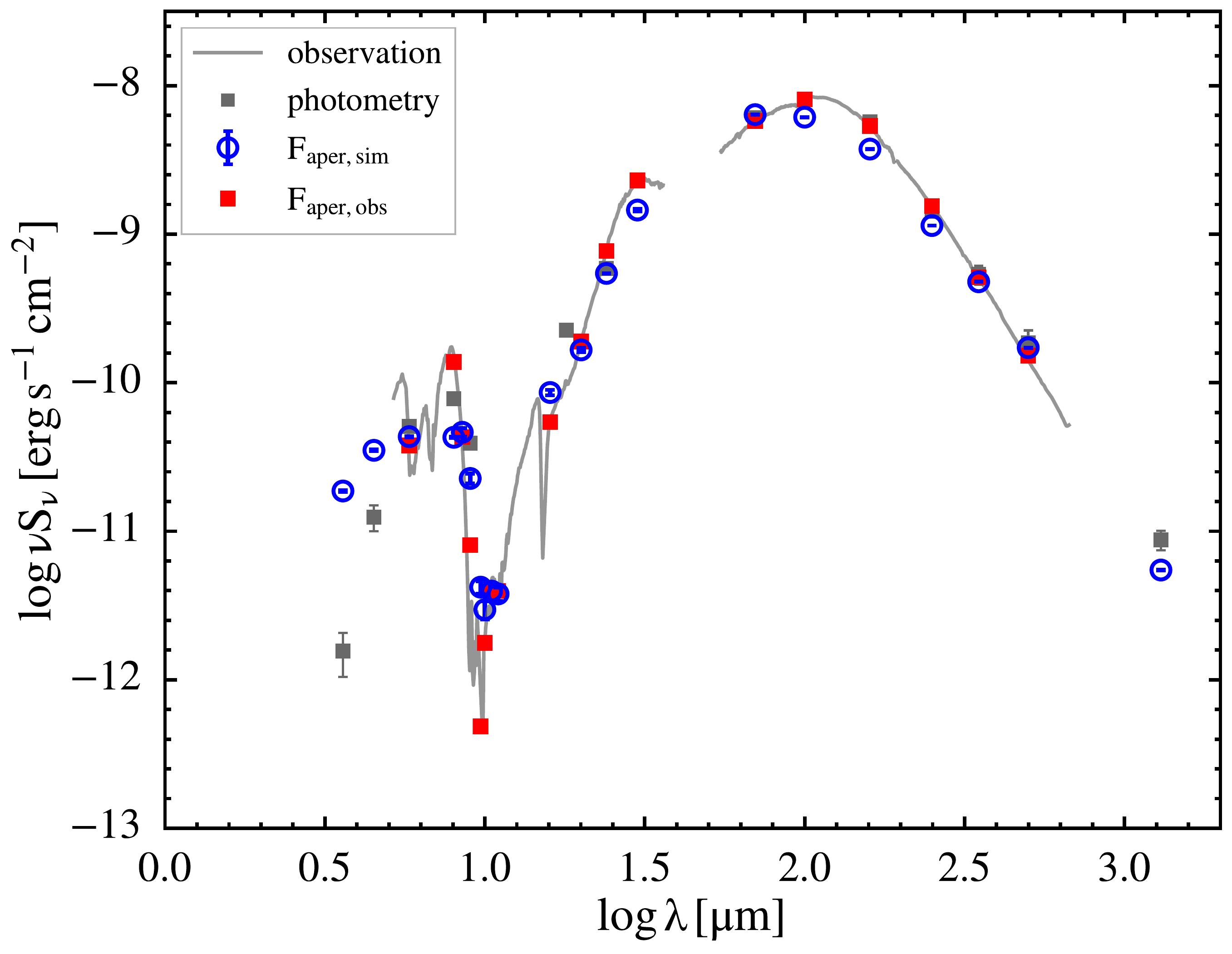}
    \caption{The best fit model compared with the observations.  The filter-convolved and beam-convolved simulated fluxes are shown in blue circles, while the observational data processed with the same procedure are shown in red squares.  The observations are shown in gray, while the photometric fluxes are shown in gray squares.}
    \label{fig:best_fit}
\end{figure}

\begin{figure*}[htbp!]
  \centering
  \includegraphics[width=0.47\textwidth]{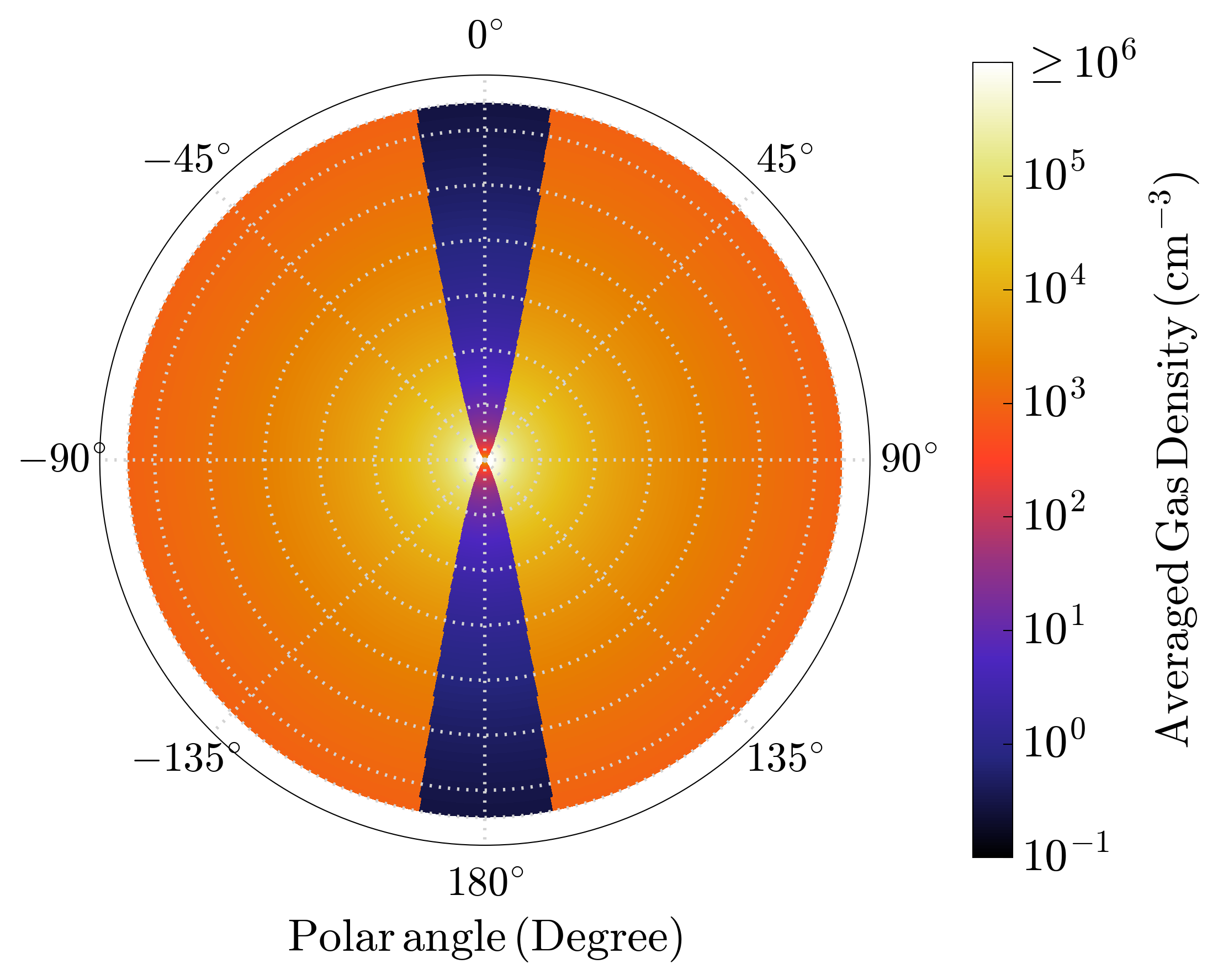}
  \includegraphics[width=0.47\textwidth]{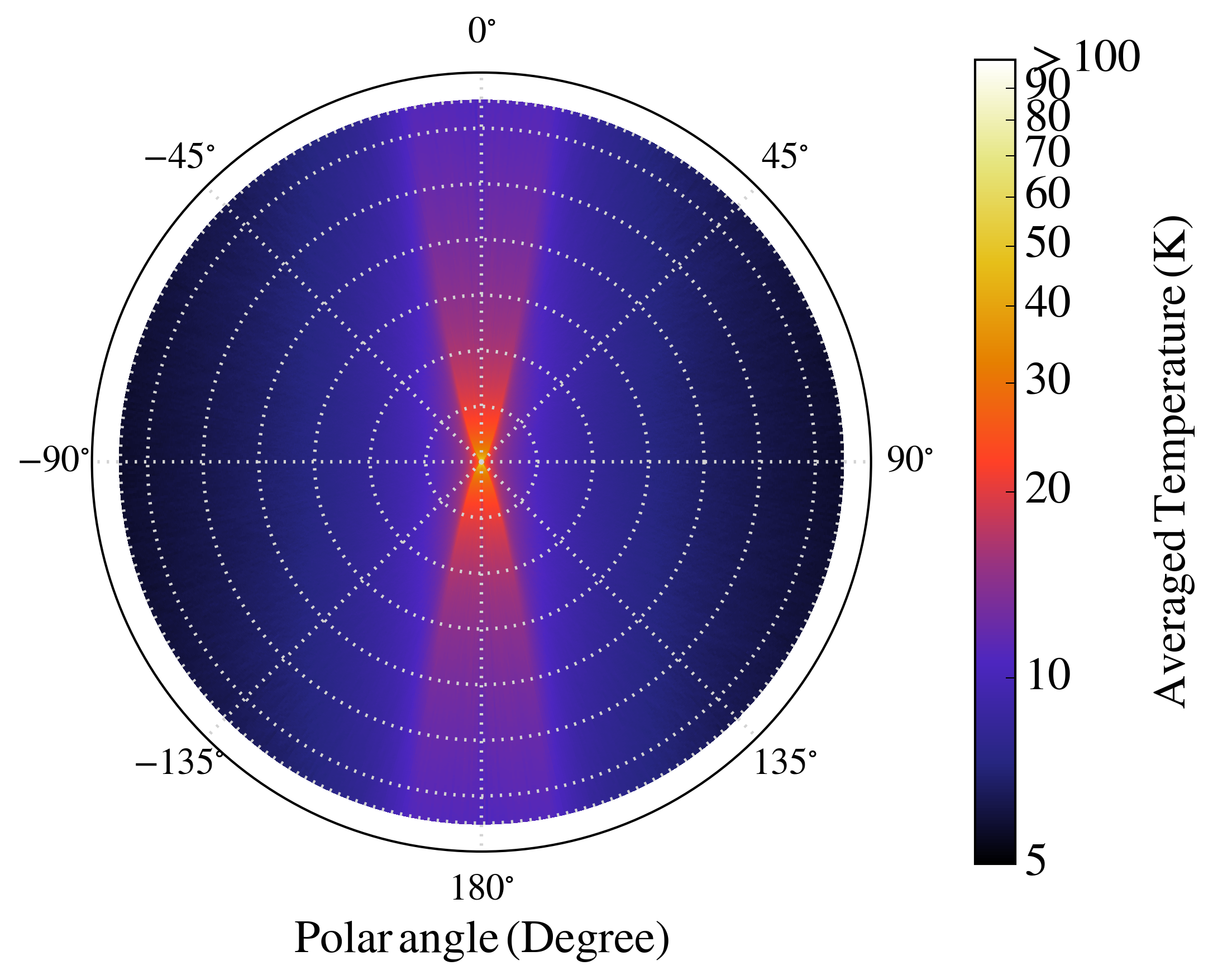}
  \caption{{\bf Left:} The gas density profile in the best fit model.  This figure shows the density averaged in azimuth for better illustration.  The gas density profile is translated into dust density with gas-to-dust ratio of 100. {\bf Right:} The temperature profile calculated from the Monte Carlo radiative transfer simulation.  The temperature in the outflow cavity is about 20~K, while the temperature in the envelope is about 10~K.  The spacing of the illustrative grids is 10000~AU.}
  \label{fig:best_t_rho}
\end{figure*}

Figure~\ref{fig:best_fit} and Figure~\ref{fig:radial_intensity} show the best fitting SED and radial intensity profile.
Table~\ref{best_fit} lists the best fitting model parameters.  Figure~\ref{fig:best_t_rho} shows the gas density profile and the dust temperature profile, with the quantities averaged over azimuth.

The overall fit to the radial profile and the SED is very good, but the exact shape of the SED from 3-5~\micron\ is not recovered with this model, suggesting that the emission from hot dust ($\sim$500~K) can reach the observer via other channels which are not included in our model.  Also note that we overestimate the observations at the shortest wavelengths, possibly due to the lack of understanding of the structures near the center and the uncertainty in the dust model.  The detailed SED in the 3-20 \micron\ region includes many ice features, which are not included in our dust opacity model, such as the CO$_{2}$ ice feature at 15.2~\micron.  The 1.3~mm flux is underestimated in our best fit model possibly due to more extended emission (as seen in our 500~\micron\ image in Figure~\ref{fig:spire500}), which is not included in our model.  Although the observed fluxes around 100-500~\micron\ are reasonably reproduced by our best fit model, the fitting may be improved with a better constraint on the initial rotational speed of the envelope.  We assume the current rotational speed measured by \citet{2008ApJ...683..862C} is the initial rotational speed in our model.  However, the rotational speed will constantly increase as the envelope collapses.  So the value we use here should be the upper limit of the initial rotational speed of the envelope.  As we discuss in Appendix~\ref{sec:env_para}, if the rotational speed is smaller, the fluxes around 100-500~\micron\ will be increased slightly and produce a better fit.

\begin{table}[htbp!]
  \ra{1.15}
  \caption{Best-Fit Model parameters}
  \centering
  \begin{tabular}{R{1in}l c}
    \toprule
    \multicolumn{3}{c}{{\bf Envelope parameters}} \\
    \hline
    $t_{\rm col}$ & 36000 years & \\
    $\csef$ & $0.37~{\rm km~s}^{-1}~^{\rm a}$ & (fixed) \\
    $\Omega_{\circ}$ & $2.5 \times 10^{-13}~{\rm s}^{-1}~^{\rm a}$ & (fixed) \\
    \multicolumn{3}{c}{\bf Disk parameters} \\
    \hline
    $M_{\rm disk}$ & 0.0875 $\rm M_{\rm \odot}~^{\rm b}$ & (fixed) \\
    $\beta$ & 1.093 & (fixed) \\
    $h_{\rm 100}$ & 8.123 AU & (fixed) \\
    \multicolumn{3}{c}{\bf Outflow cavity parameters} \\
    \hline
    $\theta_{\rm cav}$ & $20^{\circ}~^{\rm a}$ & (fixed) \\
    $\rho_{\rm cav,\circ}$ & $3 \times 10^{-20}~\rm {g~cm}^{-3}$ & \\
    $R_{\rm cav,\circ}$ & 40 AU & \\
    \multicolumn{3}{c}{\bf Other parameters} \\
    \hline
    $R_{\rm max}$ & 0.315~pc$^{\rm a}$ & (fixed) \\
    $T_{\star}$ & 6950 K & \\
    $R_{\star}$ & 3 $\rm R_{\rm \odot}~^{a}$ & (fixed) \\
    $\theta_{\rm incl.}$ & $50^{\circ}$ & \\
    \bottomrule
    \multicolumn{3}{p{3in}}{$^{\rm a}$These parameters are fixed in the search of the best fit model.} \\
    \multicolumn{3}{p{3in}}{$^{\rm b}$The disk mass is set to be 25\%\ of the total accreted mass.}
  \end{tabular}
  \label{best_fit}
\end{table}

\subsubsection{Constraining the age (submillimeter)}
\label{sec:age}
The age (\tcol) is determined by fitting the azimuthally-averaged radial intensity profile from the observation with simulations instead of fitting the SEDs.  \citet{2002ApJ...575..337S} demonstrated that the azimuthally-averaged radial intensity profile is diagnostic for the age of the envelope.
As the infall radius increases over time, the structural changes in the envelope are manifested in changes in the radial profile.  The fluxes in SED are the sum of the total flux within given apertures, which do not have the information regarding the radial profiles.  Therefore, the radial intensity profile better traces the age of the envelope.  In addition, the radial intensity profiles are normalized to their peak value so that the effect of the absolute flux calibration is insignificant.  We find that the $\chi^{2}$ values from fitting the radial intensity profiles are only sensitive to age, while the $\chi^{2}$ values from fitting the SED shows more significant correlations with other parameters.  The $\chi^{2}$ for the radial profile is evaluated by Equation~\ref{eq:chi2_rad}.

\begin{align}
  & \chi^{2} = \sum (I_{\rm sim}-I_{\rm obs})^{2} \Big/ \sigma_{\rm comb}^{2} \nonumber \\
  & \chi^{2}_{\rm reduced} = \frac{\chi^{2}}{n -1} \nonumber \\
  & \sigma_{\rm comb}^{2} = \sigma_{\rm sim}^{2} + \sigma_{\rm obs}^{2}
  \label{eq:chi2_rad}
\end{align}
where $I_{\rm sim}$ and $I_{\rm obs}$ are the intensities from simulation and observation, respectively; $\sigma_{\rm sim}$ and $\sigma_{\rm obs}$ are the corresponding uncertainties to $I_{\rm sim}$ and $I_{\rm obs}$; and $n$ is the total number of data points considered in $\chi^{2}$ calculation.

\begin{figure*}[htbp!]
    \centering
    \includegraphics[height=0.4\textwidth]{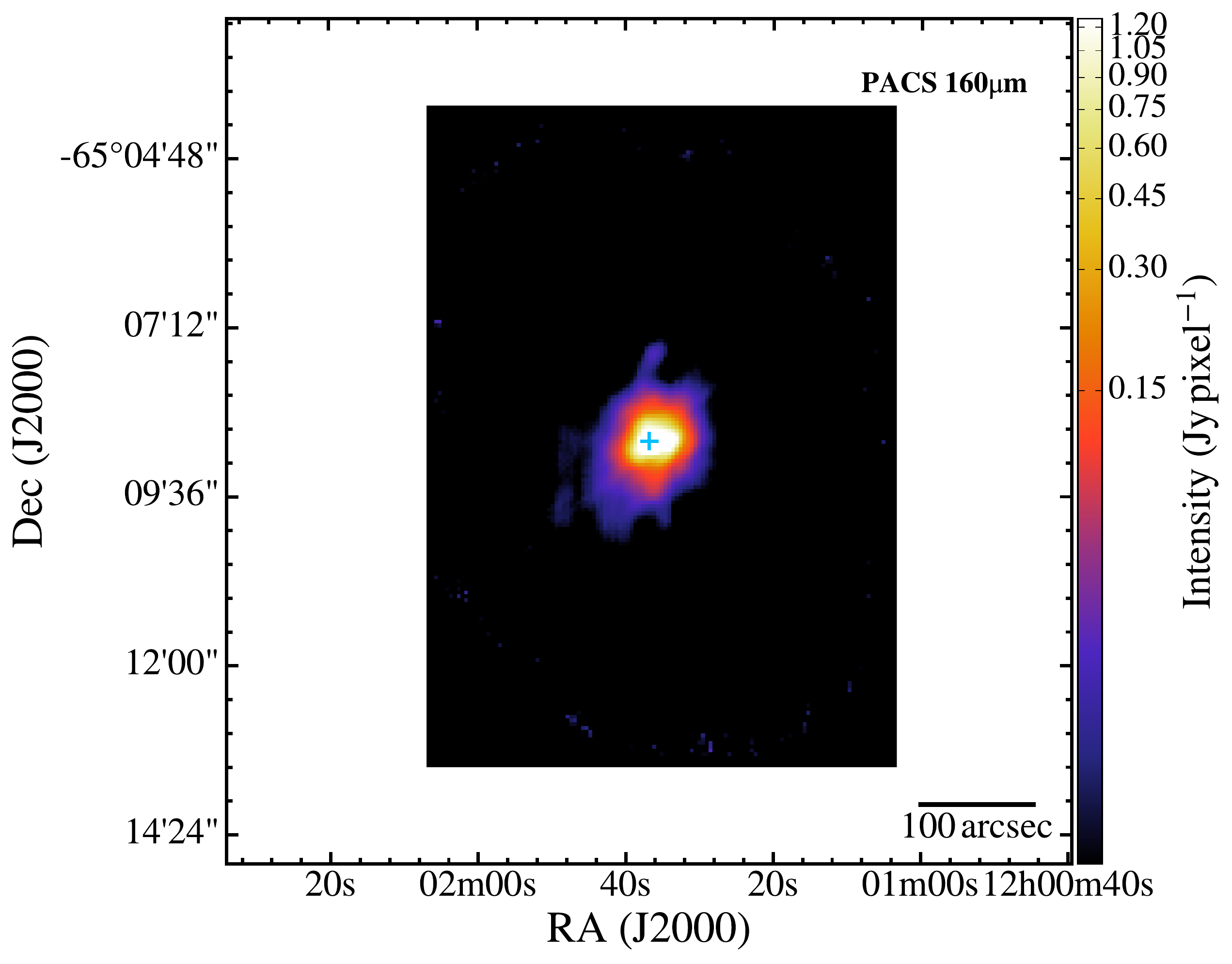}
    \includegraphics[height=0.4\textwidth]{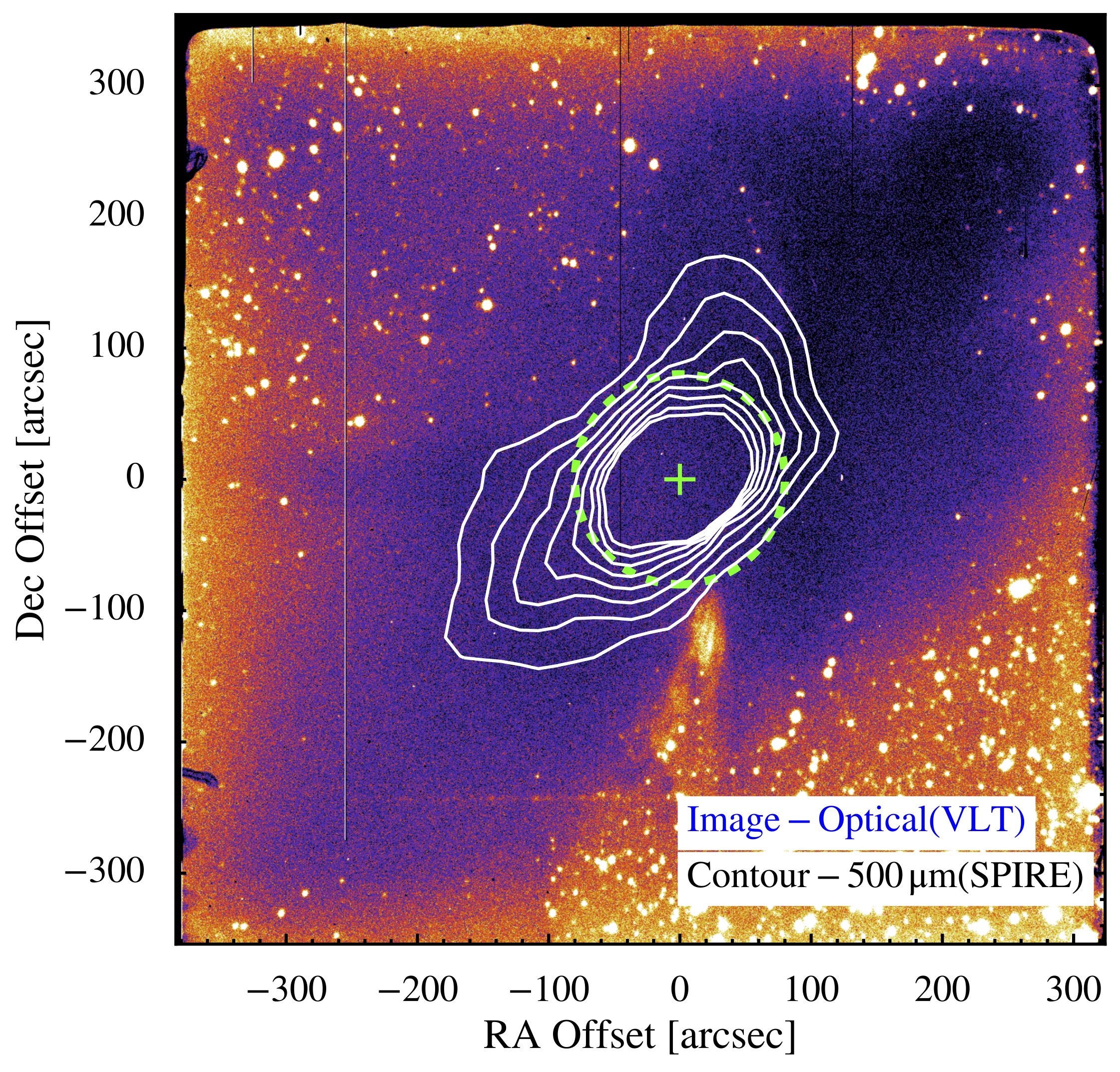}
    \caption{{\bf Left:} The \herschel\ 160~\micron\ image, plotted at the same scale as the image on the right.  {\bf Right:} The VLT-EFOSC2 image taken with Gunn $r$ filter (\#786), centering at 681.4 nm, with the contours from the \herschel-SPIRE 500~\micron\ image.  The increment between adjacent contours is 0.5$\sigma$ with the lowest contour at 1$\sigma$ above the background noise. The length of the image is about 11.8~arcmin.  The green crosses indicate the location of BHR71, and the green dashed circle indicates the region (90\arcsec\ radius) considered for radial intensity profile analysis.}
    \label{fig:spire500}
\end{figure*}

\begin{figure*}[htbp!]
    \centering
    \includegraphics[width=0.47\textwidth]{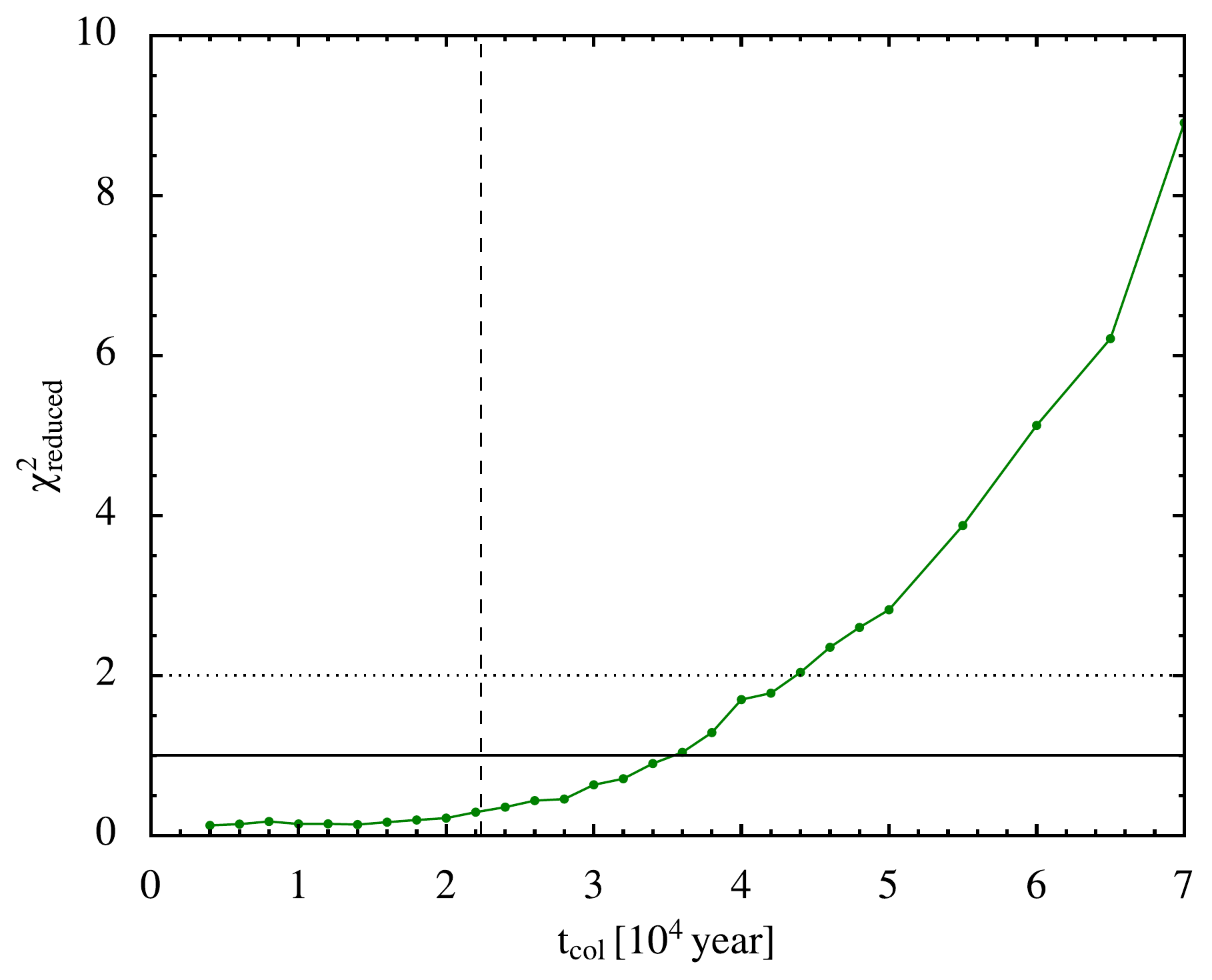}
    \includegraphics[width=0.47\textwidth]{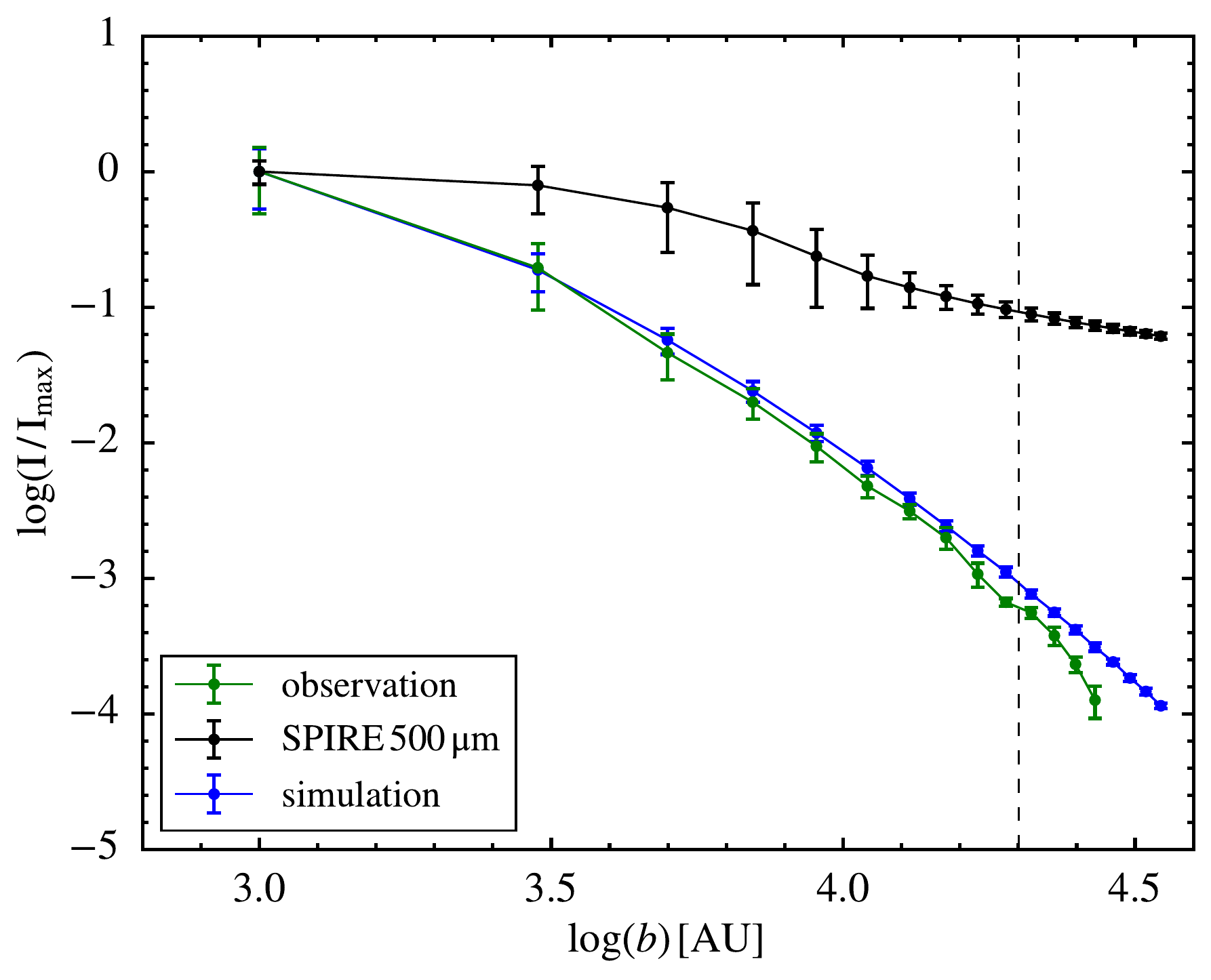}
    \caption{\textbf{Left: }The $\chi^{2}$ distribution from the radial intensity profiles as a function of \tcol.  The dashed vertical line indicates the dynamical age of the outflow, while the horizontal solid line represents $\chi^{2}=1$, and the horizontal dotted line represents $\chi^{2}=2$.
    \textbf{Right: }The azimuthal-averaged radial intensity profile at 160~\micron\ of the best fit model shown in blue, corresponding to an age of 36000 years, while the radial intensity profile extracted from \herschel-PACS 160~\micron\ image is shown in green and the profiles extracted from \herschel-SPIRE 500~\micron\ image is shown in black.  The dashed line indicates the maximum radius, 20000~AU, used for $\chi^{2}$ analysis.}
    \label{fig:radial_intensity}
\end{figure*}

The comparison of the radial intensity profiles with the observed profile is performed at 160~\micron, near the SED peak.  The key factor is to avoid any extended emission that does not belong to the collapsing envelope.  We found the emission at longer wavelengths, such as 500~\micron, is extended irregularly compared to shorter wavelength, such as 160~\micron\ (Figure~\ref{fig:spire500}).
The contour of 500~\micron\ emission shows an irregular distribution beyond 100\arcsec\ radius (Figure~\ref{fig:spire500}, right).  Also the irregular emission coincides with the region of extinction seen in the optical image obtained from ESO archive \citep{1997IAUS..182P..85C}.  The radial intensity profile extracted from 500~\micron\ shows a flatter tail due to the extended emission at larger radii, which has a different characteristic from the profile at inner radii.  Thus, we focus on the radial profile within 90\arcsec\ (18000~AU) to minimize the effect of the extended emission.  The model with an age (\tcol) of 36000 years has $\chi^{2} = 1$, while the models with smaller ages have even smaller reduced $\chi^{2}$, suggesting that the
uncertainties are overestimated.  We set a 1-$\sigma$ upper limit of 44000 years to
the age (Figure~\ref{fig:radial_intensity}).  The lower limit for the age is constrained by the outflow dynamical age, discussed in the next section.

\subsubsection{Constraining the inclination angle (near-infrared and mid-infrared)}
\label{sec:incl}
\begin{figure}[htbp!]
    \centering
    \includegraphics[width=0.47\textwidth]{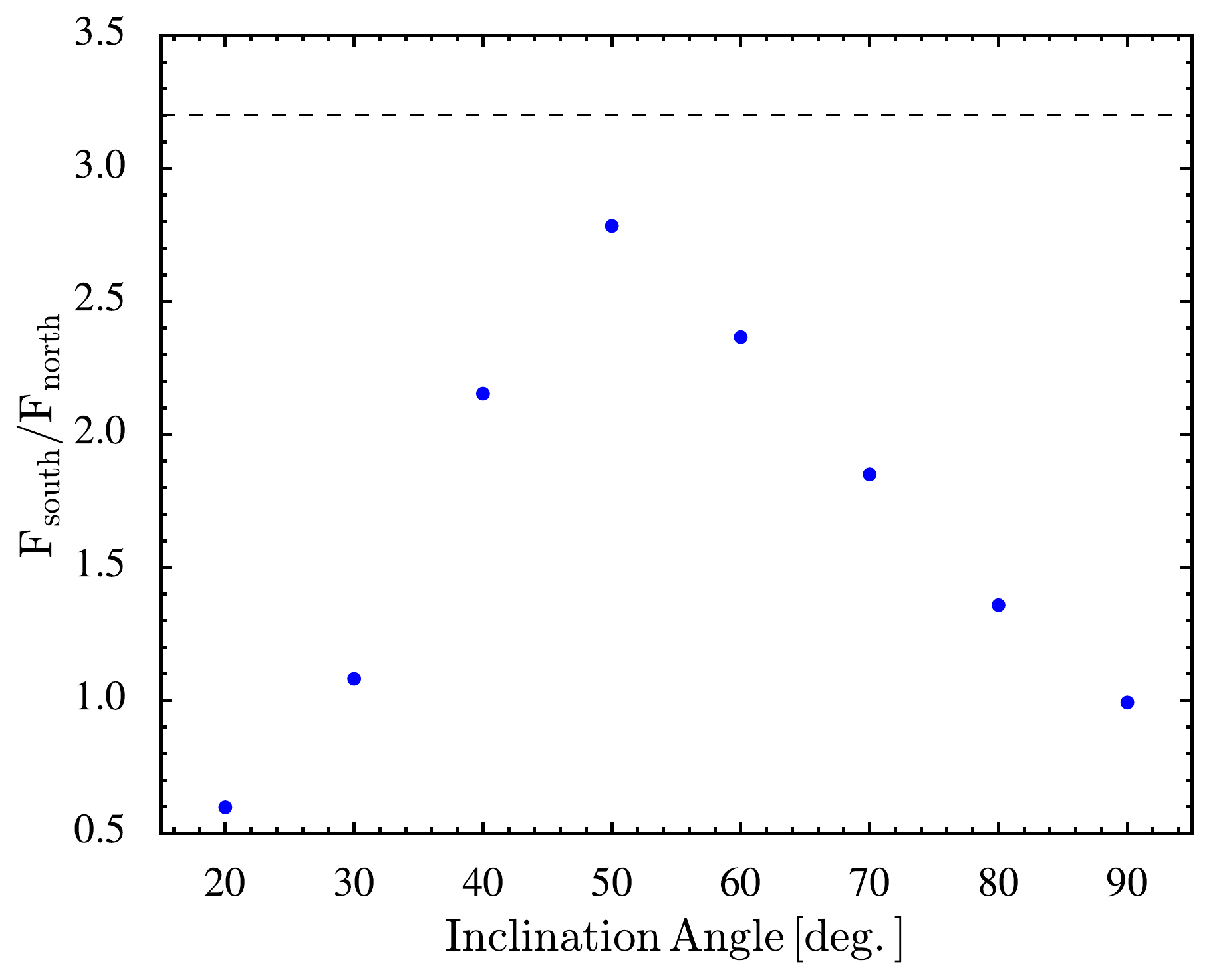}
    \caption{The flux ratio of the south and north lobes of the outflow cavity at given inclination angles, extracted from simulated images with a 100\arcsec$\times$100\arcsec\ box aperture toward the north and south directions.  Only emission at 3.6~\micron\ is considered in this figure.  The dashed line represents the flux ratio measured from BHR71 \spitzer-IRAC~1 image as a reference.}
    \label{fig:flux_NS}
\end{figure}

The inclination angle ($\theta_{\rm incl.}$) primarily affects the emission below 50~\micron.  The \spitzer-IRAC images (Figure~\ref{fig:irac}) show significant brightness differences between the north and south lobe of the outflow cavity, indicating a smaller inclination angle than found by B97.
We use the flux ratios at the north and south part of the outflow cavity as a probe for the inclination angle.  The flux of the IRAC~1 image is measured from a polygon aperture, which includes visually-identified extended features associated with outflow cavity but excludes point source contamination.  The flux ratio from simulated images is measured from two box apertures at the south and north.  Figure~\ref{fig:flux_NS} shows the flux ratio of simulated images with various inclination angles compared with the ratio measured from the \spitzer-IRAC~1 image, 3.2.  The flux ratio increases with decreasing inclination angle as we expected; however, the ratio begins to decline when the inclination angle is smaller than 50 degree, because the emission from the outflow cavity starts to appear in the opposite (north) lobe.  Despite the fact that the maximum ratio from simulated images is still smaller than the ratio from IRAC~1 image, this comparison favors an inclination angle of 50 degrees.

B97 measured the dynamical age of the outflow of BHR71 to be about 10000 years, but the dynamical age is proportional to 1/tan($\theta_{\rm incl.}$) \citep{2014ApJ...783...29D}, yielding a larger dynamical age with the smaller inclination indicated by the \spitzer\ images.  However, if we assume that outflows are only launched after the beginning of the collapse, the dynamical age of the outflow should be the lower limit of the age of envelope.  An inclination angle of 50 degree results in an outflow dynamical age of 24630 years, based on the visual extent and the characteristic velocity of the CO outflow (B97). Based on the lower limit from the outflow and inclination angle, and the upper limit from the radial profile, we choose the age of 36000 years, for which the reduced $\chi^{2}$ from the radial profile reaches unity, as the most likely value.

\begin{figure*}[htbp!]
  \centering
  \includegraphics[width=0.3\textwidth]{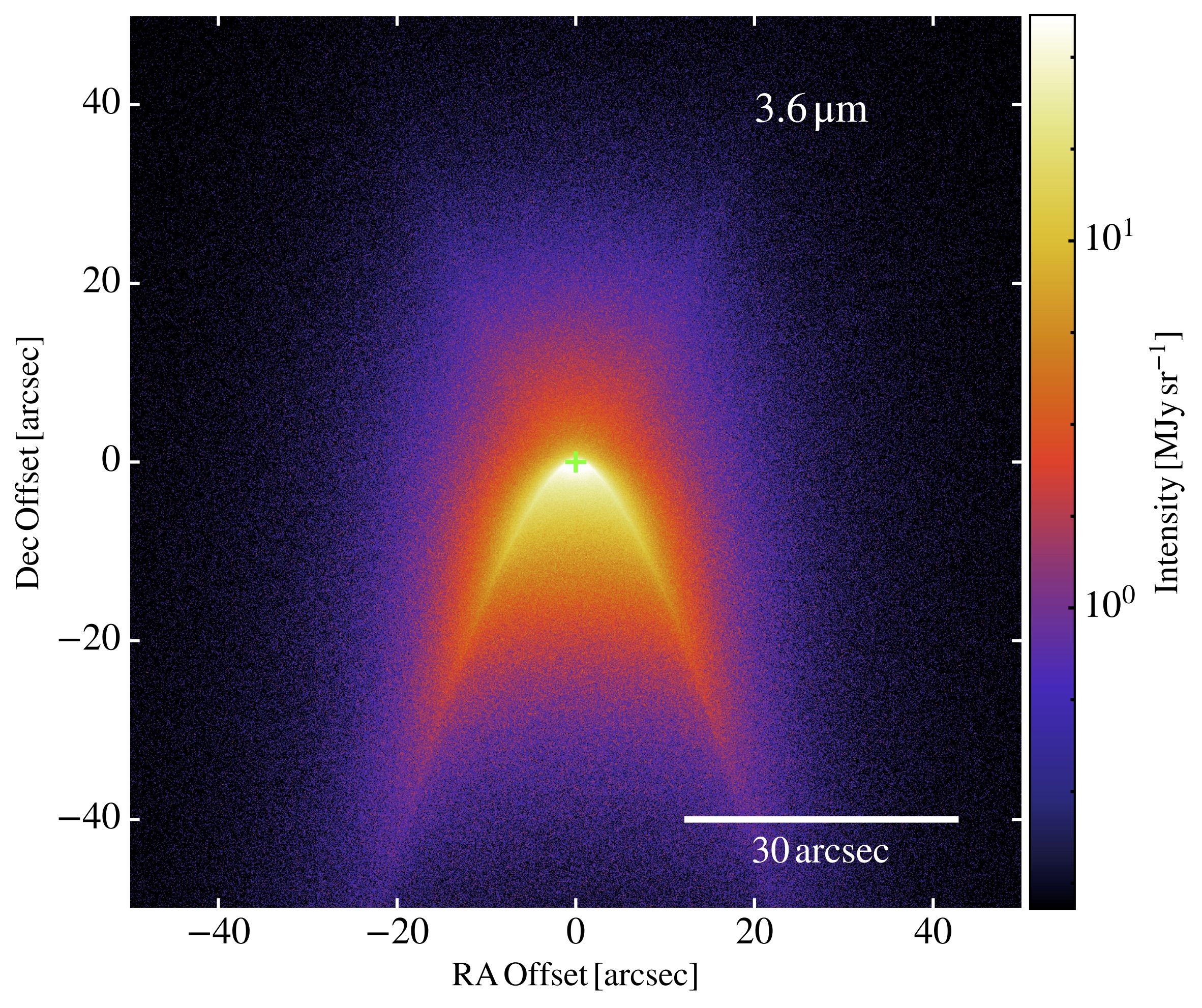}
  \includegraphics[width=0.32\textwidth]{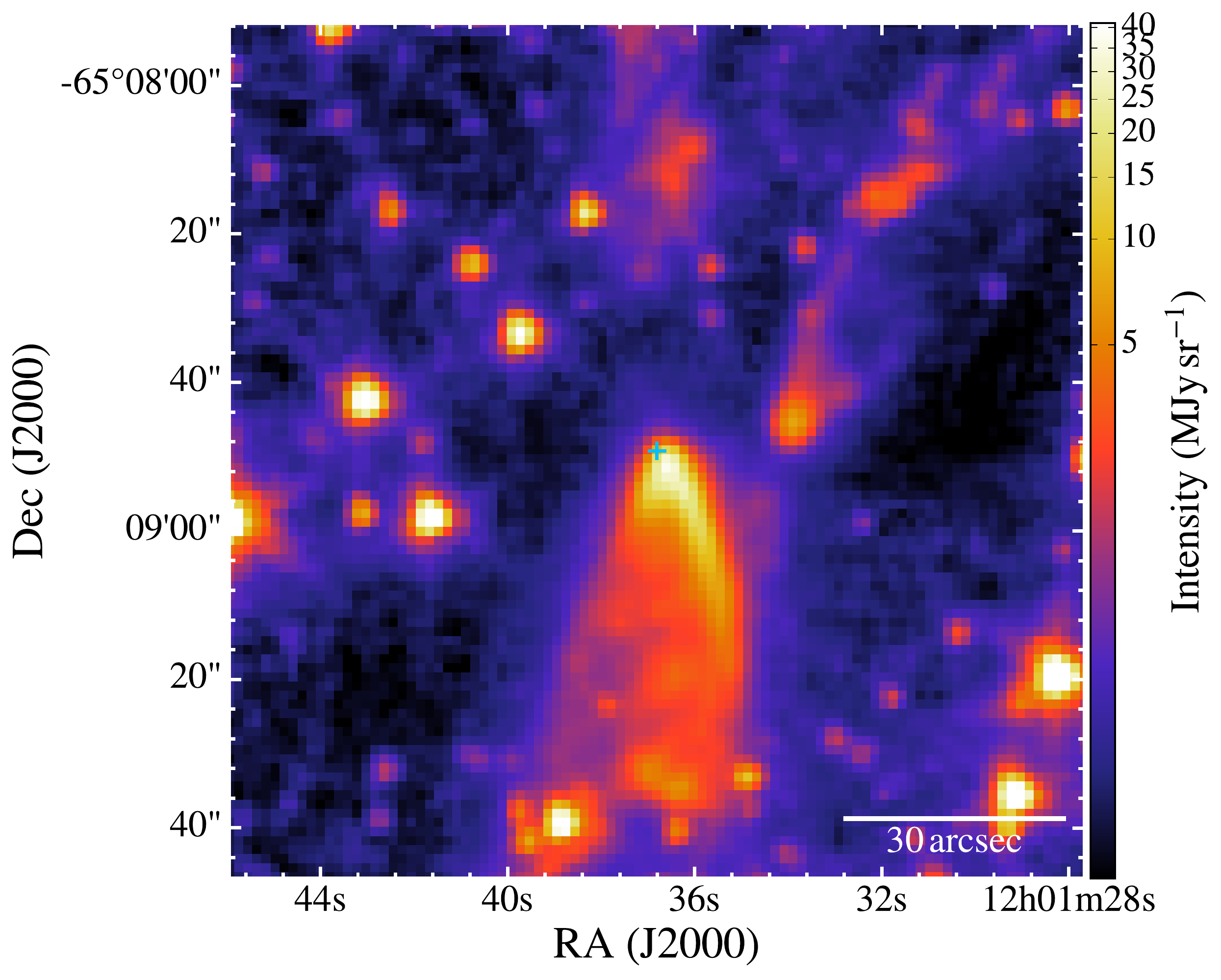}
  \includegraphics[width=0.3\textwidth]{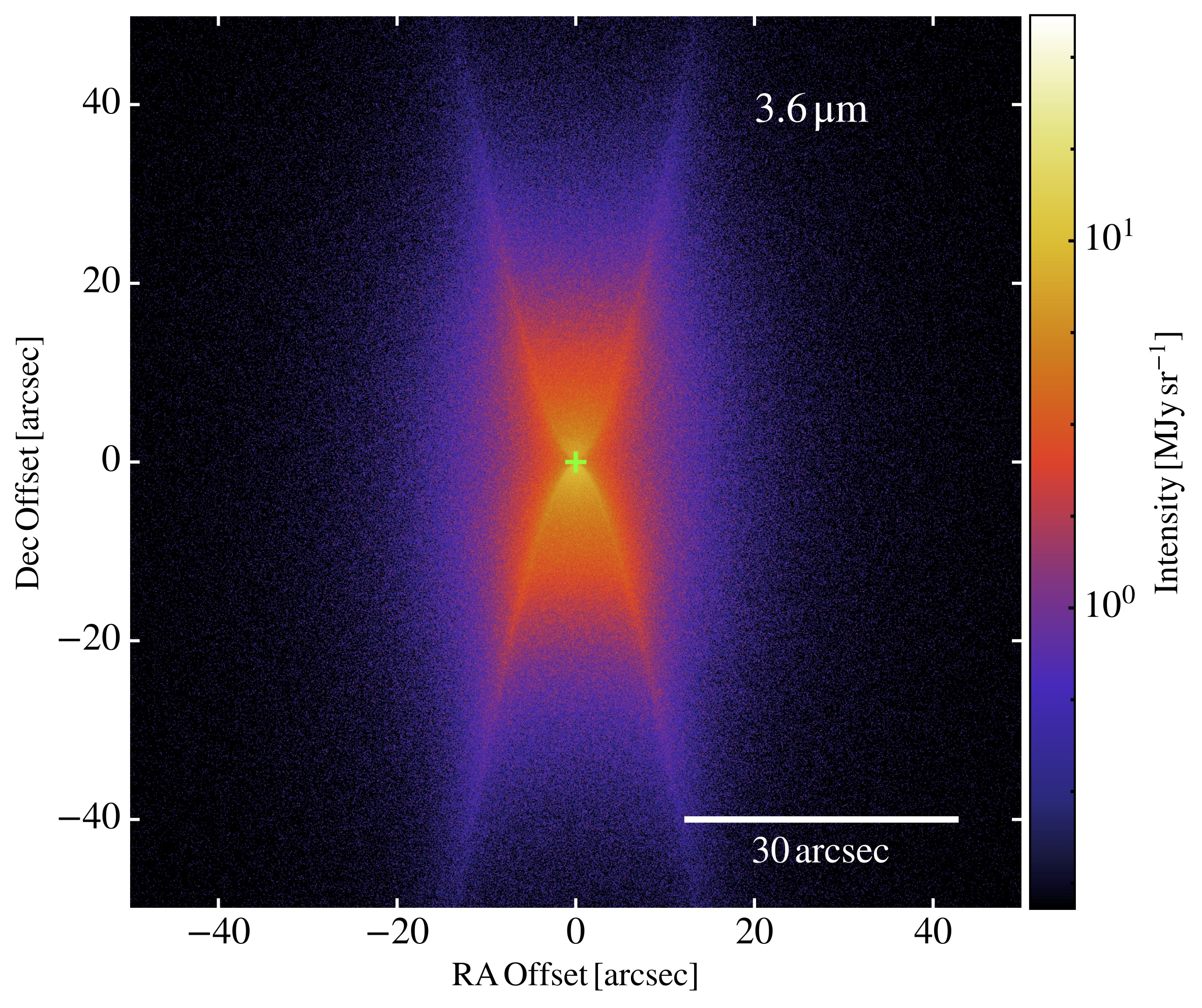}
  \caption{The simulated image at 3.6~$\mu$m of the best fit model in this study (left) and the best fit model with the geometry suggested by B97 (right), compared with the \spitzer-IRAC1 (3.6~$\mu$m) image shown in the middle.  The three images are plotted at a similar spatial scale.}
  \label{fig:flux_ratio}
\end{figure*}

Although an angle of 50 degree best reproduces the IRAC~1 image, this inclination conflicts with the close to edge-on view (84 degree) derived from the CO outflow by B97.  Figure~\ref{fig:flux_ratio} compares the IRAC~1 image with simulated images with 50 degree and 84 degree inclinations.  The morphology of the simulation with 50 degree inclination matches with IRAC~1 image.
The contrast of brightness between two outflow lobes increases with smaller inclination angles until about $50\degree$, where it best matches the observed ratio.  If we measure the outflow cavity opening angle as defined in Section~\ref{sec:cavmod}, we will find the opening angles at 10000~AU from two models are similar (17 degree and 22 degree).  Another possibility is that the extinction in the inner envelope is greater in the north than in the south, due to some deviation from symmetry. The resolution of the conflict between a smaller inclination angle needed to fit the IRAC data and an edge-on orientation based on the CO outflow awaits further observations to understand the kinematic structure of outflows and higher resolution studies of the envelope.

\subsubsection{Constraining the central luminosity}
The central luminosity source is ultimately constrained by the observed bolometric luminosity.  However, the relation depends on the inclination angle. We optimize the luminosity source at two stages, one after determining the envelope parameters, and another one after fixing the inclination angle.  In the first stage, we only tune the central luminosity to match the peak of the SED at far-infrared wavelengths.  The optimization of the central luminosity before fixing the inclination provides less confusion when constraining the viewing angle.  A smaller inclination angle produces more emission at shorter wavelengths, which can increase the observed luminosity by as much as 60\%.  With the inclination angle optimized, we then fine tune the central luminosity to best agree with the observed bolometric luminosity, 13.92~\lsun.

The best fit model requires a central luminosity, $18.8~\rm{L_{\odot}}$, which is $\sim$35\% greater than the observed bolometric luminosity.  The reason is that the asymmetric density structure, with an outflow cavity and flattened envelope, channels
the radiation out the direction of the outflow. Also, the observations do not always include all the emission, especially at longer wavelengths.  Thus, the intrinsic luminosity in the model is greater than the observed bolometric luminosity, 13.92~$\rm L_{\odot}$ for our optimum inclination angle.

\subsubsection{Calculating the $\chi^{2}$ from SED fitting}
We constrain other parameters by calculating a $\chi^2$ value for the SED. To avoid overweighting the high fluxes around the peak of the SED, we calculate the $\chi^{2}$ with the fractional difference instead of the absolute difference between the simulations and observational data.  The prescription of calculating the $\chi^{2}$  is shown in Equation~\ref{eq:chi2}.
\begin{align}
  & \chi^{2} = \sum \left(\frac{F_{\rm sim}-F_{\rm obs}}{F_{\rm obs}}\right)^{2} \Big/ \sigma_{\rm comb}^{2}, \quad F = \nu S_{\nu} \nonumber \\
  & \chi^{2}_{\rm reduced} = \frac{\chi^{2}}{n -1} \nonumber \\
  & \sigma_{\rm comb}^{2} = \left(\frac{F_{\rm sim}}{F_{\rm obs}}\right)^{2} \left(\frac{\sigma_{\rm sim}^{2}}{F_{\rm sim}^{2}} + \frac{\sigma_{\rm obs}^{2}}{F_{\rm obs}^{2}}\right)
  \label{eq:chi2}
\end{align}
where $n$ is the number of aperture-convolved fluxes and $F_{\rm sim}$ and $F_{\rm obs}$ are the aperture-convolved fluxes from the observation and simulation.  The aperture-convolved fluxes are selected by their abilities to represent the features of the simulated SED.  The fluxes at the following wavelengths, 3.6, 4.5, 8.5, 9, 9.7, 10, 10.5, 11, 16, 20, 24, 30, 70, 100, 160, 250, 350, 500, and 1300~$\mu$m, are selected and calculated by convolving with the photometric filters and beam of the instruments when the corresponding data are photometric, or averaging with the spectral resolution of the instruments when the corresponding data are spectroscopic.

\subsubsection{Constraining the cavity density profile (mid-infrared)}
\label{sec:cavdensmod}
\begin{figure}[htbp!]
  \centering
  \includegraphics[width=0.47\textwidth]{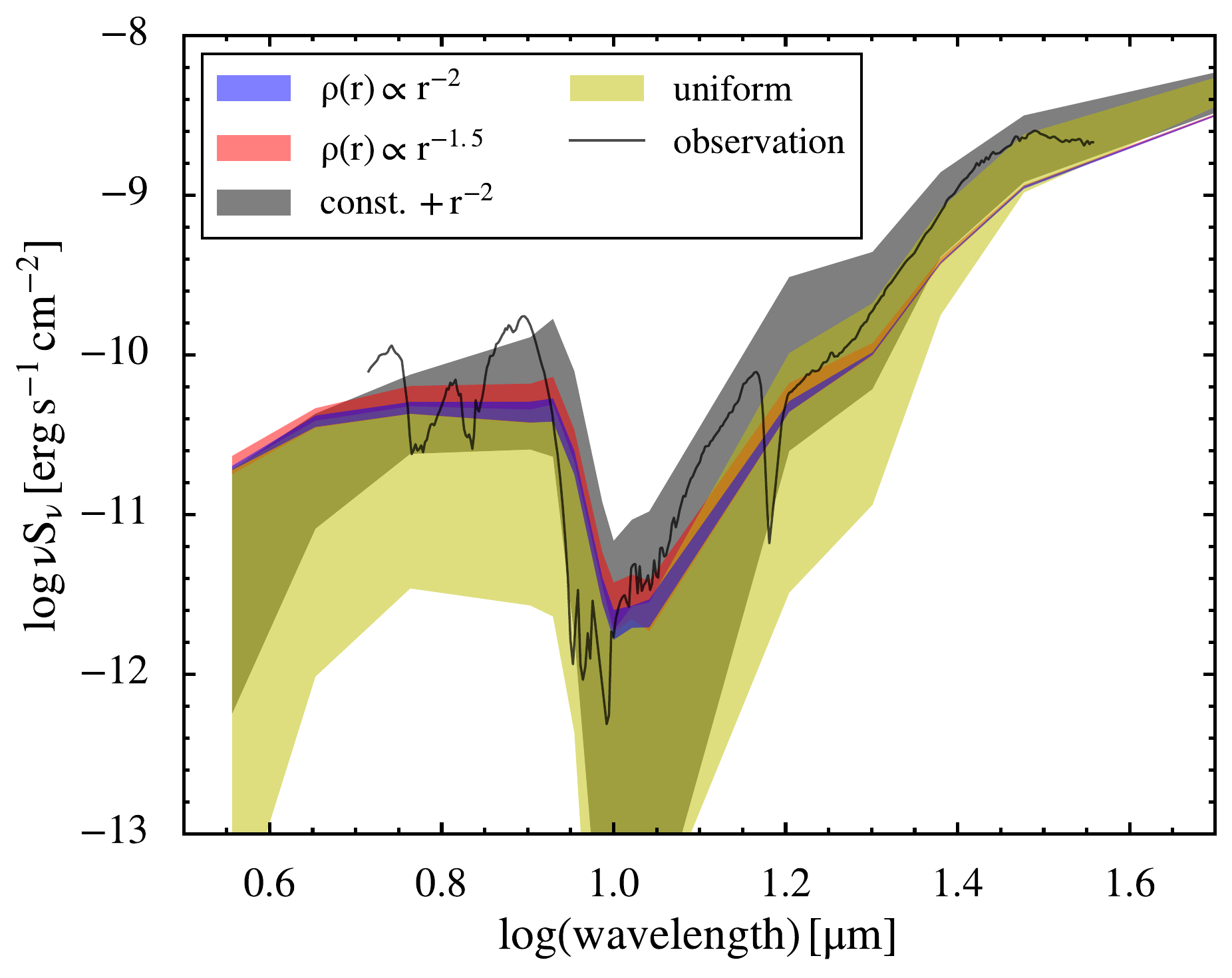}
  \caption{The simulated SEDs with cavity density profile of $r^{-2}$ (blue), $r^{-1.5}$ (red), a constant region with a $\rm r^{-2}$ tail (black), and uniform density (yellow).  The shaded area indicates the range of simulated SEDs with a same grid of density at the innermost cavity, ranging from 5\ee{-18} to 5\ee{-21}~$\rm g~cm^{-3}$ (four models with density increasing by a factor of 10).  The olive green region is the overlap between
the yellow and gray regions.  The observed SED is shown in the black solid line.}
  \label{fig:cav_struc_com}
\end{figure}

The parameters left to be determined are $\rho_{\rm cav,\circ}$ and $R_{\rm cav,\circ}$.  We find that pure power-law profiles for the cavity density fail to reproduce the observed emission at 20-40~$\mu$m (Figure \ref{fig:cav_struc_com}), suggesting that more warm dust ($\sim$100~K) is needed.  Our preferred cavity density profile consists of a constant density region at the inner radius and a region where the density declines as $r^{-2}$ (see Section~\ref{sec:cavmod} and Figure~\ref{fig:dust_radial}).
The motivation for this hybrid cavity profile arises from the discrepancy between simulations and observations at 20-40~$\mu$m from pure power-law profiles and the flux levels at both sides of the absorption feature at 10~$\mu$m.  All profiles fail to reproduce the ice absorption features at wavelengths shorter than 10~$\mu m$ and around 15~$\mu m$, because the ices causing those absorptions are not included in the dust model.  A uniform density across the entire cavity might fit the observation; however, it would produce too much emission at 20-40~$\mu m$ if we want to match the emission at near-infrared wavelengths.  Also a constant density across the entire cavity is unphysical, considering that the mass flux has to be conserved while moving into a larger area.  The underestimated flux at 20-40~$\mu$m suggests more warm dust, which usually exists in the inner region.  Therefore, we introduce a constant density region in the inner cavity to concentrate more dust toward the center compared to  simple power law profiles ($r^{-2}$ and $r^{-1.5}$).
Figure~\ref{fig:cav_struc_com} shows that this profile can successfully produce the observed emission at 20-40~$\mu$m.
A grid search of these two parameters ($\rho_{\rm cav,\circ}$ and $R_{\rm cav,\circ}$) suggests the best fitted values of 3$\rm \times 10^{-20}~g~cm^{-3}$ and 40~AU.

This wavelength region is quite diagnostic of the cavity density distribution.  The origin of this constant density region is currently unclear.  We can only speculate that it can be the result of a recent mass ejection or pseudo-disk due to magnetic field.  Observations with high angular resolution, such as ALMA, can provide more information about this hybrid profile by mapping the continuum flux distribution with sub-arcsec resolution.

\section{Discussion}
\label{sec:discussion}

\subsection{Comparison to Other Models}
\label{sec:othermods}

\begin{figure}[htbp!]
  \centering
  \includegraphics[width=0.47\textwidth]{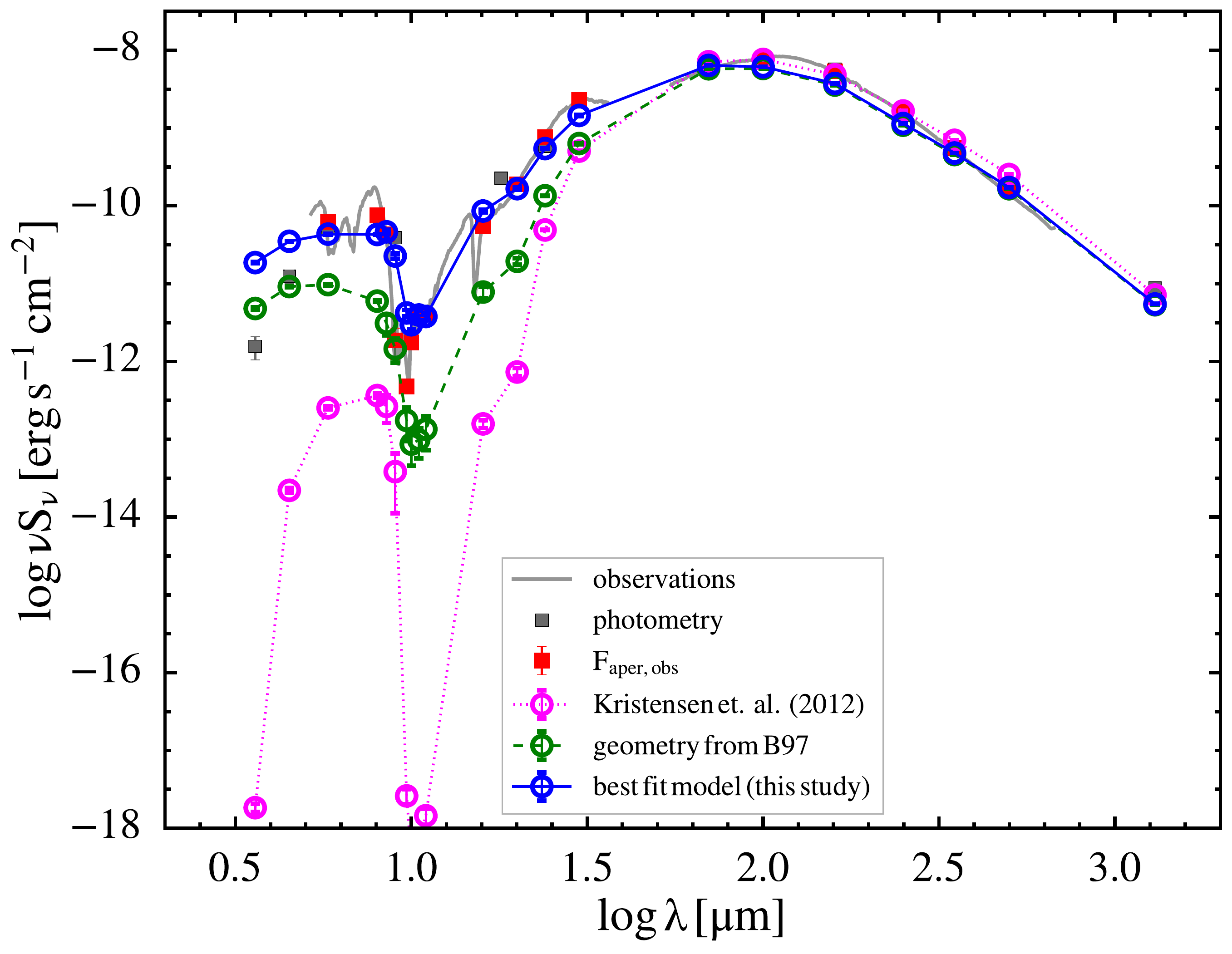}
  \caption{Simulated SEDs of our best fit model (solid blue), the model from \citet{2012A&A...542A...8K} (dotted magenta), and the model with geometry from B97 (dashed green).  The values of parameters of the best fit model are listed in Table~\ref{best_fit}.  The observations including spectra and photometry are shown in gray.  The simulated SEDs and observed spectra are both extracted with the aperture sizes used in the observations (open blue circles and red squares).}
  \label{fig:three_models}
\end{figure}

We compare the simulated SED of our best fit model to other possible models in Figure \ref{fig:three_models}.   \citet{2012A&A...542A...8K} modeled the water emission with 1-D infalling envelope, suggesting an envelope with an infall radius of 3500~AU and a mass accretion rate of $\rm 3\times 10^{-5}~M_{\odot}~yr^{-1}$, translating into an age of $\rm \sim3.3\times 10^{4}$~year with the model of \citet{1977ApJ...214..488S}, very similar to our best-fit age.  We adopt the values from \citet{2012A&A...542A...8K} for age, sound speed, and the outer radius of the envelope, while adopting other parameters from our best fit model, without a disk or outflow cavity.  We also test the geometry of the outflow cavity derived from the CO outflow map (B97), which suggests an inclination angle of 84$^{\circ}$ and an outflow cavity opening angle of 15$^{\circ}$.

The two other models are shown in Figure~\ref{fig:three_models} in magenta and green dashed lines.  All three models show a good agreement at far-infrared to submillimeter wavelengths, indicating that similar properties of the envelope are found.  The distinct difference at wavelengths below 50~$\mu$m between our best fit model and modified model of \citet{2012A&A...542A...8K} shows that an outflow cavity is required to fit the whole SED.  The lack of emission below 50~$\mu$m in the model with geometric parameters from B97 due to the large inclination angle further illustrates the disagreement between $\theta_{\rm incl}$ we derive and that derived from the distribution of CO emission lines.

\subsection{Effect of Not Including the Full TSC Model}
\label{sec:env_model}
\begin{figure}[htbp!]
  \centering
  \includegraphics[width=0.47\textwidth]{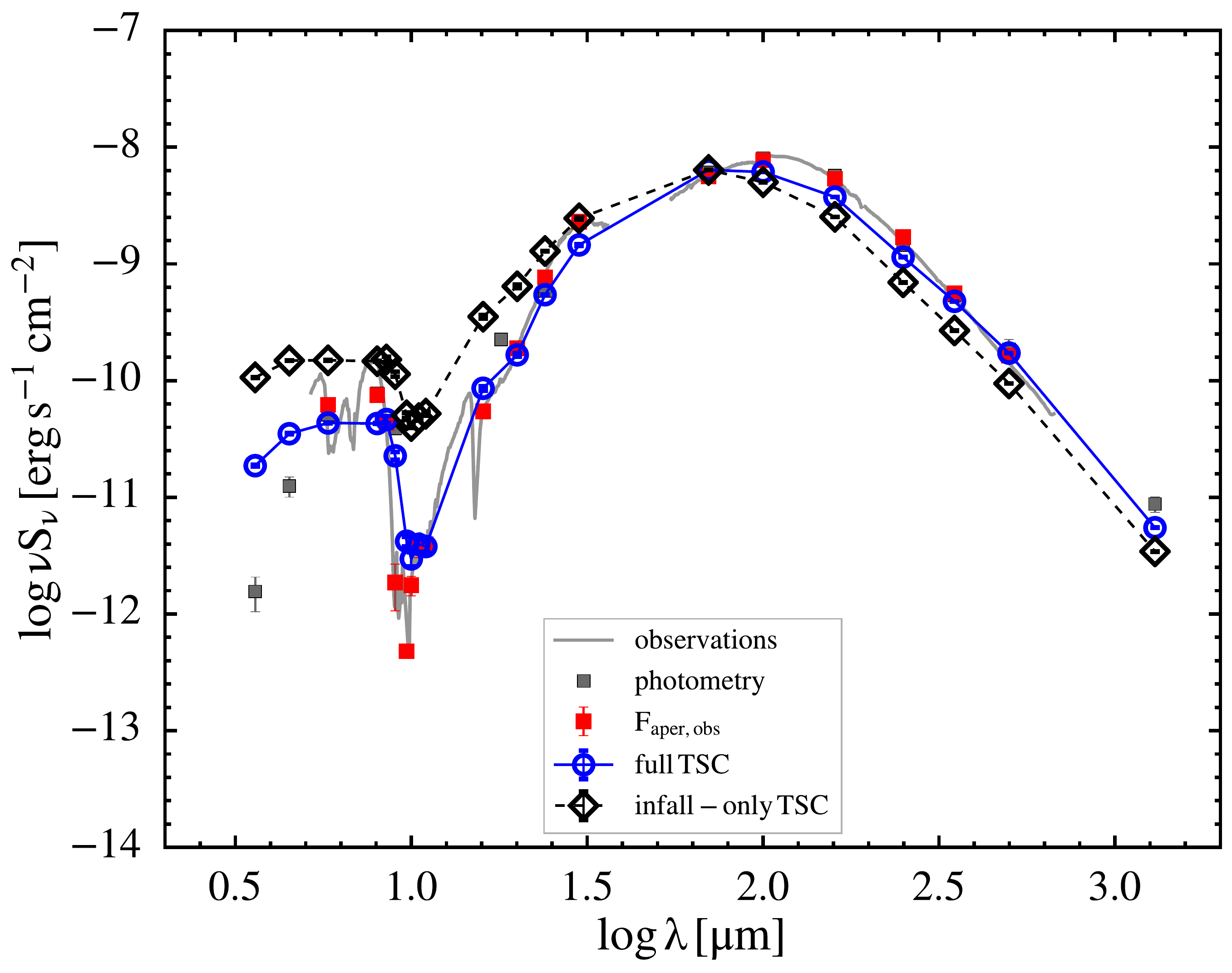}
  \caption{The simulated SEDs of the full TSC model (solid blue line/circles) and infall-only TSC model (dashed black line/diamonds).  The observation and aperture-convolved spectrophotometry are shown in light gray and red squares for comparison.}
  \label{fig:tsc_com_sed}
\end{figure}

As we mentioned in Section~\ref{sec:envmodel}, we adopt a full TSC envelope including a static outer envelope instead of an infall-only envelope.  The simulated SEDs of two types of TSC models are shown in Figure~\ref{fig:tsc_com_sed}.  The infall-only TSC model produces too much mid-infrared emission and too little submillimeter emission, while the full TSC model fits fairly well.  For young embedded protostars, the infall radius has to be smaller than the outer radius of the envelope.  Therefore, the transitional region to the isothermal profile and the outer, static envelope are required to fully describe the envelope structure.  The models in the widely used grid by \citet{2006ApJS..167..256R} do not include these regions.

\subsection{Mass Flows}
\label{sec:mass flows}

The mass infall rate ($\dot{M}_{\rm inf}$) can be compared with the rate of accretion onto the central star ($\dot{M}_{\rm acc}$) and the mass loss rate in the wind ($\dot{M}_{\rm wind}$) under several assumptions.

The mass infall rate of our best fit model, $\dot{M}_{\rm env} = 1.2\times 10^{-5}~\rm M_{\odot}~yr^{-1}$, derived by Equation~\ref{eq:tsc} agrees with that of \citet{2012A&A...542A...8K} within a factor of 2.5. The rate of accretion onto the central star can be estimated by assuming that the central luminosity is dominated by the accretion process; therefore the accretion rate can be calculated with Equation~\ref{eq:L_acc}, if the photospheric luminosity is ignored \citep{2014prpl.conf..195D}.

\begin{equation}
  \dot{M}_{\rm acc} = \frac{L_{\rm cen}R_{\star}}{G M_{\star}f_{\rm acc}}
  \label{eq:L_acc}
\end{equation}
where $L_{\rm cen}$ is the central luminosity in the best fit model, $R_{\star}$ is the radius of the protostar, $G$ is the gravitational constant, $M_{\star}$ is the total mass that has fallen into the star, and $f_{\rm acc}$ is the radiative efficiency of accretion.  The central luminosity (18.8~$\rm L_{\odot}$)  is more representative for the accretion process than the bolometric luminosity.
Assuming 25\%\ of the infallen mass remains in the disk, $M_{\star} = 0.32~\rm M_{\odot}$ derived by $M_{\star} = 0.75~\dot{M}_{\rm env}~t_{\rm col}$.  The simulation of spherical accretion shows $R_{\star} = 3~{\rm R_{\odot}}$ according to the mass of the protostar in our model (Figure~1, \citet{1992ApJ...392..667P}).
We take $f_{\rm acc} = 0.5$.
Adopting the values described above, we derive the rate of accretion onto the central star, $\dot{M}_{\rm acc} = \rm 1.1\times10^{-5}~\msun~yr^{-1}$, very close to the mass infall rate fitted by this study and only a factor of 2.5 lower than the rate fitted by \citet{2012A&A...542A...8K}.  This agreement suggests that BHR71 at this time does not show any sign of episodic accretion, which has been inferred from studies of other sources (\citealt{2014prpl.conf..195D} and references therein).

The average mass loss rate in the wind is calculated by dividing the momentum of the molecular outflow by its dynamical age and the speed of the wind.  We use the observations of the low-$J$ CO outflow (B97), but we adopt the best fitted inclination angle from this study to  calculate the momentum of the outflow and the dynamical age of the outflow.  An inclination angle of 50 degrees leads to a momentum of 15.4~$\rm \msun~km~s^{-1}$ and a dynamical age of about 24600 years.  These values imply a mass loss rate in the wind of $2.1\times 10^{-6}$  \msunyr, assuming a wind speed of 300~\kms.  This is much lower than the values found by \citet{2015ApJ...801..121N} of 4.2-53\ee{-6} \msunyr, mainly because of the smaller inclination angle in our best-fitting model.  Our mass loss rate is about 20\%\ of the rate of accretion onto the central star, similar to the factor of 0.1 that is usually assumed \citep{1994ApJ...429..781S,1996A&A...311..858B,1992ApJ...394..117P,1993ApJ...410..218W,2016ApJ...828...52W}, while the wind mass loss rate of \citet{2015ApJ...801..121N} was larger than the mass accretion rate.
In a different sample of Class 0/I protostars analyzed by \citet{1996A&A...311..858B}, the highest mass loss rate is $\sim 1.2 \times 10^{-7}~\rm M_{\odot}~yr^{-1}$, only one third of the mass loss rate of BHR71.  A high average mass loss rate suggests that BHR71 has not had much lower
mass accretion rates in the past. Unlike many Class 0 objects, there
is no evidence for episodic accretion or mass loss events.

If the \oi\ emission measures the wind mass loss rate, we can use Equation~\ref{eq:oi_massloss} to obtain an alternative estimate of the mass loss rate in the wind \citep{1985Icar...61...36H}.
\begin{equation}
  \dot{M}_{\rm wind} ({\rm 10^{-5}~M_{\odot}~yr^{-1}}) = 10L_{\rm 63}({\rm L_{\odot}})
  \label{eq:oi_massloss}
\end{equation}
where $L_{\rm 63}$ is the luminosity of the 63 \micron\ \oi\ line.
We derive a mass loss rate in the wind of 2.2$\times 10^{-7}~\rm M_{\odot}~yr^{-1}$, which is about one order of magnitude smaller than we estimated from the momentum of the outflow from CO.

The \oi~63~\micron\ line traces the the mass flux in the dissociative shocks;
however, other physical processes may contribute to the observed \oi\ line flux, such as photodissociation regions.  So the mass loss rate derived from the \oi~63~\micron\ line flux should be an upper limit of the mass loss rate in the dissociative shocks.  The ratio of the mass loss rate measured in \oi\ to the mass accretion is 0.02, which is consistent with \citet{2015ApJ...801..121N}, where they find a ratio of 0.05.  The main factors for the difference come from their assumption of stellar mass and radius.  This ratio is also consistent with the ratio predicted by the theoretical MHD jet launching model, $\sim 0.05-0.5$ \citep{2006A&A...453..785F}.

The difference between the mass loss rate derived from CO outflow momentum and \oi~63~$\mu$m could be explained by the different time scales that the two tracers probe.  The CO outflow momentum is measured from the entire outflow, therefore averaging the mass loss rate throughout the lifetime of the outflow, while the \oi~63~$\mu$m probes the mass flux in the outflow-induced shock regions, reflecting the instantaneous mass loss rate in the outflow. However,
there is no indication of episodic accretion in this source and the
current luminosity is quite high. In addition, other mechanisms can
contribute to the \oi\ line, so the weakness is especially puzzling.
This discrepancy is wide-spread, casting doubt on the \oi\ line
as a wind tracer.
With a different sample, \citet{2013A&A...552A.141K} and \citet{2015ApJ...801..121N} both find that the mass loss rates derived from CO outflow are up to two orders of magnitude higher than the mass loss rates indicated by \oi\ line flux for embedded sources.

\section{Conclusions}
\label{sec:con}

Combining \herschel\ PACS and SPIRE spectra, we find 61 lines in the 1-D spectrum and 645 lines in the full PACS and SPIRE 2-D datacube.  Among those lines, we detect \coul{4}{3} up to \coul{41}{40} with only a few non-detected levels.  The high-$J$ \co\ lines are  detected only at the center of BHR71, suggesting a shock origin within 2000~AU of the center.  Our rotational diagram analysis shows four temperature components, 43~K, 197~K, 397~K, and 1057~K from 1.09\ee{48}, 1.71\ee{49}, 4.95\ee{49}, and 2.76\ee{50} CO molecules respectively, from  the 1-D spectrum.  The wide range of temperatures indicates a range of excitation environments.

The low-$J$ and high-$J$ CO lines differ in morphology.  The low-$J$ \co\ emission distributes in a bipolar direction along the outflow, while the high-$J$ emission is compact, near the center of the source.  The low-$J$ \co\ emission lines are also found at the off-center positions.  B97 finds a similar feature in \coul{1}{0}.  The low energy water emission lines have a similar morphology to the low-$J$ \co\ contours, while the high energy water lines are usually found to be off-center on the plane normal to the outflow, with a possible extension
to IRS2.  Atomic cooling lines are detected in \oiul{1}{2}, \oiul{0}{1}, and \ciul{1}{0}.  \ciul{1}{0} emission is found to be extended across the field of view with a local minimum on IRS1.  \oiul{1}{2} emission is well centered with a slight extension toward the NE direction, while \oiul{1}{0} emission is only detected at the central spaxel.

We perform three dimensional dust radiative transfer simulations, study the effect of various parameters, and find the best fit model.  The model consists of the rotating, collapsing envelope described by \citet{1984ApJ...286..529T}, a flared disk, and an outflow cavity.  The envelope parameters can be constrained by the shape of the SED at submillimeter wavelengths, while the properties of the disk are not constrained because of the small mass of the disk.  Our parameter studies (see Appendix) reveal the degeneracy of the SEDs caused by the trade-off between the sound speed and age.  A model with a smaller sound speed and a later age can have a SED similar to a model with a larger sound speed and a earlier age.  An accurate sound speed measurement is necessary to constrain the age of the envelope.  The properties of the outflow cavity and the inclination heavily influence the mid-infrared SED.  Our model shows that a constant density region at the apex of the cavity is required in order to fit the mid- and submillimeter SED simultaneously.  The need for the constant density region in the model shows promise as a means of probing the structure in the inner region of the outflow cavity.

Several parameters are fixed prior to the model optimization based on existing observations, leaving five to fit.  We first constrain the age of the envelope, then find the best inclination angle of the outflow cavity, and finally tune the density profile of the outflow cavity that best fits the SED. The azimuthally-averaged radial intensity profile is only sensitive to the age of the source, therefore better constraining the age, while the SEDs are affected by many other factors.   A maximum age of 44000 yr is found by comparing the radial intensity profile at 160~\micron\ with observations. The best fitting inclination angle is determined from the flux ratio seen in the \spitzer-IRAC~1 image.  An inclination angle of 50 degree best reproduces the observed morphology; with this value, the minimum age from the CO outflow is 24630 yr. We choose an intermediate value of 36000 yr for the age. The smaller inclination angle required by the IRAC data  conflicts with the angle derived from the CO outflow (B97).  Further studies of the kinematic structure of the outflow may resolve this apparent discrepancy.

The best fit model has a higher central luminosity (18.8~$\rm L_{\odot}$) than the observed bolometric luminosity because of the asymmetric structure, in which radiation escapes from the center more easily along the outflow direction.  The mass accretion rate derived from the central luminosity in the model results in a mass accretion rate of $1.1\times 10^{-5}~\rm \msun~yr^{-1}$, similar to the mass infall rate of the TSC envelope, $1.2\times 10^{-5}~\rm \msun~yr^{-1}$, agreeing with the rate fitted by \citet{2012A&A...542A...8K} within a factor of 2.5. Our best fit model can well constrain the evolution of the collapse envelope, and this method can be adopted to a larger sample with known sound speeds to determine the
evolutionary sequence of  embedded protostars.
The mass outflow rate in the wind ($\dot{M}_{\rm wind}$), 2.1\ee{-6}~\msun~$\rm yr^{-1}$, is about 20\%\ of the mass accretion rate, close to the usual assumption of 10\% of the accretion rate ($\dot{M}_{\rm acc}$) if calculated from the CO outflow and lower if calculated from \oi. We suspect that the \oi\ is not measuring the wind
mass loss rate.

\acknowledgements
We thank Gibion Makiwa for helpful discussion on SPIRE spectral calibration.  JDG would like to acknowledge numerous helpful discussions with the PACS and SPIRE ICC teams, via webinars, helpdesk queries, and contacts.   Support for this work, part of the  {\it Herschel} Open Time Key Project Program, was provided by NASA through awards issued by the Jet Propulsion Laboratory, California Institute of Technology.  JDG and Y-LY acknowledge support from NASA Herschel Science Center Cycle 2 grants, and from STScI JWST Discretionary Fund (JDF).

\appendix

The appendix contains studies of how the various parameters affect the predicted
observables. These were used to constrain our exploration of parameter space, and
we present them here as a guide to other modelers.

\section{Envelope Parameters}
\label{sec:env_para}

\begin{figure*}[htbp!]
  \centering
  \includegraphics[width=\textwidth]{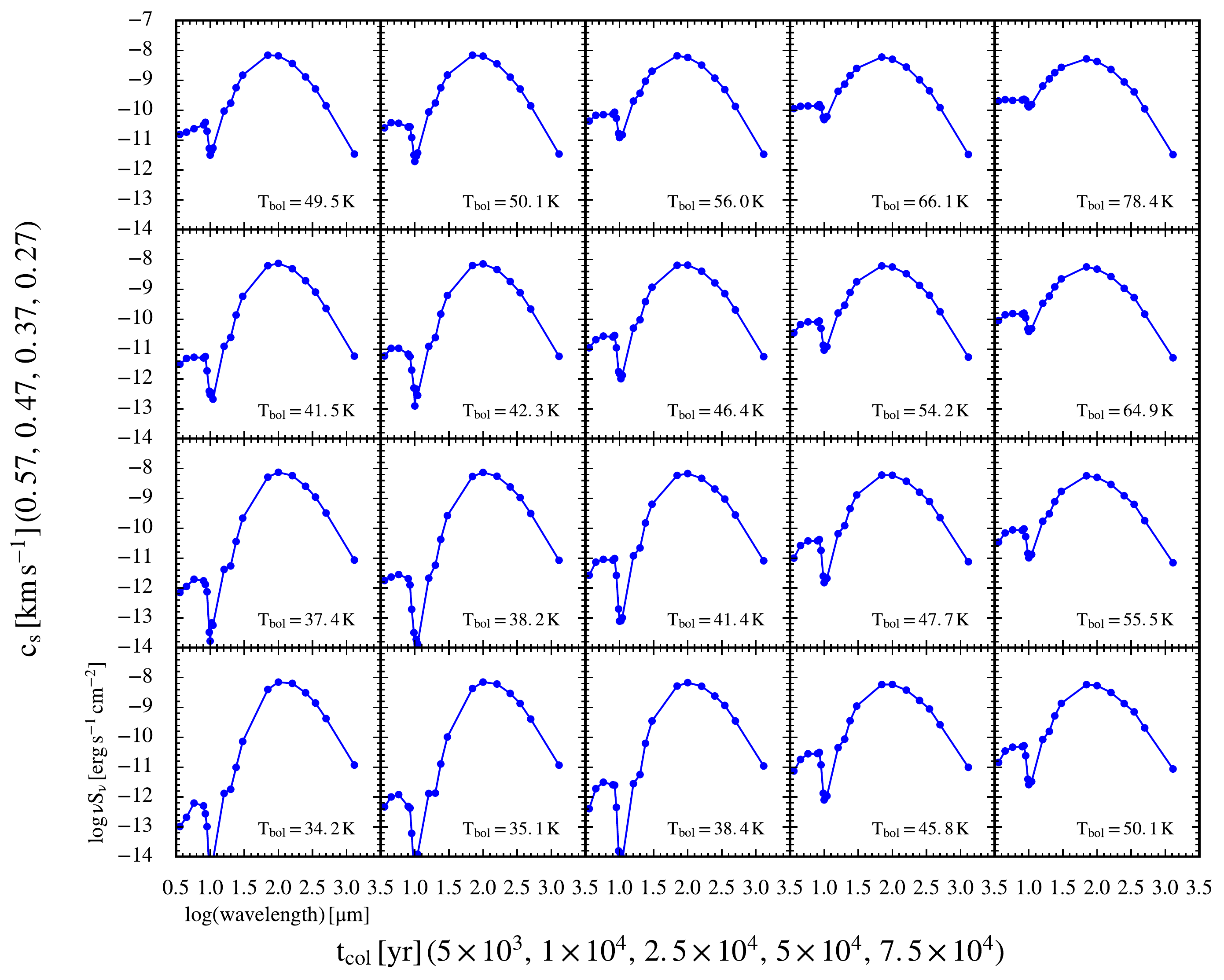}
  \caption{The simulated SEDs of the evolution sequence of model with three different sound speeds (0.27~\kms, 0.37~\kms, 0.47~\kms, and 0.57~\kms).  From top to down, each row shows a evolution sequence for given sound speed of 0.27~\kms, 0.37~\kms, 0.47~\kms, and 0.57~\kms.  From left to right, each column shows a snapshot of the evolution sequence for three different sound speed.  The age of the model increases from $\rm 5\times10^{3}$ years to $\rm 7.5\times10^{4}$ years.  The blue dot/line shows the SED with aperture-extracted photometry from simulated spectra.  Other parameters are adopted from Table~\ref{best_fit}.}
  \label{fig:sed_csage}
\end{figure*}

The properties of the envelope are mainly determined by the TSC model along with the inner and outer radius of the envelope, which are fixed as the dust sublimation radius ($\sim$ 0.1~AU with a dust sublimation temperature of 1600~K) and 0.315~pc for these parameter studies.  The three parameters of the TSC model are sound speed, age since the collapse began, and the initial rotational speed.  The TSC model provides an evolutionary model of a collapsing envelope.  Figure~\ref{fig:sed_csage} shows the evolutionary sequence from $\tcol = \rm 5\times10^{3}$ years to $\rm 7.5\times10^{4}$ years with four different sound speeds: 0.27~\kms, 0.37~\kms, 0.47~\kms, and 0.57~\kms.  The dust density profiles of these models are calculated with the setup described in Section \ref{sec:dust_setup} and the parameters in Table~\ref{best_fit} except for the two varying parameters.

As the age increases, the broad peak in the far-infrared becomes wider and moves toward shorter wavelengths.  This behavior reflects the fact that envelope mass decreases over time due to the accretion, resulting in a higher dust temperature peaking at a shorter wavelength and less extinction at mid-infrared wavelengths.  For a given age, the contrast between mid-infrared and submillimeter wavelength increases with the sound speed.  A higher sound speed provides more pressure support to prevent the gas from collapsing; therefore the initial density and the total mass of the envelope is greater with a higher sound speed, providing more extinction at shorter wavelengths and increasing the contrast (Equation~\ref{eq:isothermal}).  The bolometric temperature ($T_{\rm bol}$) of each model is shown in the corresponding panel derived with the method in \citet{1995ApJ...445..377C}.  $T_{\rm bol}$ increases with the age, a pattern that is consistent with the protostar classification using bolometric temperature \citep{2014prpl.conf..195D}.  Note that beyond a certain age, the shape of the SED remains almost the same with only subtle changes at mid-infrared wavelengths.

\begin{figure}[htbp!]
  \centering
  \includegraphics[width=0.47\textwidth]{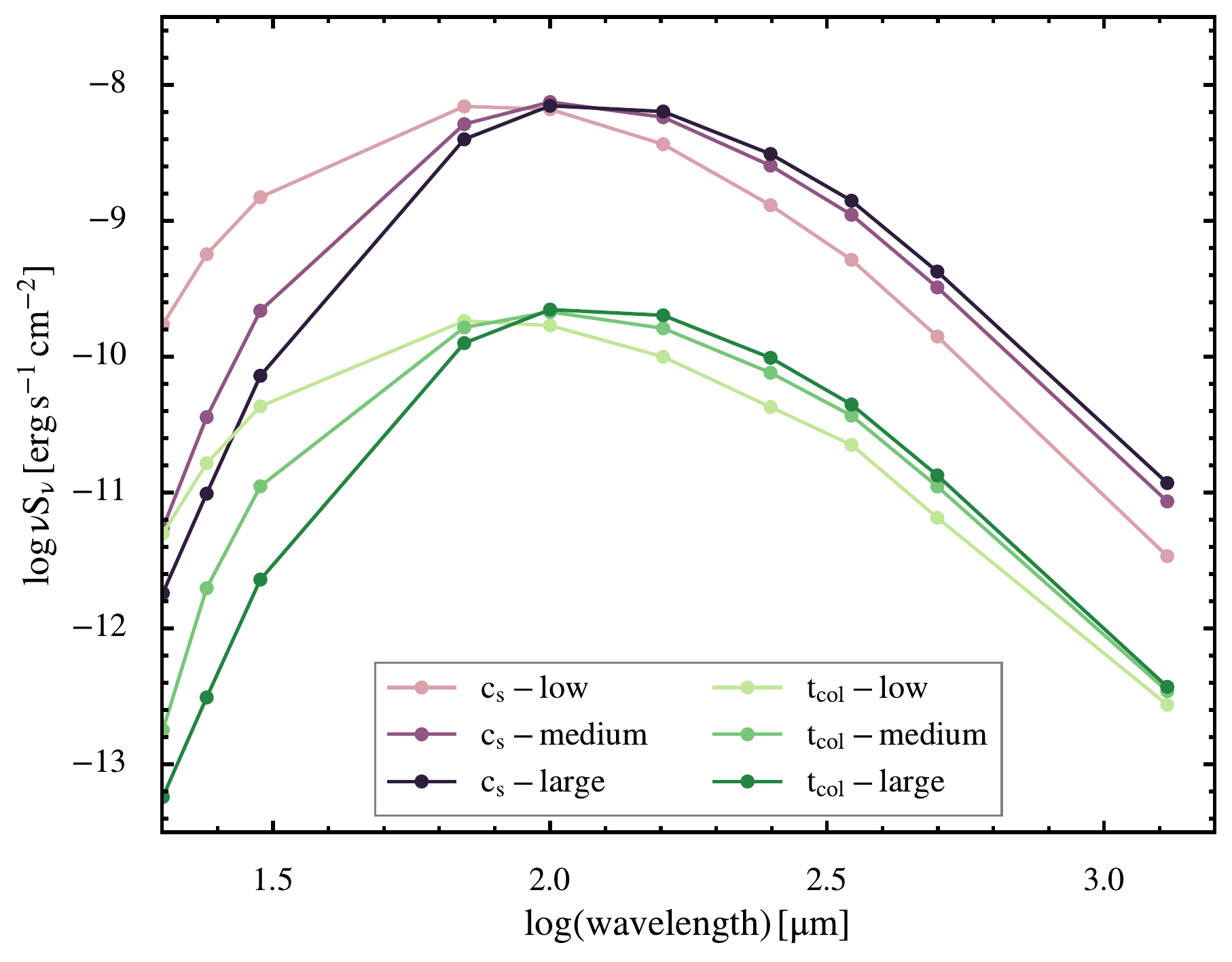}
  \caption{The simulated SEDs selected from Figure~\ref{fig:sed_csage} (the first and the last two SEDs in the first column and the first, third and fifth SEDs from the left of the last row).  The SEDs with different \tcol\ are shown in purple and the SEDs with different sound speed are shown in green, while the series of \tcol\ is offset by -1.5~dex.}
  \label{fig:csage_behavior}
\end{figure}

The sound speed and age are degenerate to first order because the infall
radius is just the product of the two. Figure~\ref{fig:sed_csage} shows the degeneracy of these two parameters: a model with higher sound speed and an earlier age can have a similar SED to a model with smaller sound speed and a later age.  For example, the SED in the bottom right panel is similar to the SED in the fourth from the left of the third row. However, envelopes with
higher sound speeds start with more mass, which is reflected best by the
emission at the longest wavelengths (e.g., 1.3 mm). Figure~\ref{fig:csage_behavior} shows the effect of sound speed and age.
The long wavelength emission is more affected by sound speed than by age.

\begin{figure}[htbp!]
    \centering
    \includegraphics[width=0.47\textwidth]{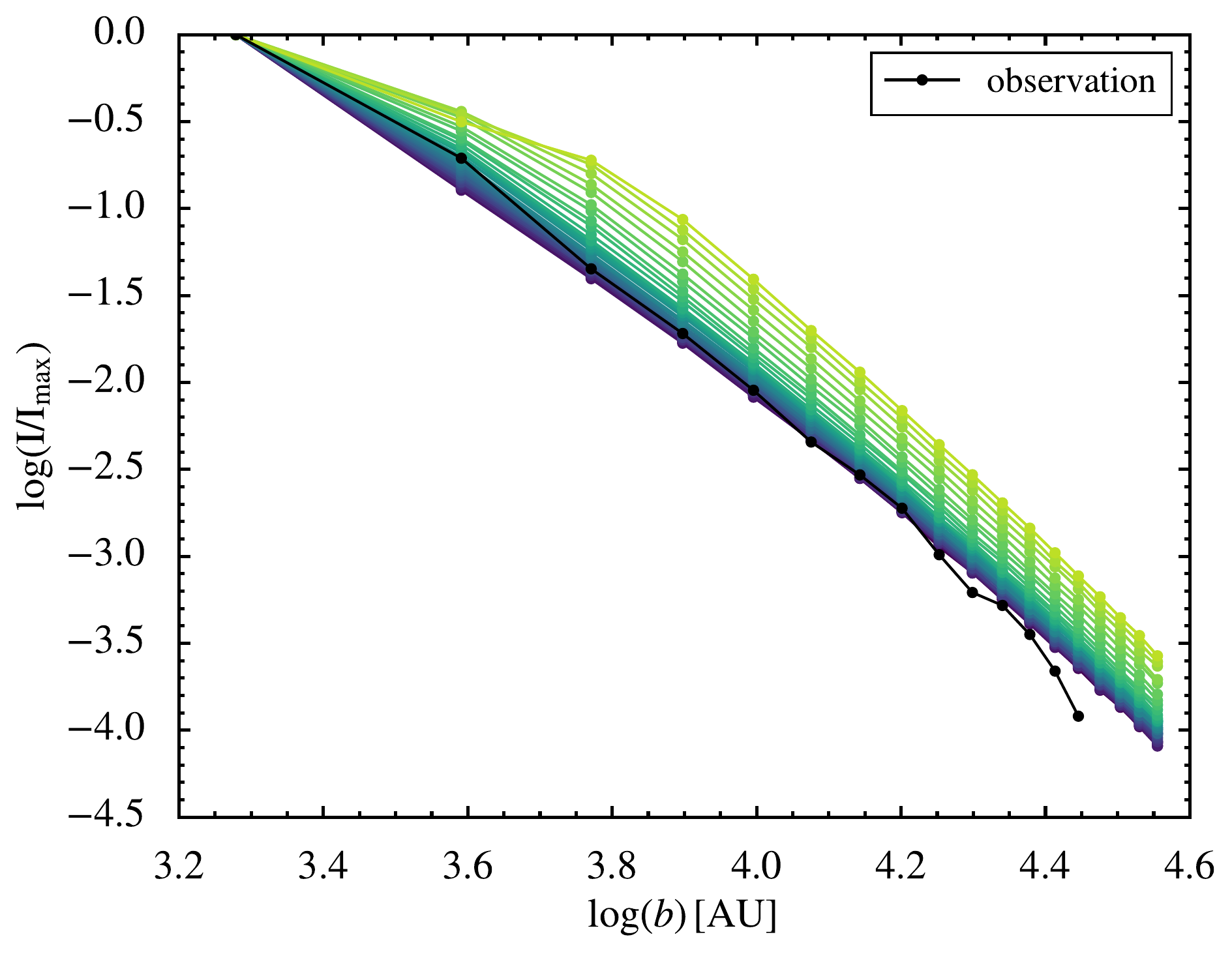}
    \caption{The simulated radial intensity profiles at 160~\micron\ are shown in sequential colors, where a lighter color corresponds to a later age.  The radial intensity profile extracted from \herschel\ 160~\micron\ image is shown in black.  The models include 34 ages, spaced by 2000 years, extending from 4000 to 50000 years, and then spaced by 5000 years, extending from 55000 to 100000 years.}
    \label{fig:rad_age}
\end{figure}

\begin{figure}[htbp!]
    \centering
    \includegraphics[width=0.47\textwidth]{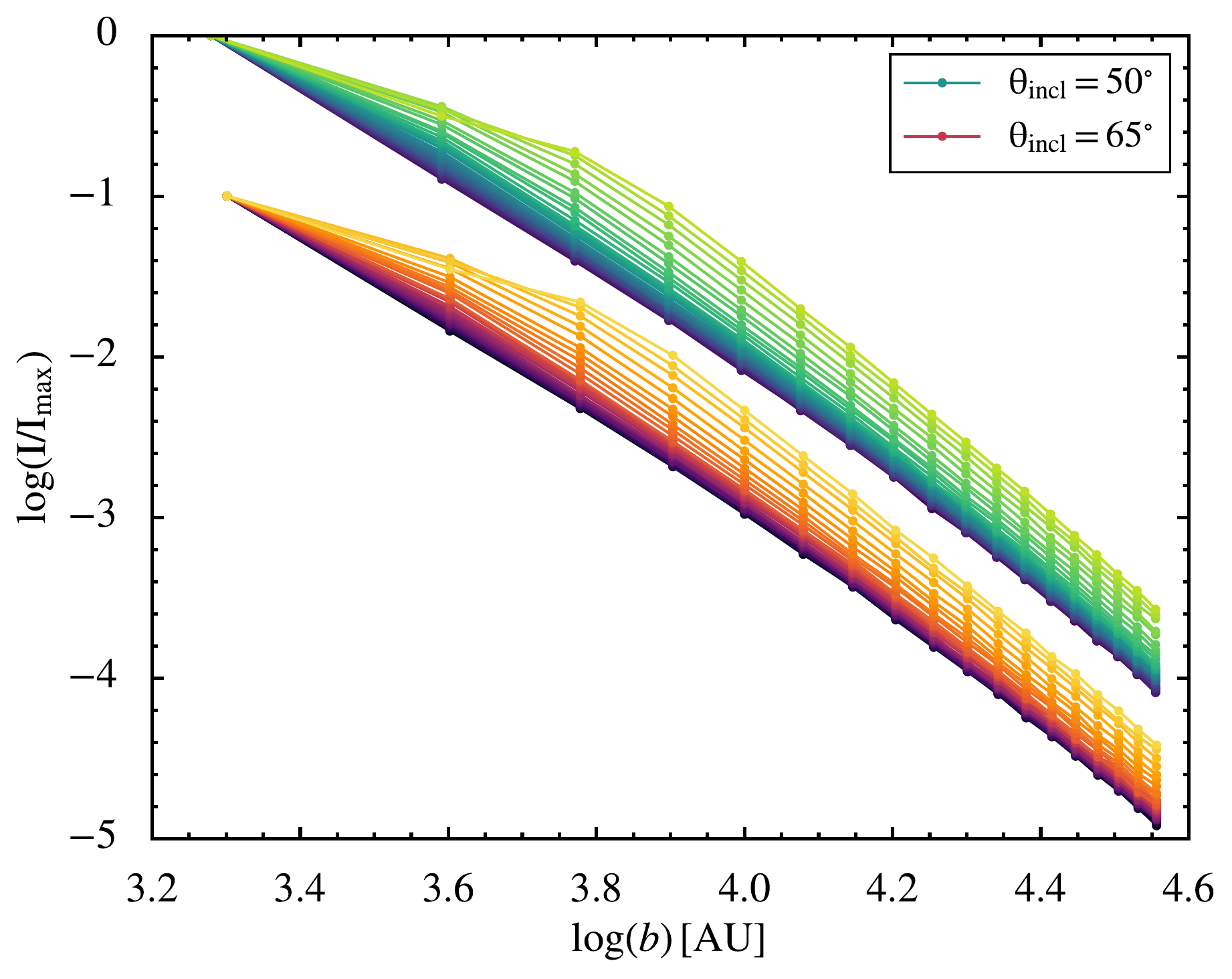}
    \caption{Two groups of simulated radial intensity profiles with the same range of age but different inclination angles, 50$^{\circ}$ (green sequence) and 65$^{\circ}$ (red sequence).
    The grid with $\theta_{\rm incl} = 65^{\circ}$ is shifted by -1.}
    \label{fig:rad_incl}
\end{figure}

The azimuthally-averaged radial intensity profile at 160~\micron\ is quite diagnostic of age (see Section~\ref{sec:age} for definition).  Figure~\ref{fig:rad_age} shows the effect of age on the radial profiles, indicating that a later age results in a flatter profile.  The emission at smaller radii increases with the age, while the shape of the radial profiles at larger radii remains the same.  As the infall radius moves out, a larger fraction of the
envelope has a flatter density profile, resulting in a flatter intensity
profile. Because we use the {\bf normalized} radial intensity profiles, the profiles are only sensitive to the structural change of the envelope, not the
overall mass.  Figure~\ref{fig:rad_incl} shows the age grid of radial intensity profiles with two different inclination angles, 53$^{\circ}$ and 61$^{\circ}$.  Both grids behave in the same way with the age of the envelope, suggesting that the effect of the inclination angle is insignificant for comparing radial intensity profiles measured at long wavelengths.

\begin{figure}[htbp!]
  \centering
  \includegraphics[width=0.47\textwidth]{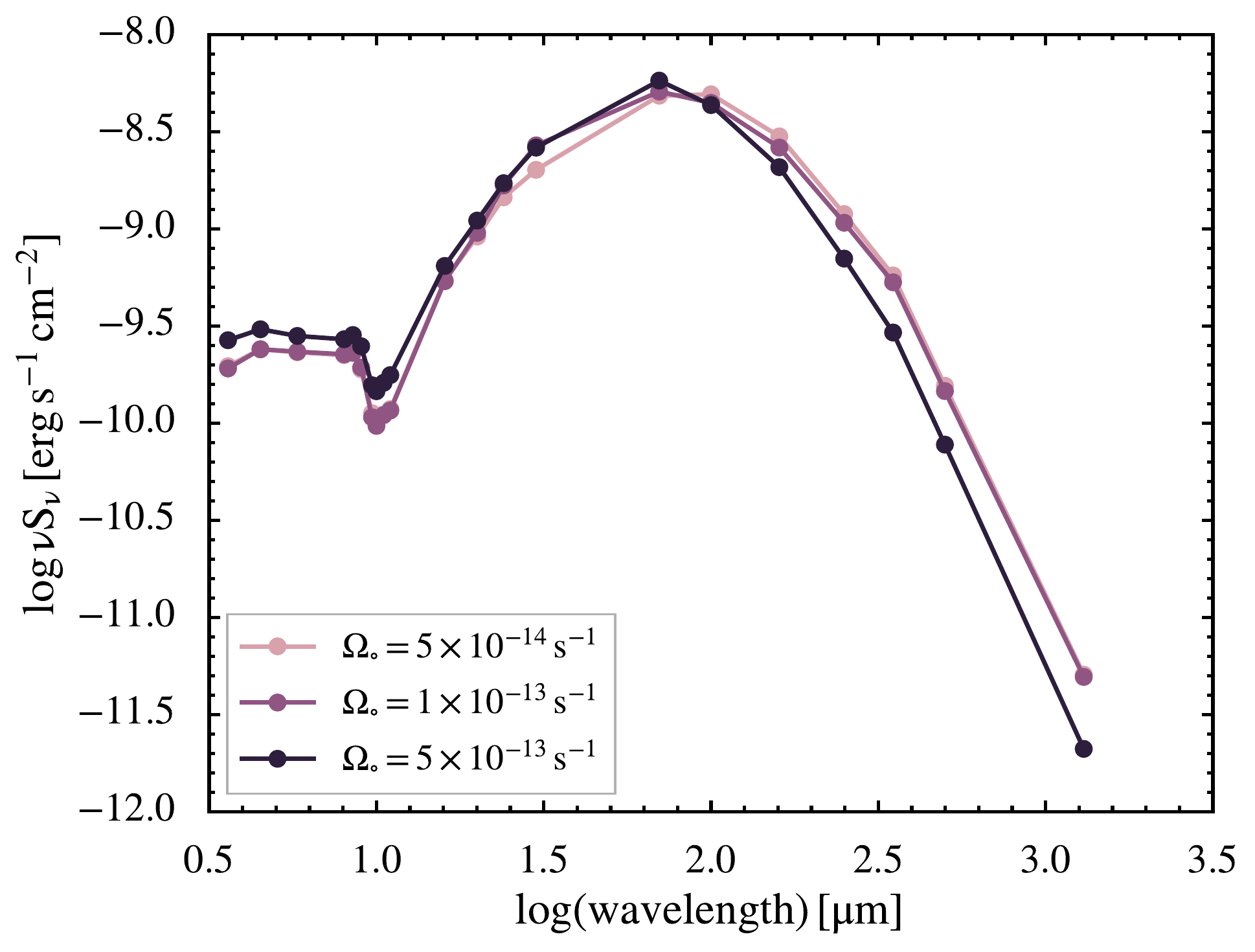}
  \caption{The simulated SEDs of TSC model with the initial rotational speed of $5\times 10^{-13}$, $10^{-13}$, and $5\times 10^{-14}~\rm s^{-1}$ with an age of 7.5$\times10^{4}$~years.    Other parameters are adopted from Table~\ref{best_fit}.}
  \label{fig:sed_omega}
\end{figure}

The third parameter of the TSC model is the initial rotational speed,
$\Omega_{\circ}$.  The initial rotational speed determines the centrifugal radius (\rcen) and thus the disk radius. A larger $\Omega_{\circ}$ leads to a density profile more flattened toward the equator.
Figure~\ref{fig:sed_omega} shows simulated SEDs with the initial rotational speed of $5 \times 10^{-13}$, $10^{-13}$, and $5 \times 10^{-14}$~s$^{-1}$, an age of $\rm 7.5\times10^{4}$~years, and other parameters from Table~\ref{best_fit}.  The corresponding centrifugal radii are 4.7~AU, 19~AU, and 470~AU.  A larger $\Omega_{\circ}$ results in more emission in the mid-infrared and less emission in the submillimeter.
A larger centrifugal radius leads to the increase of material at smaller radii due to a more flattened inner envelope; therefore, the mid-infrared excess comes from the increase of hot material.  The 1.3~mm emission, which is dominated by the total dust mass, is lower in the model with a large rotational speed due to a lower total dust mass in the envelope. Equation~\ref{eq:rot_env}, adopted from Eq. 4.43 of \citet{2009apsf.book.....H} can illustrate this effect.
\begin{equation}
    \rho(r, \theta) = \frac{\dot{M}_{\rm env}}{4\pi(GMr^{3})^{1/2}}\left(1+\frac{{\rm cos}~\theta}{{\rm cos}~\theta_{\circ}}\right)^{-1/2}\left(\frac{{\rm cos}~\theta}{{\rm cos}~\theta_{\circ}}+\frac{2{\rm cos}^{2}~\theta_{\circ}}{r/R_{\rm c}}\right)^{-1}
    \label{eq:rot_env}
\end{equation}
where $\dot{M}_{\rm env}$ is the mass infall rate in the envelope, $M$ is the total mass, $\theta_{\circ}$ is the angle between the orbital plane of mass particles and the rotation axis of the system, and $R_{\rm c}$ is the centrifugal radius, which is proportional to $\Omega_{\circ}^{2}$.
This equation represents the density for material at a given position in its collapsing trajectory, defined by $\theta$ and $\theta_{\circ}$.  Changing the rotational speed ($\Omega_{\circ}$) only, all variables in Equation~\ref{eq:rot_env} remain the same except for $R_{\rm c}$, which is proportional to $\Omega_{\circ}^{2}$.  Therefore, a larger rotational speed results in a greater $R_{\rm c}$, making the second term in the last factor in Equation~\ref{eq:rot_env} larger; since that factor is in the
denominator,  a higher rotational speed leads to a lower density, leading to a lower 1.3~mm emission.  The influence from the flattened envelope and the disk is minimal unless the resulting centrifugal radius is larger than about 100~AU.

The effects of the envelope parameters are summarized as follows:
\begin{itemize}
  \item \tcol: This controls the total amount of dust in the envelope, therefore the total amount of emission at submillimeter wavelengths.  A later age results in a less massive envelope and an SED peaking at a shorter wavelength and a flatter radial profile.
  For example, in the last row of Figure~\ref{fig:sed_csage}, the leftmost SED peaks around 10$^{2}~\mu$m, while the rightmost SED peaks around 10$^{1.75}~\mu$m.
  And the flux level at shorter wavelengths is higher at large \tcol\ (rightmost).
  \item $\csef$: This also controls the total amount of dust in the envelope.  Although its effect on the SED is similar to the effect of age, it can affect the detailed shape of the broad far-infrared peak in the SED.
  \item $\Omega_{\rm \circ}$: The effect of the initial rotational speed is minimal for a very young stellar object, $t_{\rm col}\sim {\rm 10^{4}~years}$ with modest rotation rates, but it can be important for more evolved objects (Figure~\ref{fig:sed_omega}).
\end{itemize}

\section{Disk Parameters}
\label{sec:disk_para}
\begin{figure}[htbp!]
  \centering
  \includegraphics[width=0.47\textwidth]{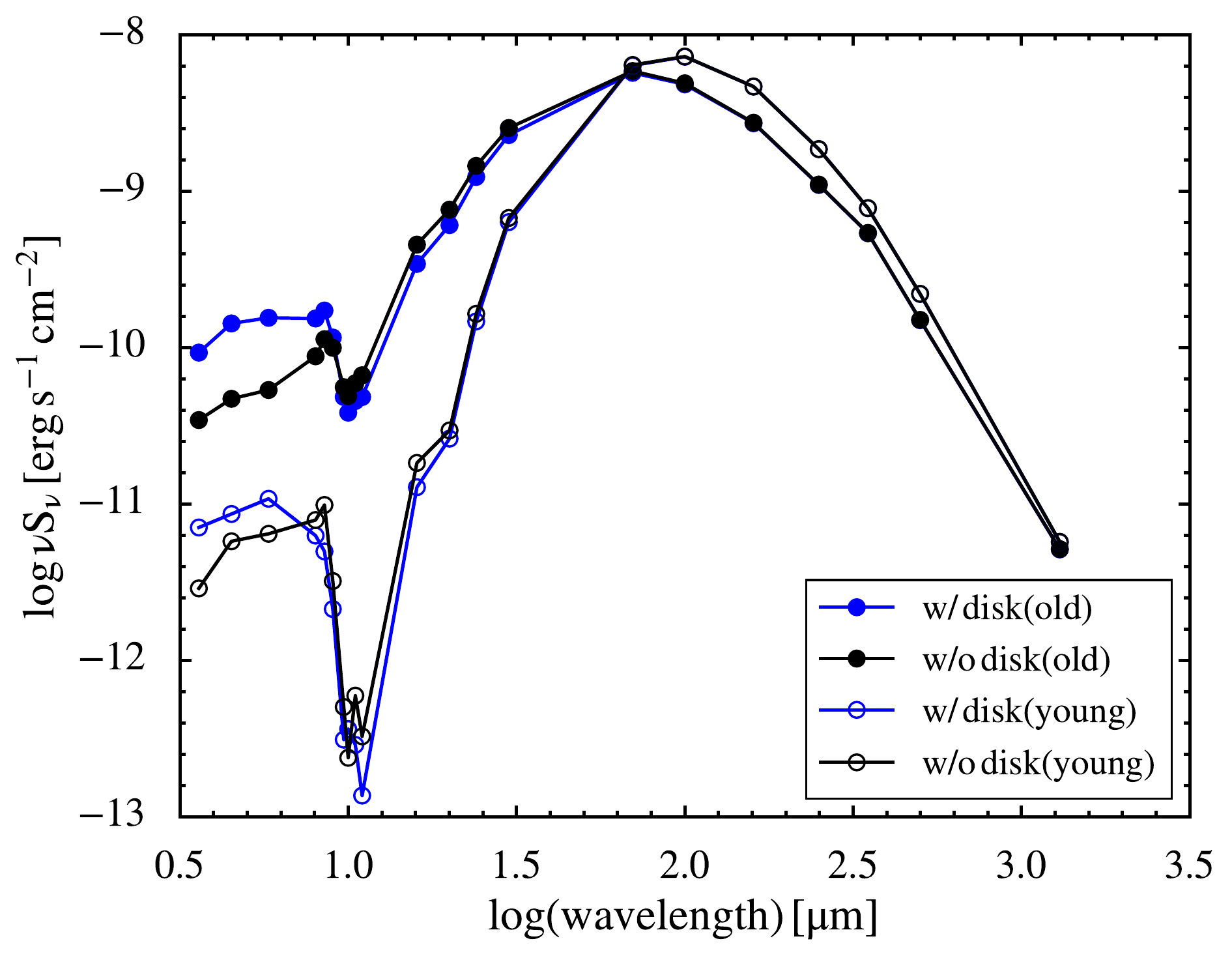}
  \caption{The simulated SEDs with and without the presence of disk for early age and late age, 10$^{4}$ and 7.5$\times 10^{4}$ years respectively (open and filled circles), while other parameters are adopted from Table~\ref{best_fit}.  The models with disk are shown in blue, while the models without disk are shown in black.}
  \label{fig:disk_com}
\end{figure}

The formation of the disk is a consequence of conservation of angular momentum.  With a small but not negligible initial rotational speed, the centrifugal barrier becomes comparable to the gravitational force as material moves toward the center.  However, the angular momentum cannot be conserved strictly; otherwise stars would spin too fast to remain stable.  Outflows and magnetic fields are the major channels for reducing the angular momentum from the system \citep{2014prpl.conf..173L}.  Observationally, disks associated with protostars are widely observed in both late stages and early stages \citep{2013A&A...559A..77F,2015ApJ...799..193Y}.  Here we demonstrate the effect of the disk around the embedded protostars on their SEDs.  Figure~\ref{fig:disk_com} shows the effect of the disk at two ages, early and late.  In both stages, early and late, the SED with a disk shows more emission at wavelengths shorter than 8~\micron\ and slightly more absorption at the 10~\micron\ silicate feature.
Although the presence of disks does affect the SED, we find the SED at the early stage is not sensitive to the disk parameters due to the small size of the disk at the early stage ($R_{\rm disk}=0.2~\rm AU$).  For objects as young as BHR71, the effects of the disk parameters on the SED are minimal.  For consistency, we still include the disk in our model setup, but its properties are unconstrained by the observations.  To emphasize the effects of the disk parameters on the SED, we show the models that have a disk radius $\sim 120$~AU ($t_{\rm col}=7.5\times 10^{4}$~years and the rest of the parameters adopted from Table~\ref{best_fit}) in the following discussion.

\begin{figure}[htbp!]
  \centering
  \includegraphics[width=0.47\textwidth]{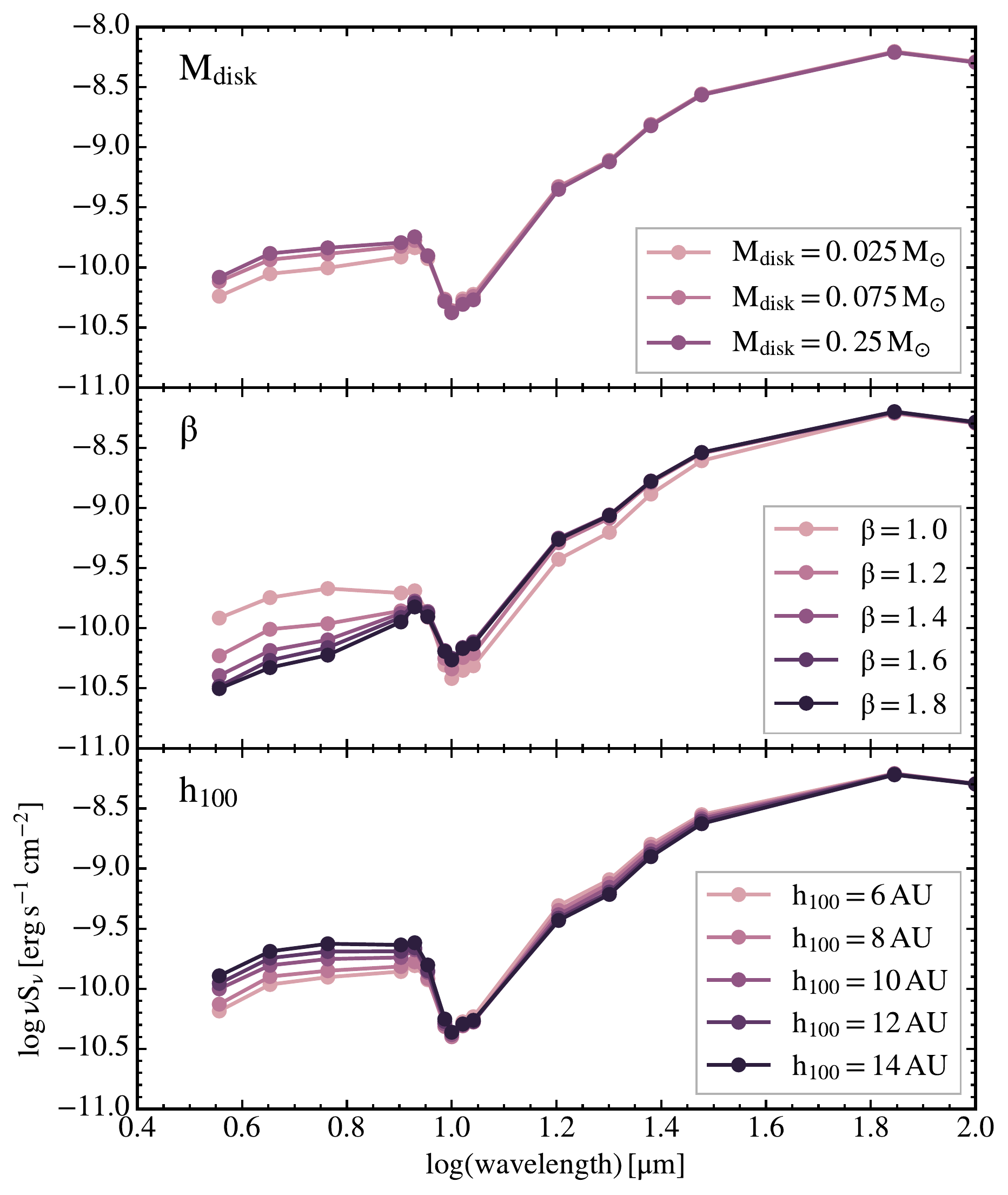}
  \caption{{\bf Top}: The simulated SEDs with disk mass of 0.025~\msun, 0.075~\msun, and 0.25~\msun (light to dark colors).  {\bf Middle}: The simulated SEDs with disk flaring power of 1.0, 1.2, 1.4, 1.6, and 1.8 (light to dark colors).  {\bf Bottom}: The simulated SEDs with disk scale height at 100~AU of 6~AU, 8~AU, 10~AU, 12~AU, and 14~AU (light to dark colors).  All other parameters are adopted from Table~\ref{best_fit}, except for a larger age of 75000 years.  The disk parameters change the SED at mid-infrared wavelength, while the rest of the SED remains the same (not shown here).}
  \label{fig:disk_summary}
\end{figure}

Figure~\ref{fig:disk_summary} (top) shows the simulated SEDs with disk masses of 0.025~\msun, 0.075~\msun, and 0.25~\msun.  The effect of the disk mass on the observed SED is primarily at near-infrared ($\lambda < 8$ \micron) wavelengths because it is close to the central protostar, resulting in a higher temperature.  A more massive disk leads to a small increase of emission at mid-infrared wavelengths.  Note that these models have an inclination angle of 53$^{\circ}$.

Figure~\ref{fig:disk_summary} (middle) shows the effect on the simulated SEDs of different disk flaring power-laws, $\beta$, of 1.0, 1.2, 1.4, 1.6, and 1.8.  The disk flaring power, $\beta$, is translated into $\alpha$, where $\alpha = \beta+1$, when calculating the disk density profile with Equation \ref{eq:disk} (also see Section~\ref{sec:diskmod}).  A larger $\beta$ results in a more flared disk, therefore the surface area exposed to the starlight increases but with a lower density.  Different disk flaring powers change the emission at mid-infrared wavelengths.  The emission below 10~$\mu$m decreases as the flaring power increases, and the emission between 10~\micron\ to 20~\micron\ increases slightly, while the rest of SED remains the same, indicating that the disk is less efficient for reproducing the radiation from the center with a more diffuse dust distribution.

Figure~\ref{fig:disk_summary} (bottom) shows the effect on the simulated SEDs of the disk scale height, $h_{\rm 100}$, of 6~AU, 8~AU, 10~AU, 12~AU, and 14~AU.  Note that $h_{\rm 100}$ is defined as the scale height at 100~AU.  A larger scale height results in more emission at near-infrared wavelengths, and slightly less emission at mid-infrared wavelengths.  Because of the increase of the disk thickness, the irradiated area at the disk inner radius is larger, resulting in more dust with a higher temperature, which increases the emission in the near-infrared.  While more radiation is collected by the dust at the disk inner radius, less radiation can be received by the dust at a larger radius, leading to a slight decrease of the emission at mid-infrared wavelengths.

The effect of disk parameters can be summarized as follows:
\begin{itemize}
  \item $M_{\rm disk}$: Increasing the disk mass increases the emission at near-infrared wavelengths, which can be seen from the SEDs below 10~$\mu$m in the top panel of Figure~\ref{fig:disk_summary}.
  \item $\beta$: Increasing the flaring power decreases the emission at near-infrared wavelength, and slightly increases the emission at mid-infrared wavelengths (the middle panel of Figure~\ref{fig:disk_summary}).
  \item $h_{\rm 100}$: Increasing the disk scale height increases the emission at near-infrared wavelengths, and decreases the emission at mid-infrared wavelengths (the bottom panel of Figure~\ref{fig:disk_summary}).
\end{itemize}

\section{Outflow Cavity Parameters}
\label{sec:cav_para}
The existence of the outflow cavity allows the radiation to escape from the center in bipolar directions.  We use a power-law profile with a power of $-1.5$ for the shape of the outflow cavity (Section \ref{sec:cavmod}). Therefore the only parameter that scales the shape of outflow cavity is the cavity opening angle.   Observing a 3-D object is a projection onto a plane so that the net effect of both cavity opening angle and inclination is to change the projected area of the outflow cavity, resulting in variations in mid-infrared emission.

\begin{figure}[htbp!]
  \centering
  \includegraphics[width=0.47\textwidth]{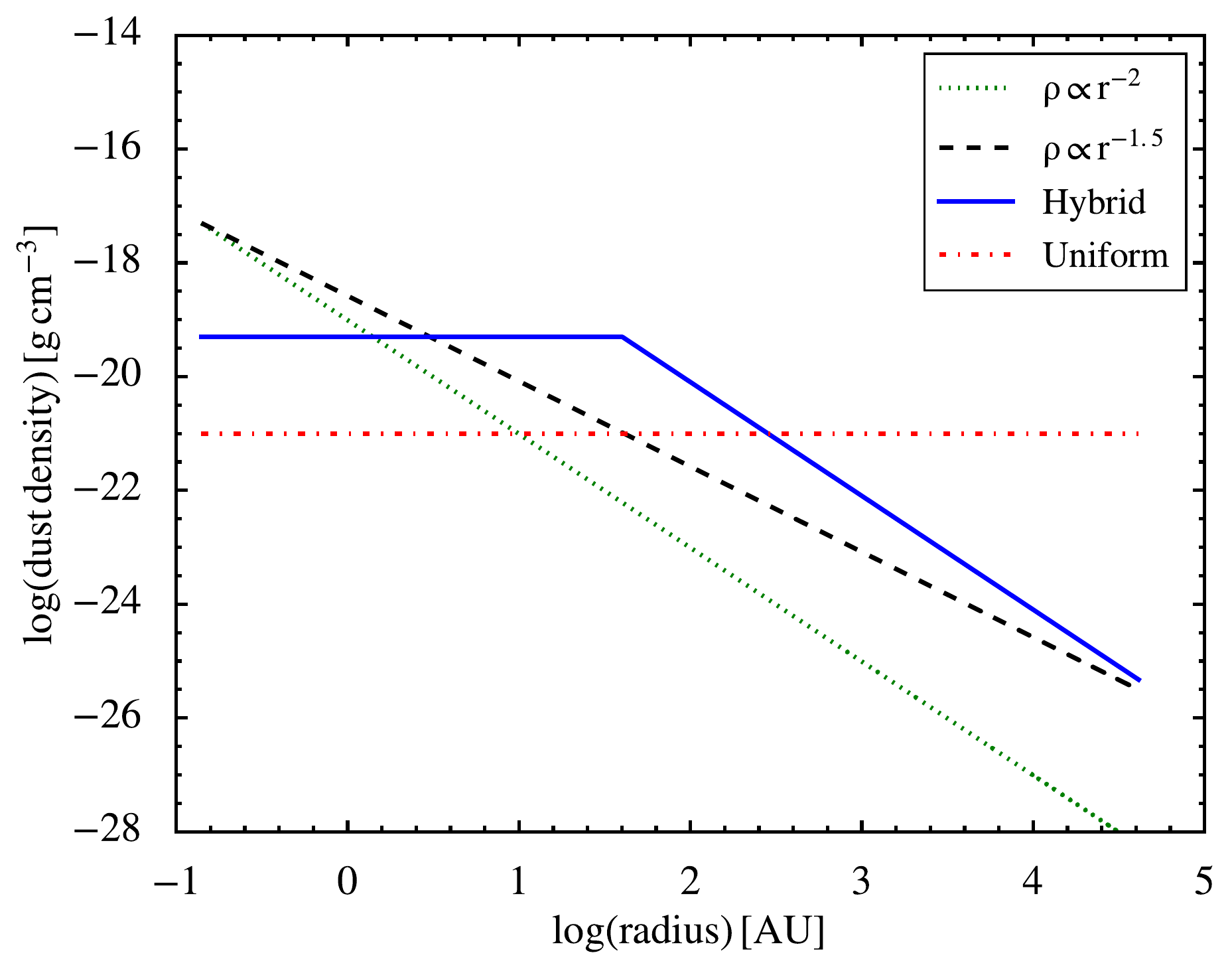}
  \caption{The radial density profiles with prescriptions of two power laws ($\rm r^{-2}$ and $\rm r^{-1.5}$) shown in black dahsed and green dotted lines, uniform density shown in red dot-dashed line, and the hybrid profile, which consists of a constant density region at the center followed by a power law decrease, shown in blue solid line.}
  \label{fig:cav_struc_density}
\end{figure}

We have tested several simple cavity density profiles of a uniform density, $\rm \rho(r)\propto r^{-2}$, $\rm \rho(r)\propto r^{-1.5}$, and a hybrid profile with a uniform density out to some radius, followed by a power law decrease (Figure~\ref{fig:cav_struc_density}).  The hybrid profile can be described with three parameters, the outer radius of the constant density region ($R_{\rm cav,\circ}$), the dust density in the constant region ($\rho_{\rm cav,\circ}$), and the cavity opening angle ($\theta_{\rm cav}$).
The simulated SEDs with the simple power-law cavity profiles ($r^{\rm -2}$ and $r^{\rm -1.5}$)  show only minimal changes.  We find that the ratio of the flux at 8~$\mu$m to the flux at 16~$\mu$m is most diagnostic for characterizing the effect of cavity profiles.  For the SEDs with a lower density at the center ($\rho_{\rm cav,\circ} = 10^{-21}~\rm g~cm^{-3}$), the flux ratio is approximately the same, while the ratio for the SEDs with a higher density at the center ($\rho_{\rm cav,\circ} =  10^{-18}~\rm g~cm^{-3}$) is larger.
This feature is seen for  the power law cavity profile with $r^{-2}$, but not in the profile with $r^{-1.5}$.  On the other hand, the hybrid profile can produce more emission at 10~\micron\ to 20~\micron, while having a similar emission at shorter wavelengths.

\begin{figure*}[htbp!]
  \centering
  \includegraphics[width=\textwidth]{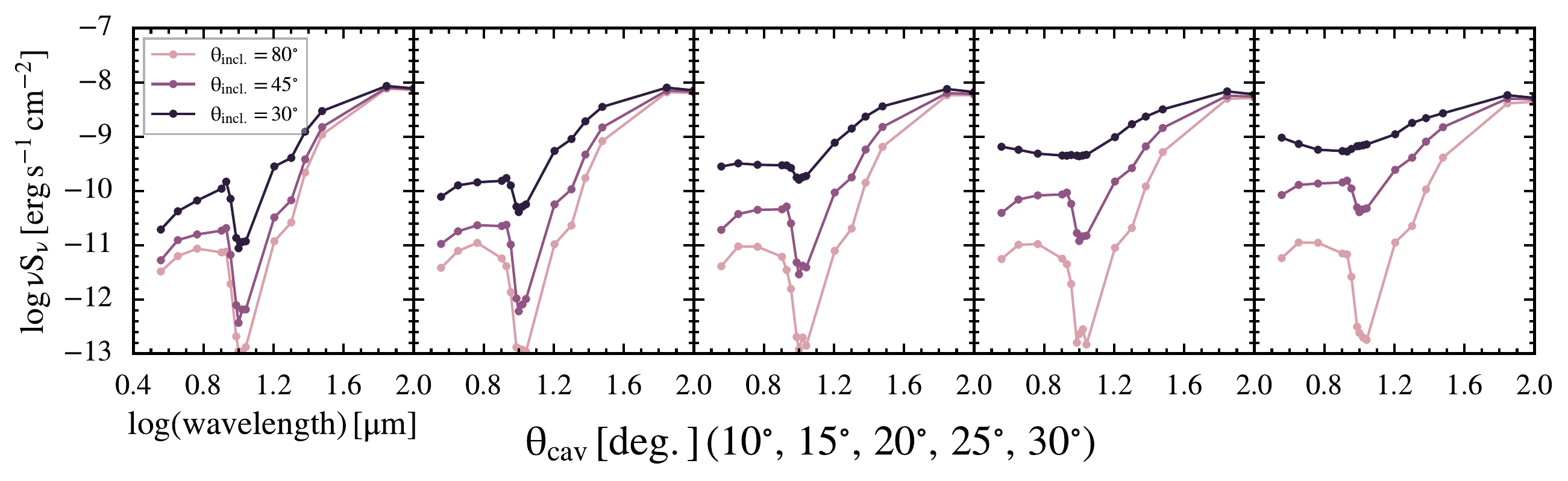}
  \caption{The simulated SEDs with the grid of cavity opening angle and the inclination angle parameters.  The purple dot/line shows the SED with aperture-extracted photometry from simulated spectra with the inclination angle of $80^{\circ}$, $40^{\circ}$, and $30^{\circ}$ shown in different transparency. The cavity opening angles are $10^{\circ}$, $15^{\circ}$, $20^{\circ}$, $25^{\circ}$, and $30^{\circ}$ from left to right.  Other parameters are adopted from Table~\ref{best_fit}.}
  \label{fig:sed_cav_incl}
\end{figure*}

The major effects of the outflow cavity parameters are at mid-infrared wavelengths, while submillimeter emission is unaffected; therefore we plot the SED out to only 100 \micron\ in the figures discussing the outflow cavity parameters.  Figure~\ref{fig:sed_cav_incl} shows a grid of SEDs with three inclinations ($80^{\circ}$, $40^{\circ}$, and $30^{\circ}$) and five cavity opening angles ($10^{\circ}$, $15^{\circ}$, $20^{\circ}$, $25^{\circ}$, and $30^{\circ}$).
The definition of inclination follows the convention in \hy, where $0^{\circ}$ means face-on and $90^{\circ}$ means edge-on.  For a certain opening angle, decreasing the inclination angle reduces the contrast between the mid-infrared and submillimeter emission and the depth of the silicate absorption feature.  At a given inclination, the amount of emission in IRAC bands (3.6~$\mu$m, 4.5~$\mu$m, 5.8~$\mu$m, and 8.0~$\mu$m) increases with opening angles, because the low density region is larger in a wider cavity and more emission escapes from the center with less attenuation.  Note that the SED at wavelengths from 20~$\mu$m to 40~$\mu$m is sensitive to the cavity opening angle, especially when the inclination angle is small (left two panels in Figure~\ref{fig:sed_cav_incl}).

The images at IRAC bands are direct probes of the inclination angle of  the outflow cavity.  \citet{2003ApJ...598.1079W} used two dimensional radiative transfer simulations to show that the contrast between two lobes of an outflow cavity is greater with a smaller inclination.  The same effect is also found in our simulated images.  We compare the flux ratio of the south and north lobe from simulations at various inclinations (Figure~\ref{fig:flux_NS}).  The flux ratio is a maximum at intermediate inclination angles.

\begin{figure*}[htbp!]
  \centering
  \includegraphics[width=0.8\textwidth]{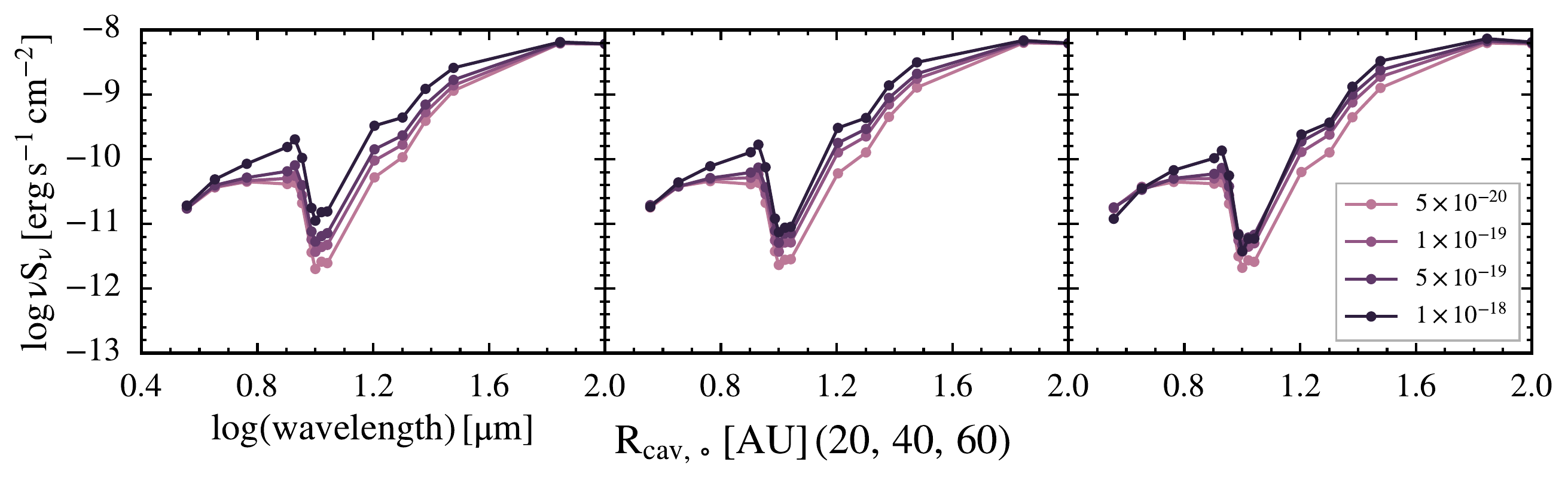}
  \caption{The simulated SEDs with the grid of the dust density in the inner cavity and the extent of the inner cavity region parameters.  The purple dot/line shows the SED with aperture-extracted photometry from simulated spectra with the innermost constant density of $5\times 10^{-20}$, $1\times 10^{-19}$, $5\times 10^{-19}$, and $1\times 10^{-18}~\rm g~cm^{-3}$ shown in different transparency.  The size of the innermost constant density region increases from the left to right, 20~AU, 40~AU, and 60~AU.  Other parameters are adopted from Table~\ref{best_fit}.}
  \label{fig:sed_centeredge}
\end{figure*}

Figure~\ref{fig:sed_centeredge} shows the simulated SEDs with a grid of the size of the constant density region in the cavity, $R_{\rm cav,\circ}$, and the density of the constant density region, $\rho_{\rm cav,\circ}$.
The models with $\rho_{\rm cav,\circ}$ of $5\times 10^{-20}$, $1\times 10^{-19}$, $5\times 10^{-19}$, and $1\times 10^{-18}~\rm g~cm^{-3}$ are shown in lines with different transparency, and the models with $R_{\rm cav,\circ}$ of 20~AU, 40~AU, and 60~AU are shown from left to right.
Both $\rho_{\rm cav,\circ}$ and $R_{\rm cav,\circ}$ change the emission from 5-40~\micron, a region where none of other parameters discussed so far have a significant effect.  These two parameters together define the total mass at the constant density region, which affects the emission at 5-40~\micron.  It is shown that a higher $\rho_{\rm cav,\circ}$ increases the overall flux at 5-40~\micron, while $R_{\rm cav,\circ}$ has the same effect but is less significant.  Note that the fluxes at 3.6~\micron\ and 4.5~\micron\ are not affected by these two parameters.

The effects of outflow cavity parameters can be summarized as follows:
\begin{itemize}
  \item $\theta_{\rm cav}$:  Increasing the cavity opening angle increases the emission at wavelengths from 5~$\mu$m to 40~$\mu$m.  However, the inclination angle must be considered at the same time.  Increasing the inclination angle decreases the emission at wavelengths from 5~$\mu$m to 40~$\mu$m.
  \item $\rho_{\rm cav, \circ}$: It changes the absolute level of the SED at wavelengths from 5~$\mu$m to 40~$\mu$m.  A larger $\rho_{\rm cav, \circ}$ leads to a higher flux level at mid-infrared wavelengths.
  \item $R_{\rm cav, \circ}$: It also changes the absolute level of the SED at wavelengths from 5~$\mu$m to 40~$\mu$m, but it is less effective than modifying $\rho_{\rm cav, \circ}$.
\end{itemize}

\section{Other parameters}
\label{sec:other_para}

\begin{figure}[htbp!]
  \centering
  \includegraphics[width=0.47\textwidth]{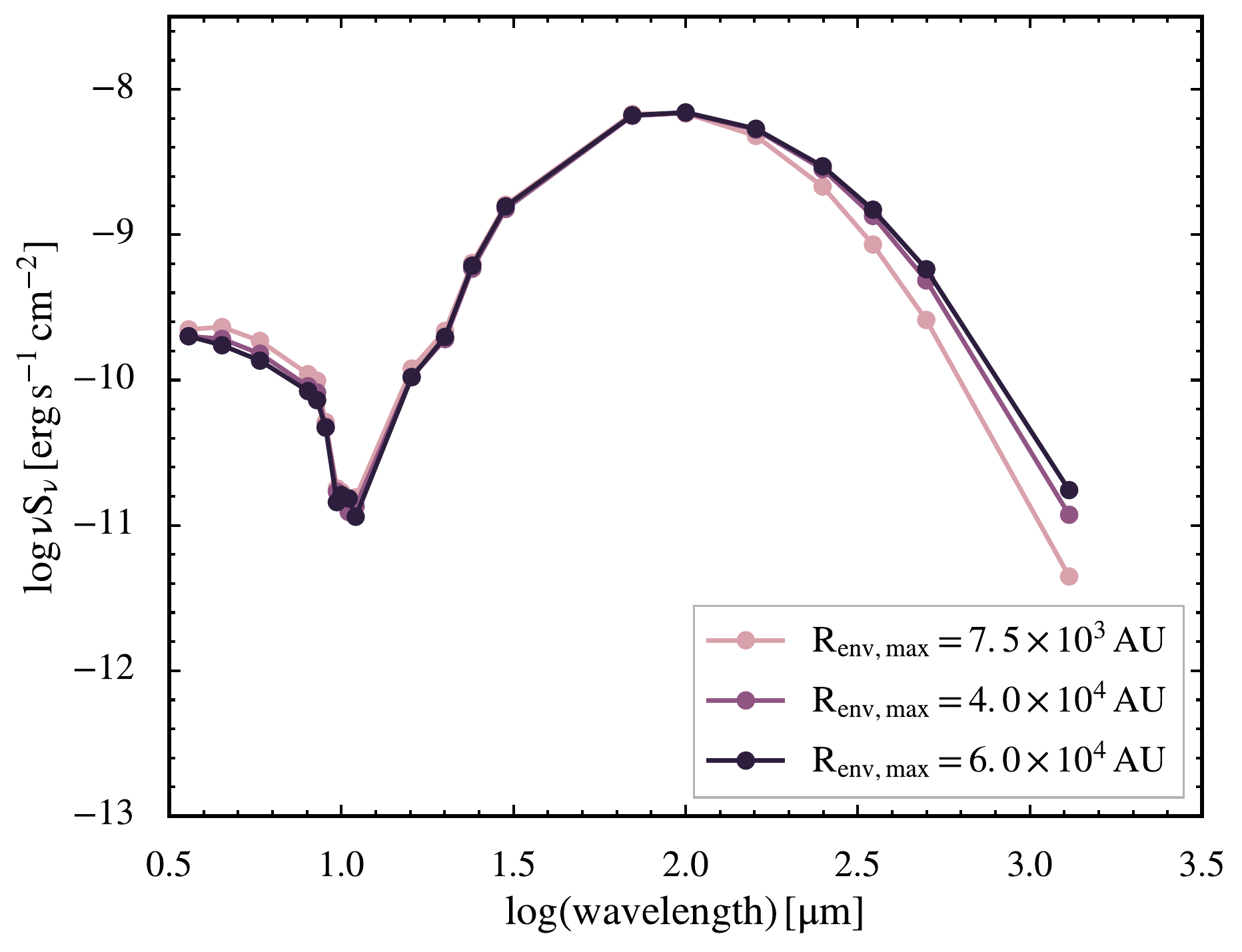}
  \caption{
  The simulated SEDs with outer envelope radius of $7.5\times 10^{3}$ AU, $1\times 10^{4}$ AU, and $2.5\times 10^{4}$ AU.  The blue dot/line shows the SED with
  the extracted photometry from the simulated spectra \textit{without} a specific aperture.    Other parameters are adopted from Table~\ref{best_fit}.}
  \label{fig:sed_rmax}
\end{figure}

Model parameters in this category are less constrained by the observations in this study.  Figure~\ref{fig:sed_rmax} shows the effect of the envelope outer radius.  Increasing the size of the envelope adds extinction, resulting in less emission at mid-infrared wavelengths.  On the other hand, the emission at submillimeter wavelengths increases with the envelope outer radius, probing the total mass in the envelope.  Note that the simulated spectra shown in Figure~\ref{fig:sed_rmax} are
from the full model, rather than values within apertures to better emphasize the effect of the envelope outer radius. If the apertures are smaller than the source
size the effect of increasing the outer radius on the SED is much less noticeable.

\begin{figure}[htbp!]
  \centering
  \includegraphics[width=0.47\textwidth]{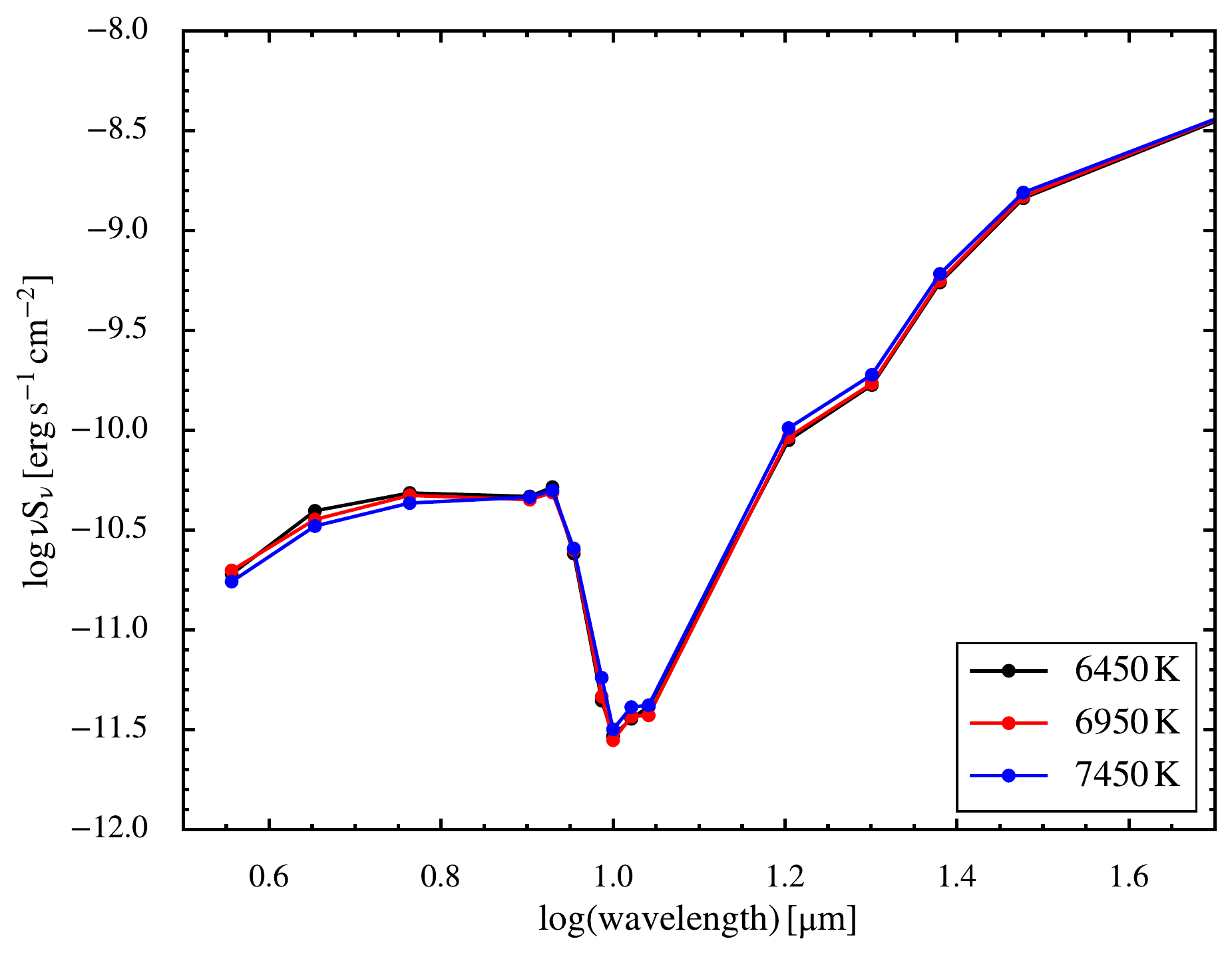}
  \caption{The simulated SEDs with the central star temperature of 6450~K, 6950~K, and 7450~K, while the central luminosity remains the same, 18.8~\lsun.  The protostar radius is calculated accordingly to satisfy the same central luminosity.  The simulated SEDs show no difference beyond 50~\micron.}
  \label{fig:sed_lstar}
\end{figure}

The other two parameters, $T_{\star}$ and $R_{\star}$, are coupled in determining the central luminosity assuming a pure blackbody spectrum (Equation \ref{eq:central_lun}).  The protostar radius also enters the disk density profile (Equation \ref{eq:disk}).  Since the central protostar is highly embedded within the envelope, there is no direct measurement of its physical properties.
In our model, these two parameters are degenerate except that the luminosity, which is calculated from $T_{\star}$ and $R_{\star}$, is required to make the luminosity from the simulated observations fit the observed luminosity.  Figure~\ref{fig:sed_lstar} shows the simulated SEDs with protostar temperatures of 6450~K, 6950~K, and 7450~K, while the luminosity is fixed.  The protostar radius is calculated accordingly.  We find that the simulated SEDs show a subtle increase at wavelengths below 10~$\mu$m as we decrease the stellar temperature and a slight increase at wavelengths between 20~$\mu m$ to 40~$\mu m$, suggesting that a small fraction of emission at near-infrared may be scattered light from the central luminosity source; therefore a lower stellar temperature leads to more stellar emission at near-infrared wavelength, resulting in a slight increase of emission at near-infrared wavelengths of the simulated SEDs.


The effect of other parameters can be summarized as follows:
\begin{itemize}
  \item $R_{\rm max}$: Increasing the outer radius of the envelope decreases the emission at wavelengths from 5~$\mu$m to 15~$\mu$m and increases the total emission at the longest wavelengths.
  \item $T_{\star}$:  Increasing the temperature of the star slightly decreases the emission at wavelengths below 10~$\mu$m, when the emission at 20~$\mu m$ to 40~$\mu m$ shows a slight increase.
\end{itemize}

\end{document}